\documentclass[11pt,prd,groupedaddress,showpacs,doublecolumn,nofootinbib]{revtex4-1}

\usepackage{graphicx}
\usepackage{dcolumn}
\usepackage{bm}
\usepackage{amssymb}
\usepackage{amsmath}
\usepackage{epsfig}
\usepackage{hyperref}
\usepackage{hhline}
\usepackage{color}
\usepackage{multirow}
\usepackage[normalem]{ulem}
\usepackage{slashed}
\usepackage{slashed}
\usepackage{tikz}
\usetikzlibrary{snakes}
\usepackage{blindtext}
\usepackage{hyperref}
\usepackage[section]{placeins}

\definecolor{Zsug}{RGB}{0, 145, 33} 
\definecolor{Zcor}{RGB}{210, 0, 210}
\definecolor{Zque}{RGB}{0, 180, 190} 
 
\definecolor{jd}{rgb}{0.858, 0.188, 0.478}

\def\lapp{\mathrel{\rlap{\raise.5ex\hbox{$<$}}
                    {\lower.5ex\hbox{$\sim$}}}}
\def\gapp{\mathrel{\rlap{\raise.5ex\hbox{$>$}}
                    {\lower.5ex\hbox{$\sim$}}}}

{\newcommand{\lsim}{\mbox{\raisebox{-.6ex}{~$\stackrel{<}{\sim}$~}}}
{\newcommand{\gsim}{\mbox{\raisebox{-.6ex}{~$\stackrel{>}{\sim}$~}}}

\newcommand{\bmt}{\begin{pmatrix}}
\newcommand{\emt}{\end{pmatrix}}
\newcommand{\ba}{\begin{array}{c}}
\newcommand{\ea}{\end{array}}
\newcommand{\be}{\begin{equation}}
\newcommand{\ee}{\end{equation}}
\newcommand{\bea}{\begin{eqnarray}}
\newcommand{\eea}{\end{eqnarray}}

\newcommand{\bi}{\begin{itemize}}
\newcommand{\ei}{\end{itemize}}

\newcommand{\baz}{\begin{array}{cc}}

\newcommand{\mathsym}[1]{{}}

\newcommand{\bt}{\begin{tabular}}
\newcommand{\et}{\end{tabular}}

\newcommand{\benu}{\begin{enumerate}}
\newcommand{\eenu}{\end{enumerate}}

\newcommand{\bav}{\begin{array}{cccc}}

\begin{document}

\renewcommand*{\thefootnote}{\fnsymbol{footnote}}

\begin{center}
 {\Large\bf Fermion Dark Matter with Scalar Triplet at Direct and Collider Searches}
 \\
 \vskip .5cm
 {
 Basabendu Barman$^{a,}$\footnote{bb1988@iitg.ac.in},
 Subhaditya Bhattacharya$^{a,}$\footnote{subhab@iitg.ac.in},
 Purusottam Ghosh$^{a,}$\footnote{p.ghosh@iitg.ac.in},
 Saurabh Kadam$^{b,}$\footnote{saurabh.kadam@students.iiserpune.ac.in},
 Narendra Sahu$^{c,}$\footnote{nsahu@iith.ac.in}
 }\\[3mm]
 {\it{
 $^a$ Department of Physics, Indian Institute of Technology Guwahati, Assam 781039, India \\
 $^b$Department of Physics,Indian Institute of Science Education and Research, Dr. Homi Bhabha Road, Pashan, Pune-411008, India}\\
 $^c$ Department of Physics, Indian Institute of Technology Hyderabad, Kandi, Sangareddy, Telangana State-502285, India 
 }
 \end{center}
\vspace{1cm}

\begin{center}
 {\bf{Abstract}}
\end{center}

\begin{abstract}
Fermion dark matter (DM) as an admixture of additional singlet and doublet vector like fermions provides an attractive and allowed framework by relic density and 
direct search constraints within TeV scale, although limited by its 
discovery potential at the Large Hadron Collider (LHC) excepting for a displaced vertex signature of charged vector like lepton. An extension of the 
model with scalar triplet can yield neutrino masses and provide some cushion to the 
direct search constraint of the DM through pseudo-Dirac mass splitting. This in turn, allow the model to live in a larger region of the parameter 
space and open the door for detection at LHC through hadronically quiet dilepton channel, 
even if slightly. We also note an interesting consequence to the hadronically quiet four lepton signal produced by the doubly charged scalar 
belonging to the triplet, in presence of additional vector like fermions as in our model. The model however can see an early discovery at International Linear Collider 
(ILC) without too much of fine-tuning. The complementarity of LHC, ILC and direct search prospect of this framework is studied in this paper. 
\end{abstract}



\maketitle
\flushbottom


\setcounter{footnote}{0}
\renewcommand*{\thefootnote}{\arabic{footnote}}

\section{Introduction}
\label{sec:intro}

The existence of dark matter (DM) on a larger scale ($>$ a few kpc) is irrefutably shown by many evidences, such as galaxy rotation curve, gravitational lensing, existence of large scale structure of the Universe, cosmic microwave background {\it etc} (See for a review~\cite{Jungman:1995df,Bertone:2004pz,Conrad:2014tla,Gaskins:2016cha}). In fact, the satellite borne experiments, such as WMAP~\cite{Hinshaw:2012aka} and PLANCK~\cite{Akrami:2018odb}, which study the temperature fluctuations in the cosmic microwave background, precisely measure the current relic density of DM in terms of a dimensionless parameter $\Omega_{\rm DM}h^2= 0.1199\pm 0.0027$, where $\Omega_{\rm DM}=\rho_{\rm DM}/\rho_c$; $\rho_c$ being the critical density of the Universe and $h\approx 0.73$ is a parameter which defines the current Hubble scale of expansion $H_0=100 h$ km/s/Mpc. However, the above mentioned evidences are based on gravitational interaction of DM and pose a challenge for particle physicists to probe it on an earth-based laboratory where the DM density is extremely low in comparison to baryonic matter. Of many possibilities, a weakly interacting massive particle (WIMP)~\cite{Jungman:1995df,Kolb:1990vq} is an elusive candidate for DM \footnote {The other possible candidates for DM may also come from feebly interacting massive particle (FIMP)~\cite{Hall:2009bx}, or strongly interacting massive particle (SIMP)~\cite{Hochberg:2014dra} with limited experimental probe.}. Due to the additional weak interaction property,  WIMPs can interact with the standard model (SM) particles at a short distance and can thermalise in the early Universe at a temperature above its mass scale. As the Universe expands and cools down, the WIMP density freezes out at a temperature below its mass scale. In fact, the freeze-out density of WIMP matches to a good accuracy with the experimental value of relic density obtained by PLANCK. The weak interaction property of WIMP DM is currently under investigation at direct search experiments such as LUX~\cite{Akerib:2017kat}, PANDA~\cite{Zhang:2018xdp}, XENON1T~\cite{Aprile:2018dbl} as well as collider search experiments such as~\cite{Lowette:2014yta,Ahuja:2018bbj}. 

At present the SM of particle physics is the best theory to describe the fundamental particles and their interactions in nature. After 
the Higgs discovery, the particle spectrum of the SM is almost complete. However, the SM does not possess a candidate that can mimic the 
nature of DM inferred from astrophysical observations. Moreover, the SM does not explain the sub-eV masses of the active left-handed neutrinos which is required to explain observed solar and atmospheric oscillation 
phenomena~\cite{ PhysRevD.98.030001}. Therefore, it is crucial to explore physics beyond the SM to incorporate at least non-zero masses of active neutrinos as well as dark matter content of the Universe. It is quite possible 
that the origin of DM is completely different from neutrino mass. However, it is always attractive to find a simultaneous solution for non-zero neutrino mass and dark matter in a single platform with a minimal extension of the 
SM~\cite{Ma:2006km,Ma:2018zuj}. 

Till date, the only precisely measured quantity related to DM known to us is its relic density. The microscopic nature of DM is hitherto not known. Amongst many possibilities to accommodate DM in an extension of SM, 
a simple possibility is to extend the SM with two vector-like fermions: $\chi^0 (1,1,0)$ and $\psi(1,2,-1)$, where the numbers inside the parentheses are the quantum numbers under the SM gauge group 
$SU(3)_C\times SU(2)_L \times U(1)_Y$~\cite{Bhattacharya:2015qpa,Bhattacharya:2017sml,Bhattacharya:2018cgx,Bhattacharya:2018fus}. The lightest component in the mixture of the neutral component 
of the doublet $\psi$ and the singlet $\chi^0$ gives rise to a viable DM candidate. The stability of the lightest component can be ensured by an added $\mathcal{Z}_2$ symmetry. The singlet-doublet mixing, 
defined by $\sin \theta$, plays an important role in probing the DM at direct and collider search experiments. A large singlet-doublet mixing ($\sin \theta > 0.1$) introduces a larger doublet component and hence 
strongly constrained by the Z-mediated DM-nucleon scattering at direct search experiments, while small mixing ($\sin \theta < 10^{-5}$) leads to over production of DM after big bang nucleosynthesis (BBN) by the 
decay of the next-to-lightest-stable particle (NLSP) $\psi^\pm$, the charged component of doublet $\psi$. Therefore, the singlet-doublet mixing in a range: $10^{-5} < \sin \theta < 0.05$~\cite{Bhattacharya:2015qpa} 
is appropriate to give rise to correct relic density of the DM while being compatible with the latest bound from direct search experiments such as. It is important to note that due to the small mixing, the annihilation cross-section 
of the DM is not enough to acquire correct relic density, which requires contribution from co-annihilation with NLSP resulting to a small mass splitting between NLSP and DM. The collider search of such a framework 
is therefore narrowed down to only a displaced vertex signature of the NLSP: $\psi^\pm$. 

In this paper we study the detector accessibility of the singlet-doublet DM in presence of a scalar triplet $\Delta (1,3,2)$, where the quantum 
numbers are with respect to the SM gauge group $SU(3)_C\times SU(2)_L \times U(1)_Y$. We demand that the scalar triplet should not acquire any 
explicit vacuum expectation value (vev) as in the case of type-II seesaw~\cite{Goh:2004fy,Caetano:2012qc}. However, after the electroweak phase transition the $\Delta$ can acquire an induced vev of sub-GeV order in order to be 
compatible with precision electroweak data $\rho\simeq1$ in the SM. As a result the symmetrical coupling of $\Delta$ with the SM lepton doublet $L$ can give rise to sub-eV Majorana masses for the active neutrinos. Moreover, 
we show that the scalar triplet widens up the allowed parameter space through pseudo-Dirac splitting of the DM, which makes the direct search through inelastic $Z$ mediation harder. Aided by that, the model can acquire correct 
density and still obey direct search constraints for larger singlet doublet mixing as well as with larger mass splitting between NLSP and DM. This can yield leptonic signature excess through hadronically quiet opposite sign dilepton 
(OSD) at LHC. The model also has the advantage of searching for the NLSP $\psi^\pm$ decaying to DM through the same OSD channel at the ILC. The Complementarity of the discovery potential of the model at the LHC and the 
ILC, in comparison to that of direct search, is analyzed in detail in this paper.

The doubly charged Higgs belonging to the scalar triplet is known to produce hadronically quiet four lepton signal at LHC and ILC~\cite{Ghosh:2017pxl,Agrawal:2018pci}. Therefore, in addition to hadronically quiet dilepton channel, four lepton will 
also be a signature of the model in presence of a scalar triplet. Here we point out that whenever the scalar triplet mass is heavier than twice of the vector like fermion mass, a sizeable branching fraction of the doubly charged scalar allows it to decay to two charged vector like lepton and thereafter produce hadronically quiet four lepton (HQ4l) signature through charged lepton decay. 
This feature not only allows the four lepton signature of our model distinguishable from SM background, but also it segregates our case from the usual four lepton event rates of scalar triplet in Type II seesaw scenario.

The paper is organized as follows: in Sec.~\ref{sec:model}, we discuss the important aspects of the model. Sec.~\ref{sec:constraint} deals with the constraints on the model parameters. Then we discuss the DM phenomenology in Sec.~\ref{sec:dmpheno}, where we demonstrate the model parameter space compatible with the observed relic density and latest direct search experiments. Sec.~\ref{sec:collider} is then devoted to find relevant collider signatures. In sectio~\ref {sec:complmntr}, we discuss the Complementarity of the discovery potential of the model at the LHC and the ILC while being compatible with DM constraints. Finally we conclude in Sec.~\ref{sec:conclusion}. 

\section{The Model}\label{sec:model}

\subsection{Fields and interactions}
\label{sec:lagrange}
We extend the Standard Model (SM) by introducing two vector like fermions (VLF):  one singlet ($\chi^0$) and a doublet $\psi$. In addition to that we 
introduce a scalar triplet ($\Delta$) with hypercharge $Y=2$. A discrete $\mathcal{Z}_2$ symmetry is imposed on top of the SM gauge symmetry, under which 
the VLFs are odd, while other fields, including $\Delta$, are even to stabilize the DM from decay. The charges of the new particles as well as that of the 
SM Higgs under $SU(3)_c\times SU(2)\times U(1)_Y \times \mathcal{Z}_2$  are given in Table \ref{tab:charges}. 
\begin{table}[htb]
 \begin{center}
 \begin{tabular}{|c| c| c| c|c|} 
 \hline
 Particles & $SU(3)_c$ & $SU(2)$ & $U(1)_Y$ & $\mathcal{Z}_2$ \\ [0.5ex] 
 \hline\hline
 $\psi^T:\left(\psi^0,\psi^{-}\right)$ & 1 & 2 & -1 & -1 \\ 
 \hline
 $\chi^0$ & 1 & 1 & 0 & -1 \\
 \hline
 $\Delta$ & 1 & 3 & 2 & +1 \\
 \hline
 $H$ & 1 & 2 & 1 & +1 \\
 \hline
 \hline
 \end{tabular}
\end{center}
\caption{Relevant particle content of the model and their charges under $\rm SM\times \mathcal{Z}_2$.}
\label{tab:charges}
\end{table}
The Lagrangian for this model is given as:
\bea
\mathcal{L} = \mathcal{L}_{SM}+\mathcal{L}_{f}+\mathcal{L}_{s}+\mathcal{L}_{yuk}~,
\label{eq:lagrang}
\eea
where $\mathcal{L}_f$ is the Lagrangian for the VLFs, $\mathcal{L}_s$ involves the SM doublet and the additional triplet scalar, and $\mathcal{L}_{yuk}$ contains the Yukawa interaction terms. 
The interaction Lagrangian for the VLFs is given by~\cite{Bhattacharya:2017sml,Bhattacharya:2018cgx}:

\bea
\mathcal{L}_f =  \bar{\psi}\slashed{D}\psi + \bar{\chi^0}\slashed{\partial}\chi^0 - M_{\psi}\bar{\psi}\psi - M_{\chi}\bar{\chi^0}\chi^0~,
\label{eq:lfermi}
\eea

where $D_{\mu}$ is the covariant derivative under $SU(2)\times U(1)$ and is given by:

\bea
\begin{split}
D_{\mu}\psi &= \partial_{\mu}\psi-i g \frac{\sigma^{a}}{2} W_{\mu}^a\psi + i \frac{g^{'}}{2} B_{\mu}\psi~,
\label{eq:covderiv}
\end{split}
\eea

where $g$ and $g^{'}$ are the gauge couplings corresponding to $SU(2)$ and $U(1)_Y$ and $a=1,2,3$, for the generators of $SU(2)$. 
$W_{\mu}$ and $B_{\mu}$ are the gauge bosons corresponding to SM $SU(2)$ and $U(1)_Y$ gauge groups. Similarly, DM realization as an admixture of fermion singlet and triplet~\cite{Choubey:2017yyn}, as well as a doublet and triplet have also been addressed~\cite{Dedes:2014hga}. 
Lagrangian of the scalar sector involving SM Higgs doublet ($H$) and the additional scalar triplet ($\Delta$) can be written as~\cite{Arhrib:2011uy}:

\bea
\mathcal{L}_s = \left(D^{\mu}H\right)^{\dagger}\left(D_{\mu}H\right) + Tr \left[\left(D^{\mu}\Delta\right)^{\dagger}\left(D_{\mu}\Delta\right)\right]-V(H,\Delta).
\label{eq:lscalar}
\eea

The covariant derivatives of the scalars are:
\bea
D_{\mu}H &=&\partial_{\mu}H- i g\frac{\sigma^{a}}{2} W_{\mu}^a H - \frac{i g^{'}}{2} B_{\mu}H \nonumber,\\
D_{\mu}\Delta &=& \partial_{\mu}\Delta - i g\left[\frac{\sigma^{a}}{2} W_{\mu}^a ,\Delta\right] - \frac{i g^{'}}{2} Y_\Delta B_{\mu}\Delta ~.
\label{eq:covderivtrip}
\eea
$\Delta$ is written in the adjoint representation of $SU(2)$ as follows:
\bea
\Delta =   \quad
\begin{pmatrix} 
\Delta^{+}/\sqrt{2} & \Delta^{++}\\
\Delta^{0} & -\Delta^{+}/\sqrt{2}
\end{pmatrix}.
\quad
\label{eq:triplet}
\eea

The most general scalar potential for this model with scalar triplet ($\Delta$) of hypercharge $Y=2$ can be written as~\cite{Arhrib:2011uy}:
\bea
\begin{split}
V (H,\Delta) &= -\mu_H^2 H^{\dagger}H + \frac{\lambda}{4}\left(H^{\dagger}H\right)^2+\mu_{\Delta}^2 Tr\left(\Delta^{\dagger}\Delta\right)+ \left[\mu\left(H^T i \sigma^2 \Delta^{\dagger}H\right)+h.c.\right]\\ & \ + \lambda_{1} \left(H^{\dagger}H\right)Tr\left(\Delta^{\dagger}\Delta\right) + \lambda_2 \left(Tr[\Delta^{\dagger}\Delta]\right)^2+\lambda_3 ~Tr [\left(\Delta^{\dagger}\Delta\right)^2] + \lambda_4 ~\left(H^{\dagger}\Delta\Delta^{\dagger}H \right). 
\end{split}
\label{eq:pot}
\eea

Finally, the Yukawa interaction is given by~\cite{Bhattacharya:2017sml}:
\bea
-\mathcal{L}_{yuk} = \frac{1}{\sqrt{2}}\left[(y_L)_{ij}\bar{L^c}_{i}i \sigma^2\Delta L_{j}+y_{\psi} \bar{\psi^c}i \sigma^2\Delta\psi+h.c.\right]+\left(Y\bar{\psi}\widetilde{H}\chi^0+h.c.\right),
\label{eq:lyuk}
\eea
where in the first parenthesis we have the interaction between the triplet scalar ($\Delta$) with the SM lepton doublet ($L$) proportional to $y_L$ where the indices ($i,j$) run over three families 
and also the Yukawa interaction with the VLF doublet ($\psi$) proportional to $y_\psi$. In the second parenthesis we have the VLF-SM Higgs Yukawa interaction proportional to the coupling strength $Y$, 
where $\widetilde{H}=i\sigma^2 H^{*}$.  

The electroweak symmetry breaking (EWSB) occurs when the SM Higgs acquires a VEV ($v_d$) given by: 
\begin{equation}
\langle H \rangle= 
\begin{pmatrix} 
0  \\
\frac{v_d}{\sqrt{2}}
\end{pmatrix}.
\end{equation}

We assume that $\Delta$ does not acquire any explicit vev. However, the vev of SM Higgs induces a small vev to the scalar triplet $\Delta$ ($v_t$) given by:
\begin{equation}
\langle\Delta\rangle = 
\quad
\begin{pmatrix} 
0 & 0 \\
\frac{v_t}{\sqrt{2}} & 0
\end{pmatrix}\,.
\end{equation}

The alignment of the two vevs may not be same. Therefore, it is convenient to define $v=\sqrt{v_d^2+2 v_t^2}=246~\rm GeV$. After minimization of 
the potential in Eq.~\ref{eq:pot}, one arrives at the following necessary conditions~\cite{Arhrib:2011uy}:
\bea
\begin{split}
\mu_{\Delta}^2 &=\frac{2\mu v_d^2-\sqrt{2}\left(\lambda_1+\lambda_4\right)v_d^2 v_t-2\sqrt{2}\left(\lambda_2+\lambda_3\right)v_t^3}{2\sqrt{2}v_t} \\ & \mu_H^2=\frac{\lambda v_d^2}{4}-\sqrt{2}\mu v_t+\frac{\left(\lambda_1+\lambda_4\right)v_t^2}{2}.
\end{split}
\label{eq:minim1}
\eea

\subsection{Mixing of the doublet and triplet scalar}
\label{sec:scalarmix}

In the scalar sector, masses of the doubly and singly-charged fields corresponding to the triplet can be found in ~\cite{Arhrib:2011uy} and are as follows: 
\bea
m_{H^{\pm \pm}}^2= \frac{\sqrt2 \mu v_d^2 - \lambda_4 v_d^2 v_t-2 \lambda_3 v_t^2}{2 v_t}, ~~ m_{H^\pm}^2=\frac{(v_d^2+2v_t^2)(2\sqrt2 \mu -\lambda_4 v_t)}{4 v_t}
\eea

The neutral scalar sector consists of  CP-even and CP-odd mass matrices as: 

\bea
\mathcal{M}^2_{CP_{even}}=
\quad
\begin{pmatrix} 
P & Q \\
Q & R
\end{pmatrix},
\quad~\mathcal{M}^2_{CP_{odd}}=
\quad
\begin{pmatrix} 
2v_t & -v_d \\
-v_d & v_d^2/2v_t
\end{pmatrix},
\quad
\label{eq:scalarmass}
\eea

where
\bea
P=\frac{\lambda}{2}v_d^2,~ Q=v_d\left(-\sqrt{2}\mu\left(\lambda_1+\lambda_2\right)v_t\right)~{\rm and}~R=\frac{\sqrt{2}\mu v_t^2+4\left(\lambda_2+\lambda_3\right)v_t^3}{2 v_t}.
\eea

The CP-even mass matrix is diagonalized using the orthogonal matrix:

\bea
\mathcal{U} = 
\quad
\begin{pmatrix} 
\cos\theta_0 & \sin\theta_0  \\
-\sin\theta_0  & \cos\theta_0 
\end{pmatrix},
\quad
\label{eq:umatscalar}
\eea

where $\theta_0$ is the mixing angle. Upon diagonalization, we end up with the following physical CP-even eigenstates:

\bea
H_1 = \cos\theta_0 h + \sin\theta_0 \zeta^0,~H_2 = \sin\theta_0 h + \cos\theta_0 \zeta^0,
\label{eq:physcal}
\eea

where $h$ and $\zeta^0$ are the real parts of $H^0$ and $\Delta^0$ fields, shifted by their respective VEVs as:

\bea
H^0 = \frac{1}{\sqrt{2}}\left(v_d+h+i \eta_1\right),~\Delta^0 = \frac{1}{\sqrt{2}}\left(v_t+\zeta+i \eta_2\right).
\label{eq:scalarfield}
\eea
As it is evident from Eq.~\ref{eq:physcal}, under small mixing approximation, $H_1$ acts like SM Higgs, while $H_2$ behaves more like a heavy Higgs. We call $H_2$ heavy as we have not observed any such 
neutral scalar in experiments yet and is therefore limited by a lower mass limit as we discuss next in the constraints section. The mixing angle in the CP-even scalar sector is given by:

\bea
\tan 2\theta_0 = \frac{2 Q}{P-R}.
\eea
The CP-odd mass matrix, on diagonalization, gives rise to a massive physical pseudoscalar ($A_0$) with mass:
\bea
m^2_{A_0} = \frac{\mu\left(v_d^2+4 v_t^2\right)}{\sqrt{2} v_t},
\eea
and another massless Goldstone boson. 
Therefore, after EWSB, the scalar spectrum contains seven massive physical Higgs bosons: two doubly charged ($H^{\pm\pm}$), two singly charged ($H^{\pm}$), two CP-even neutral Higgs ($H_1,H_2$) and a CP-odd Higgs ($A_0$). 
All the couplings, which can be casted in terms of the physical masses appearing in the scalar potential are listed in~\ref{sec:cplings}.

\subsection{Mixing of the VLFs}
\label{sec:vlfmix}

The neutral components of the doublet ($\psi^0$) and singlet ($\chi^0$) mix after EWSB thanks to the Yukawa interaction (Eq.~\ref{eq:lyuk}). 
The mass matrix can be diagonalized in the usual way using orthogonal rotation matrix to obtain the masses in the physical basis $(\psi_1,\psi_2)^T$:

\bea
\quad
\begin{pmatrix} 
M_{\psi_1} & 0 \\
0 & M_{\psi_2} 
\end{pmatrix}
\quad = ~\mathcal{U^T}\quad
\begin{pmatrix} 
M_{\psi} & m \\
m & M_{\chi} 
\end{pmatrix}
\quad \mathcal{U},
\label{eq:massmatrix}
\eea
where the non-diagonal mass term is obtained by $m=Y v_d/\sqrt{2}$, from Eq.~\ref{eq:lyuk} and the rotation matrix is given by
$\mathcal{U}=\begin{pmatrix} 
\cos\theta & \sin\theta \\
-\sin\theta  & \cos\theta 
\end{pmatrix}
\quad$. The mixing angle can be related to the mass terms as:
\bea
\tan 2\theta=\frac{2 m}{M_{\psi}-M_{\chi}}.
\label{eq:mixangle}
\eea

Therefore, the physical eigenstates (in mass basis) are the linear superposition of the neutral weak eigenstates and are given in terms of the mixing angle:
\bea
\psi_1 &= \cos\theta\chi^0+\sin\theta\psi^0,~
\psi_2 = -\sin\theta\chi^0+\cos\theta\psi^0.
\label{eq:vlfmix}
\eea
The lightest electromagnetic charge neutral $Z_2$ odd particle is a viable DM candidate of this model and we choose it to be $\psi_1$. 
The charged component of the VLF doublet $\psi^{\pm}$ acquires a mass as (in the small mixing limit):
\bea
M_{\psi^{\pm}} = M_{\psi_1}\sin^2\theta+M_{\psi_2}\cos^2\theta\approx M_{\psi_2}.
\eea

From Eq.~\ref{eq:mixangle}, we see that the VLF Yukawa is related to the mass difference between two physical eigenstates and is no more an independent parameter:
\bea
Y = \frac{(M_{\psi_2}-M_{\psi_1})\sin 2\theta}{\sqrt{2} v_d} = \frac{\Delta M\sin 2\theta}{\sqrt{2} v_d}.
\label{eq:vlfyuk}
\eea

Therefore, to summarize the model section, we see that the model provides with a fermion DM ($\psi_1$) which is an admixture of the doublet and singlet VLFs, with additional charged and neutral heavy fermions which all have Yukawa and gauge interactions with SM. On the other hand, the scalar sector is more rich with the presence of additional triplet which not only provides additional charged and neutral heavy scalar fields but also, have interactions to the dark sector through the Yukawa coupling. The model has several independent parameters and they are as follows:
\bea
\{~M_{\psi_1}, ~\Delta M, ~\sin\theta, ~y_L, ~ y_{\psi}, ~m_{H_2}, ~m_{A}, ~m_{H^\pm}, ~m_{H^{\pm\pm}},~\sin\theta_0~\}
\eea

We vary some of these relevant parameters to find relic density and direct search allowed parameter space of the model to proceed further for discovery potential of the framework at collider. 

\section{Constraints on model parameters}
\label{sec:constraint}

In this section we will discuss the possible constraints appearing on the parameters of this model from various theoretical and experimental bounds. 

\subsection*{Stability}
\label{sec:stability}

In order the potential to be bounded from below, the quartic couplings appearing in the potential must satisfy the following co-positivity condition~\cite{Kannike:2012pe,Arhrib:2011uy}:

\bea
\nonumber
\lambda>0,~~\lambda_2+\lambda_3>0,~~\lambda_2+\frac{\lambda_3}{2}>0 \\ 
\nonumber
\lambda_1+\sqrt{\lambda\left(\lambda_2+\lambda_3\right)}>0,~~\lambda_1+\sqrt{\lambda\left(\lambda_2+\frac{\lambda_3}{2}\right)}>0 \\ 
\left(\lambda_1+\lambda_4\right)+\sqrt{\lambda\left(\lambda_2+\lambda_3\right)}>0,~~
\left(\lambda_1+\lambda_4\right)+\sqrt{\lambda\left(\lambda_2+\frac{\lambda_3}{2}\right)}>0
\eea

\subsection*{Perturbativity}
\label{sec:perturb}

The quartic couplings ($\lambda_i$) and the Yukawa couplings appearing in the theory need to satisfy the following conditions in order to remain within perturbative limit:

\bea
|\lambda_i|<4\pi,~|y_{\psi}|<\sqrt{4\pi},~|Y|<\sqrt{4\pi},
\eea

where $\lambda_i=\lambda,\lambda_{1,2,3,4}$.

\subsection*{Electroweak precision observables (EWPO)}
\label{sec:ewpo}

$T$-parameter puts the strongest bound on the mass splitting between $m_{H^{\pm\pm}}$ and $m_{H^{\pm}}$, requiring: $|m_{H^{\pm\pm}}-m_{H^{\pm}}|\lsim 50~\rm GeV$~\cite{Ghosh:2017pxl}. Here we have assumed a conservative mass difference of 10 GeV. 

\subsection*{Experimental bounds}
\label{sec:exptbound}

Since the addition of scalar triplet can modify the $\rho$-parameter, hence a bound on the triplet Higgs VEV can appear from the measurement of the $\rho$ parameter $\rho=1.0008^{+0.0017}_{-0.0010}$~\cite{PhysRevD.98.030001}. Theoretically this can be expressed as:

\bea
\rho\simeq1-\frac{2 v_t^2}{v_d^2}=1+\delta\rho,
\eea

which further translates into: $v_t\leq 3~\rm GeV$ assuming $v=\sqrt{v_d^2+2 v_t^2}=246~\rm GeV$, which enters into the expression for the known SM gauge boson masses. For a small triplet VEV $v_t\lsim10^{-4}~\rm GeV$, stringent constraint on $m_{H^{\pm\pm}}$ has been placed by CMS searches: $m_{H^{\pm\pm}}>820~\rm GeV$ at 95 \%\ C.L.~\cite{Agrawal:2018pci} and also by ATLAS searches: $m_{H^{\pm\pm}}>870~\rm GeV$ at 95 \%\ C.L.~\cite{Aaboud:2017qph}. For $v_t\lsim10^{-4}~\rm GeV$, direct search bound from LHC also constraints other non-standard Higgs masses: $m_{H^+}>365~\rm GeV$ and $m_{H_2,A_0}>150~\rm GeV$~\cite{Das:2016bir}. For a larger triplet VEV, however, these constraints are significantly loosened~\cite{Melfo:2011nx,Dev:2018kpa}. In our analysis we have kept $v_t=1~\rm GeV$, where all these bounds can be overlooked~\cite{Melfo:2011nx}. We have still maintained a particular mass hierarchy amongst different components of the triplet:

\bea\nonumber
m_{H^{\pm\pm}}>m_{H^{\pm}}>m_{H_2,A},
\eea
which is dubbed as ``Negative scenario''~\cite{Ghosh:2017pxl}. The mixing between the CP-even scalar states is also constrained from Higgs decay measurement. As obtained in~\cite{Bhattacharya:2017sml}, $\sin\theta_0\lsim 0.05$ is consistent with experimental data of $H_1\to W W^*$ with $m_{H_1}=125~\rm GeV$.

\subsection*{Neutrino mass constraint}
\label{sec:numass}
Light neutrino mass is generated due to the coupling of the SM leptons with the scalar triplet through Yukawa interaction. As the triplet gets a non-zero VEV, one can write from Eq.~\ref{eq:lyuk}~\cite{Ma:1998dx}:

\bea
\left(m_{\nu}\right)_{ij}=\frac{1}{2} (y_L)_{ij} \langle\Delta\rangle \simeq (y_L)_{ij}~\frac{\mu v_d^2}{2\sqrt{2}\mu_{\Delta}^2}, 
\label{eq:lightnumass}
\eea

where $\{i,j\}=\{1,2,3\}$ are the family indices. We can then generate small neutrino masses through a small value of triplet VEV, i.e. by having a large triplet scalar mass through type II seesaw. Interestingly, the triplet scalar 
also interacts with the VLFs via Yukawa coupling $y_{\psi}$ as described in Eq.~\ref{eq:lyuk}. Thus, the VEV of $\Delta$ induces a Majorana mass term ($m$) for the VLFs on top of the Dirac mass term as follows:

\bea
m = \frac{1}{2} y_\psi \sin^2\theta\langle\Delta\rangle.
\label{eq:majoranamass}
\eea

If we trade $\langle\Delta\rangle$ from Eq.~\ref{eq:lightnumass}, then from Eq.~\ref{eq:majoranamass} we obtain the following relation between 
light neutrino mass and Majorana mass term for the DM:

\bea
\left(m_{\nu}\right)_{ij}=\left(\frac{(y_L)_{\alpha\beta}}{y_{\psi}\sin^2\theta}\right) m.
\label{eq:nubound}
\eea

Now, due to the introduction of the Majorana mass, the Dirac state $\psi^0$ splits into two pseudo-Dirac states with a mass difference $\delta=2m$. 
This plays a very important role in direct search of the DM, which we shall explore further in subsec.~\ref{sec:velo}. We shall show, in order to avoid $Z$-mediated direct detection of the DM, $\delta\gsim \mathcal{O}(100)~\rm keV$. Therefore, if we consider, the light neutrino mass $\sim\mathcal{O}(0.1~\rm eV)$ and the Majorana mass $\sim\mathcal{O}(100~\rm keV)$ to forbid $Z$-mediation, then from Eq.~\ref{eq:nubound} we immediately get:

\bea
\mathcal{R}=\left(\frac{(y_L)_{\alpha\beta}}{y_{\psi}\sin^2\theta}\right)\lsim\rm 10^{-6}.
\label{eq:nubound1}
\eea

This shows that the coupling of the scalar triplet to the SM sector is highly suppressed compared to the DM sector. Although we have chosen $y_{\psi}=1$ for our analysis in order to have contribution from the triplet, but the constraint from Eq.~\ref{eq:nubound1} has also been followed in order to ensure that the model also addresses correct neutrino mass. It is important to note that unlike the usual type-II seesaw scenario, where the correct neutrino mass predicts very heavy triplet scalars beyond any experimental reach, the presence of VLFs alter the situation significantly by allowing the triplet scalar within experimental search while addressing correct light neutrino masses.

\subsection*{Relic abundance constraint}
\label{sec:relic1}

The PLANCK-observed relic abundance puts a stringent bound on the DM parameter space as it suggests, for CDM: $\Omega_{DM}h^2=0.1199\pm 0.0027$~\cite{Ade:2015xua}. the effect of this constraint on the parameter space of the model will be explored in detail in our analysis.

\subsection*{Invisible decay constraints}
\label{sec:invdecay}

When the DM mass is less than half of Higgs or $Z$ Boson mass, they can decay to a pair of the VLF DM ($\psi_1$). Higgs and $Z$ invisible decays are however well constrained at the LHC~\cite{Khachatryan:2016whc, PhysRevD.98.030001}, which therefore constrains our DM model in such a mass limit. Both Higgs and $Z$ invisible decay to DM is proportional to VLF mixing angle $\sin\theta$ (These have been explicitly calculated and tabulated in~\ref{sec:invdecay}). We will show later that DM direct search constraint limits the mixing to small $\sin\theta$ regions which therefore naturally evades the invisible decay width limits.

\section{Dark matter phenomenology}
\label{sec:dmpheno}

As mentioned earlier, $\psi_1$ is the DM candidate in this model and in the following subsections we shall analyze the parameter space allowed by observed relic abundance of DM and also from direct detection bounds. Relic density and direct search outcome of the VLF DM as an admixture of singlet-doublet has already been studied elaborately before~\cite{Bhattacharya:2015qpa}. The case in presence of scalar triplet has also been studied briefly~\cite{Bhattacharya:2017sml}. We would therefore elaborate on the effect of scalar triplet in the DM scenario. 

\subsection{Relic abundance of DM}
\label{sec:relic}

Relic abundance of $\psi_1$ DM is determined by its annihilation to SM particles and also to scalar triplet, if the DM is heavier than the triplet. 
Such processes are mediated by SM Higgs, gauge bosons and scalar triplet. As the dark sector has charged fermions ($\psi^{\pm}$) and a heavy neutral fermion ($\psi_2$), the freeze-out of the DM will also be affected by the co-annihilation of the additional dark sector particles. This important feature makes this model survive the strong direct search limits, as we will demonstrate. All the Feynman graphs for freeze-out are shown in \ref{sec:tripdiagram}. Relic density can then be calculated by:

 \bea
 \frac{dn}{dt} + 3 H n = -{\langle \sigma v\rangle}_{eff} \Big(n^2-n_{eq}^2\Big),
 \eea
 where
\bea
{\langle \sigma v\rangle}_{eff}&&= \frac{g_1^2}{g_{eff}^2} {\langle \sigma v \rangle}_{\overline{\psi_1}\psi_1}+\frac{2 g_1 g_2}{g_{eff}^2} {\langle \sigma v \rangle}_{\overline{\psi_1}\psi_2}\Big(1+\frac{\Delta M}{M_{\psi_1}}\Big)^\frac{3}{2} e^{-x \frac{\Delta M}{M_{\psi_1}}} \nonumber \\
&&+\frac{2 g_1 g_3}{g_{eff}^2} {\langle \sigma v \rangle}_{\overline{\psi_1}\psi^-}\Big(1+\frac{\Delta M}{M_{\psi_1}}\Big)^\frac{3}{2} e^{-x \frac{\Delta M}{M_{\psi_1}}} \nonumber \\
&& +\frac{2 g_2 g_3}{g_{eff}^2} {\langle \sigma v \rangle}_{\overline{\psi_2}\psi^-}\Big(1+\frac{\Delta M}{M_{\psi_1}}\Big)^3 e^{- 2 x \frac{\Delta M}{M_{\psi_1}}} \nonumber \\
&& +\frac{g_2^2}{g_{eff}^2} {\langle \sigma v \rangle}_{\overline{\psi_2}\psi_2}\Big(1+\frac{\Delta M}{M_{\psi_1}}\Big)^3 e^{- 2 x \frac{\Delta M}{M_{\psi_1}}} \nonumber \\
&& +\frac{g_3^2}{g_{eff}^2} {\langle \sigma v \rangle}_{{\psi^+}\psi^-}\Big(1+\frac{\Delta M}{M_{\psi_1}}\Big)^3 e^{- 2 x \frac{\Delta M}{M_{\psi_1}}}, 
\label{eq:vf-ann}
\eea

with $n=n_{\psi_1}+n_{\psi_2}+n_{\psi^\pm}$. In above equation, $g_{eff}$ is defined as effective degrees of freedom, given by:
\bea
g_{eff}=g_1 + g_2 \Big(1+\frac{\Delta M}{M_{\psi_1}}\Big)^\frac{3}{2} e^{-x \frac{\Delta M}{M_{\psi_1}}} + g_3\Big(1+\frac{\Delta M}{M_{\psi_1}}\Big)^\frac{3}{2} e^{-x \frac{\Delta M}{M_{\psi_1}}} ,
\eea
where $g_1 ,~ g_2 \rm ~and~ g_3$ are the degrees of freedom of $\psi_1, ~\psi_2 \rm ~and~ \psi^-$ respectively and $x=x_f=\frac{M_{\psi_1}}{T_f}$, where $T_f$ is the freeze out temperature. For the numerical analysis we implemented the model in {\tt LanHEP}~\cite{Semenov:2008jy} and the outputs are then fed into {\tt MicrOmegas}~\cite{Belanger:2001fz} 
to obtain relic density.

\begin{figure}[htb!]
$$
 \includegraphics[scale=0.4]{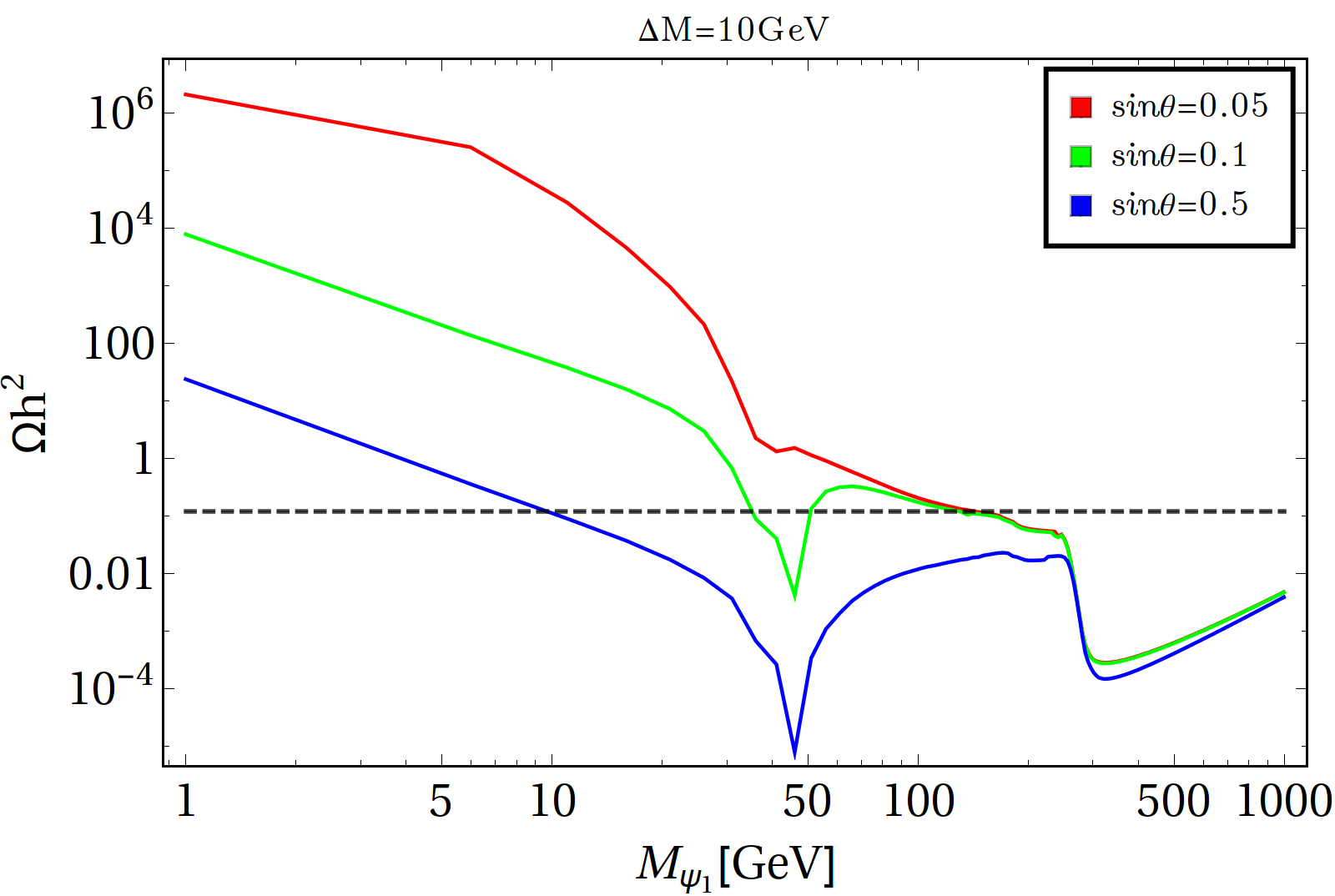}
 \includegraphics[scale=0.4]{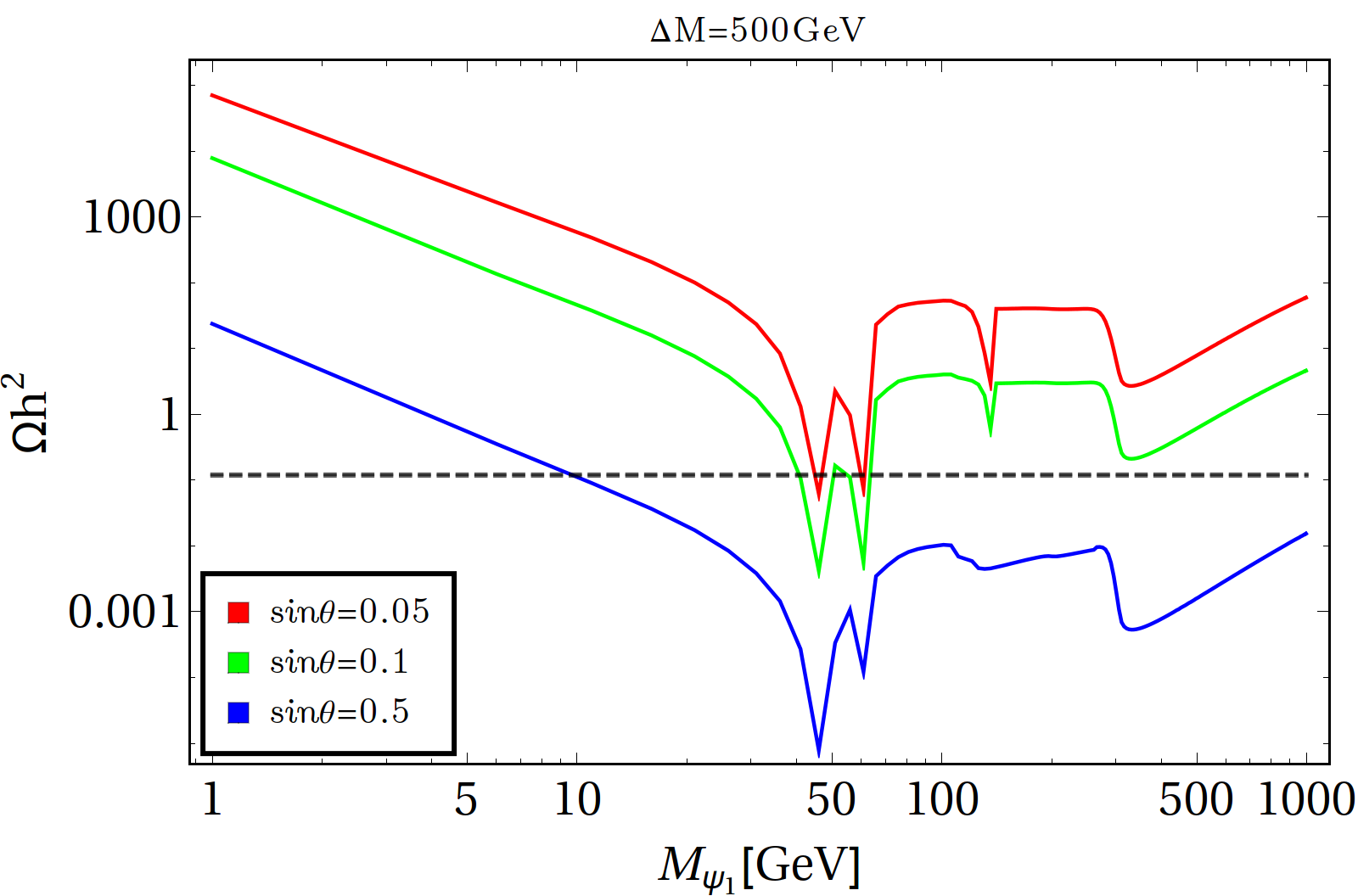}
 $$
 $$
 \includegraphics[scale=0.4]{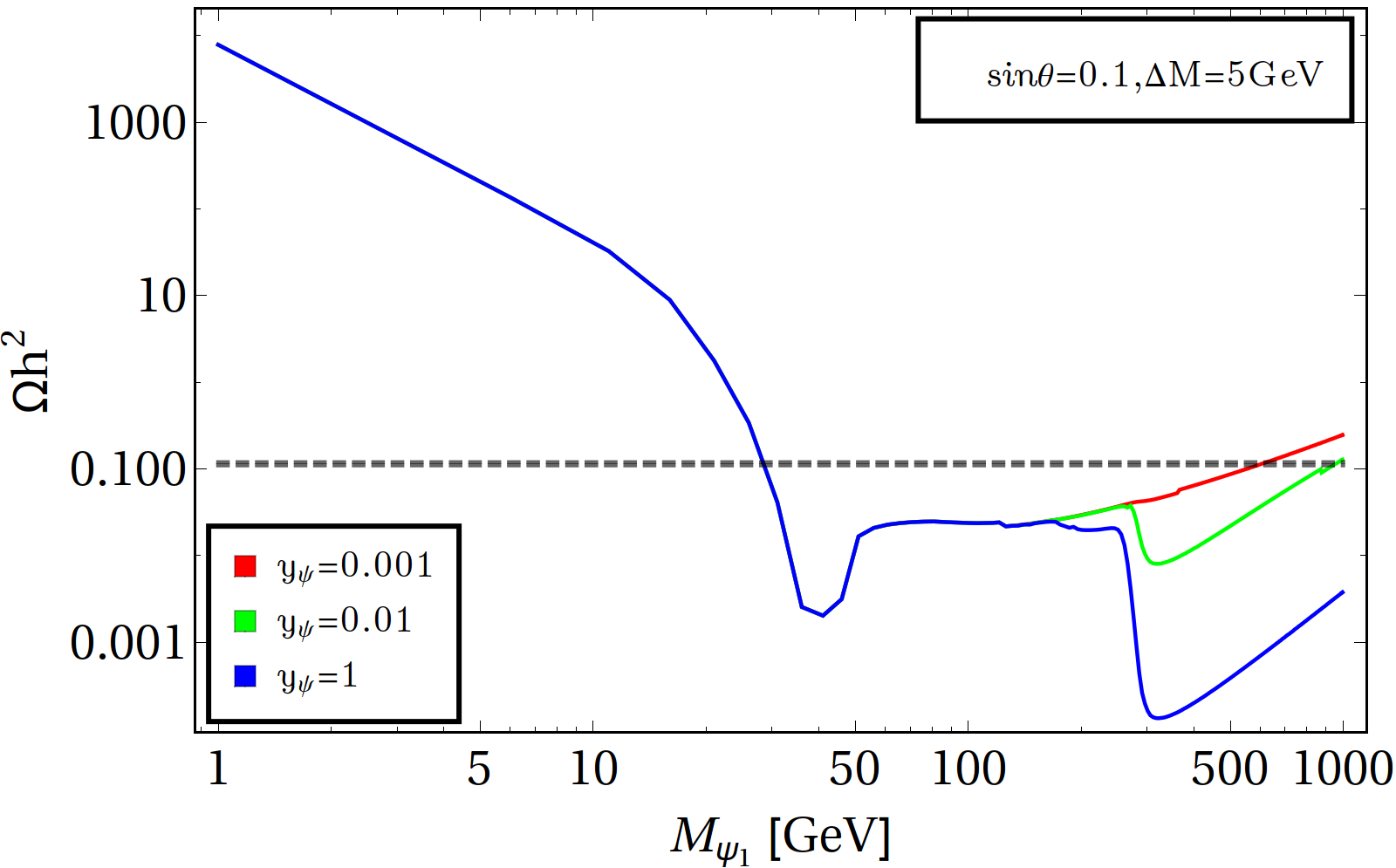}
 \includegraphics[scale=0.4]{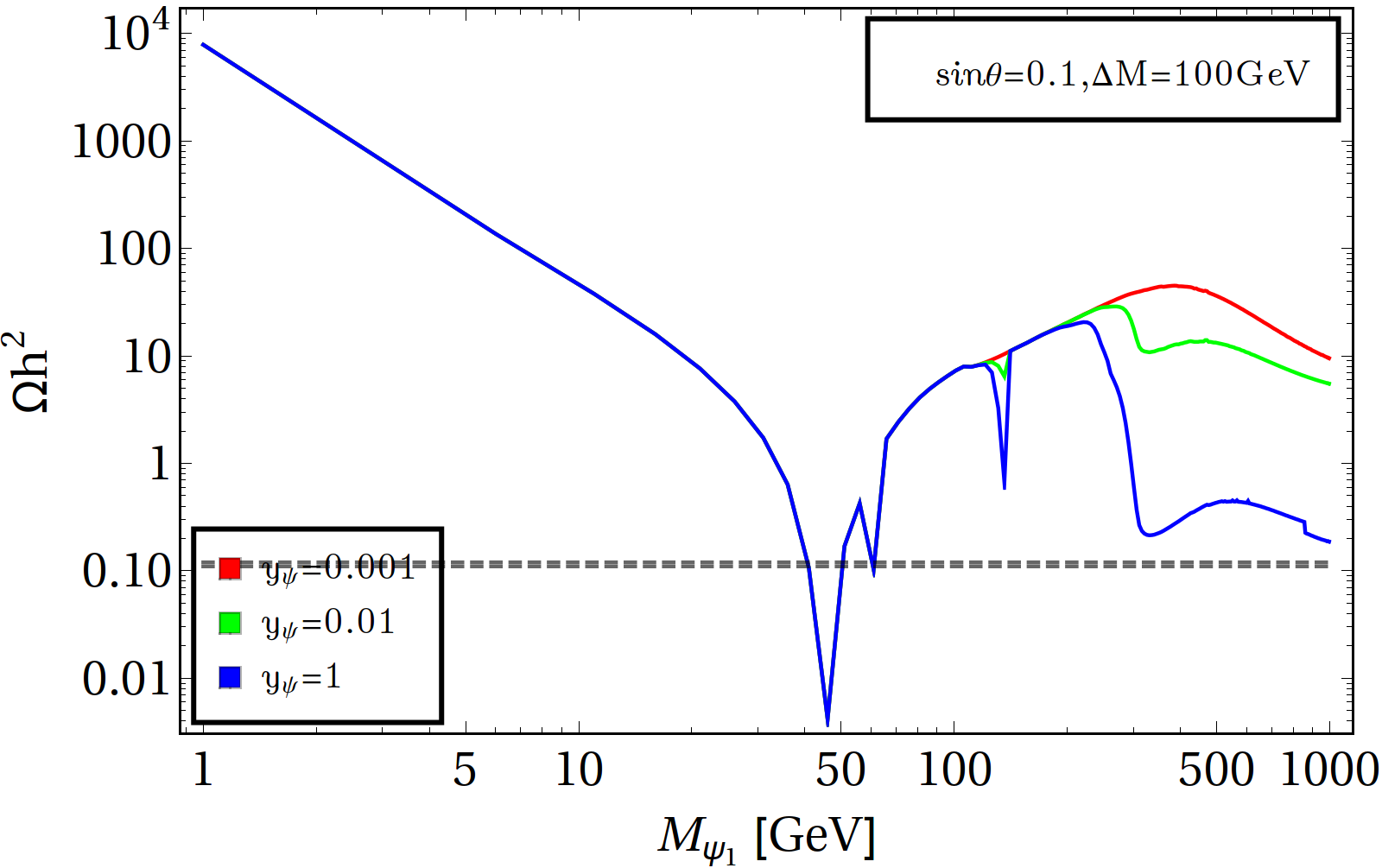}
 $$
 $$
 \includegraphics[scale=0.4]{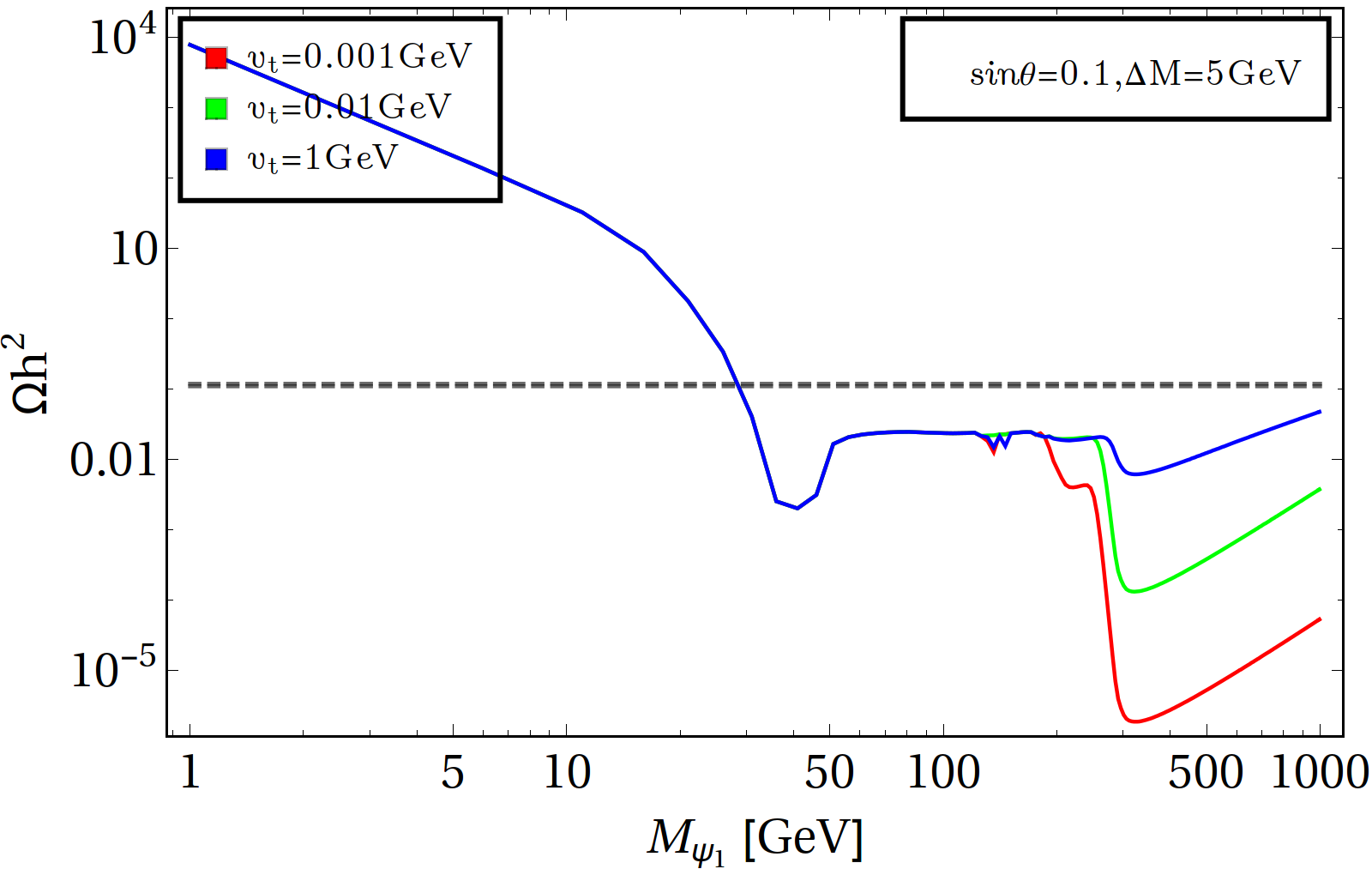}
 \includegraphics[scale=0.4]{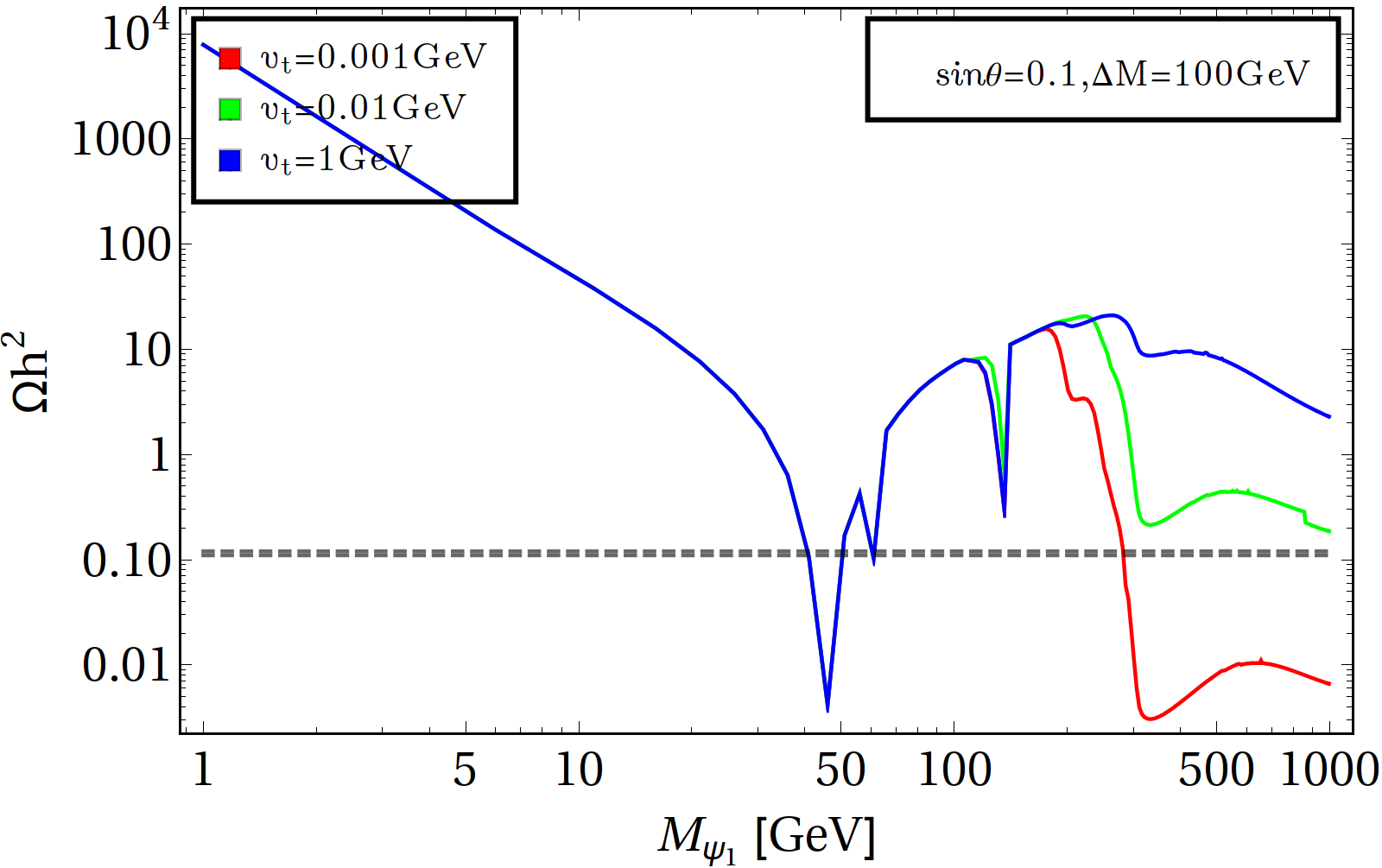}
 $$
 \caption{{\it Top Left:} Variation of relic abundance of $\psi_1$ with its mass $M_{\psi_1}$ for different singlet-doublet mixing: $\sin\theta=0.05$ (blue), $\sin\theta=0.1$ (orange) and $\sin\theta=0.5$ (green) keeping $\Delta M=10~\rm GeV$. {\it Top Right:} Same with $\Delta M=500~\rm GeV$. In both cases $y_{\psi}=1.0$ and $v_t=1~\rm GeV$. {\it Middle left:} Variation of relic abundance with DM mass for three different choices of the Yukawa $y_{\psi}:\{0.01,0.1,1.0\}$ in red, green and blue respectively for $\Delta M=5~\rm GeV$ and $\sin\theta=0.1$. {\it Middle right:} Same with $\Delta M=100~\rm GeV$ and $\sin\theta=0.1$. {\it Bottom Left:} Variation of relic abundance of the DM with DM mass for different choices of the VLF-triplet Yukawa coupling $y_{\psi}$ with the triplet VEV $v_t=1~\rm GeV$. {\it Bottom Right:} Same for three different values of the triplet VEV for $y_{\psi}=1.0$. In each case the black dashed line shows the right order of observed relic abundance.}
 \label{fig:mn1vsrelic}
\end{figure}

In the top panel of Fig.~\ref{fig:mn1vsrelic} we have shown how the relic abundance of the DM varies with its mass for some chosen singlet-doublet VLF mixings. In the LHS of the top panel, $\Delta M$ is fixed at 10 GeV, while in the RHS it is kept fixed at a larger value 500 GeV. First of all we see three different kinds of resonance drops: one at half of the $Z$ mass $\sim 45$ GeV, the second at half of the Higgs mass $\sim 62.5$ GeV and the third at the half of the triplet scalar mass $\sim 150$ GeV (the triplet scalar masses are kept fixed around $\sim$ 300 GeV). The first resonance is prominent, the second one is mild, while the third one is only visible for smaller $\sin\theta$ and large $\Delta M$ (right hand side of the top panel). Finally at around 300 GeV, a new annihilation channel to the triplet scalar opens up and correspondingly we observe a drop in relic density. Importantly, for small $\Delta M$, co-annihilation plays an important role. This can be seen on the top left panel, where the relic density drops, particularly for small $\sin\theta$, while for large $\Delta M$ such effect is subdominant.  With the increase in DM mass, the relic density finally increases suggesting decrease in annihilation cross section due to unitarity. Note that the relic density decreases, {\it i.e}, the annihilation cross-section rises with larger $\sin\theta$ (for a fixed $\Delta M$) due to larger gauge ($Z$) mediated contribution. We have kept $y_\psi=1, v_t=1~\rm GeV$ for plots in the top panel, while for all the plots in Fig.~\ref{fig:mn1vsrelic} other physical masses are kept fixed at: $m_{H^{\pm\pm}}=310~\rm GeV$, $m_{H^{\pm}}=300~\rm GeV$ and $m_{A,H_2}=280~\rm GeV$. In the middle panel of  Fig.~\ref{fig:mn1vsrelic} we have illustrated how the relic abundance behaves with the triplet-VLF coupling $y_{\psi}$ for a fixed $\sin\theta =0.1$ and $\Delta M$ (5 GeV in the left panel and 100 GeV in the right panel). The effect of $y_{\psi}$ is only observed in the annihilation to triplet final state (i.e. for DM mass $>$ triplet mass which is kept at 300 GeV). As we increase $y_{\psi}$, more annihilation to triplet state is expected, which causes the relic density to further decrease. Again the effect of co-annihilation is apparent for small $\Delta M$ in the left panel where relic density drops due to such effects, which, for large $\Delta M$ is not visible in the right hand panel. Lastly, we show the effect of triplet VEV $v_t$ as a function of DM mass in the bottom panel of Fig.~\ref{fig:mn1vsrelic} for two different choices of $\Delta M$. Again, the effect can be realised for DM annihilation to triplet final states and therefore lies in the region where DM mass $\gsim$ triplet mass. As the triplet final state (charged or neutral) diagrams are proportional to $(y_{\psi}/v_t)^2$ (see~\ref{sec:cplings}), for a fixed $y_{\psi}=1$, increasing $v_t$ reduces the annihilation cross-section, resulting in over-abundance. 
 

Now, once we have identified the important physics aspects of the variation of relic abundance with different parameters, we are in a position to find the relic density allowed parameter space. The independent DM parameters that we vary for this model are:

\bea
\{M_{\psi_1},\Delta M,\sin\theta\},
\eea

while the effects of triplet scalar parameters like $M_\Delta, y_\psi, v_t$ are also important, which we have kept at fixed values. We have scanned the relic density allowed parameter space in the following region: 

\bea
M_{\psi_1}:\{10-1000\}~\rm GeV,~\Delta M=\{1-1000\}~\rm GeV,~\sin\theta=\{0.01-0.5\}.
\eea

We would like to remind once more that, other parameters are kept fixed throughout the scan at the following values: 
\begin{center}
 $y_{\psi}=1.0$, $v_t=1~\rm GeV$, $m_{H^{\pm\pm}}=310~\rm GeV$, $m_{H^{\pm}}=300~\rm GeV$, $m_{A,H_2}=280~\rm GeV$, 
\end{center}
which evade the constraints discussed in Sec.~\ref{sec:constraint}.

\begin{figure}[htb!]
$$
 \includegraphics[scale=0.38]{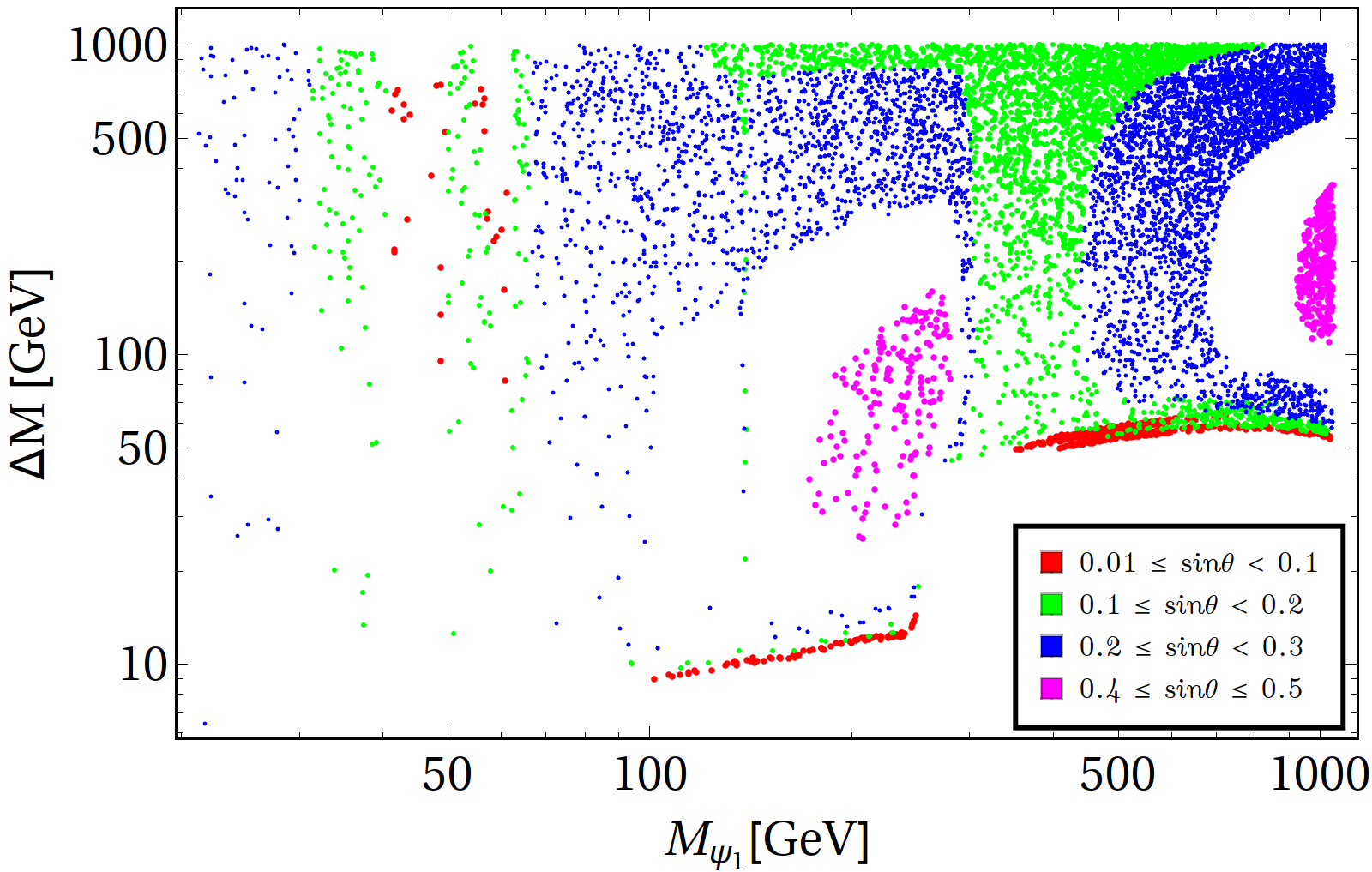}
  \includegraphics[scale=0.38]{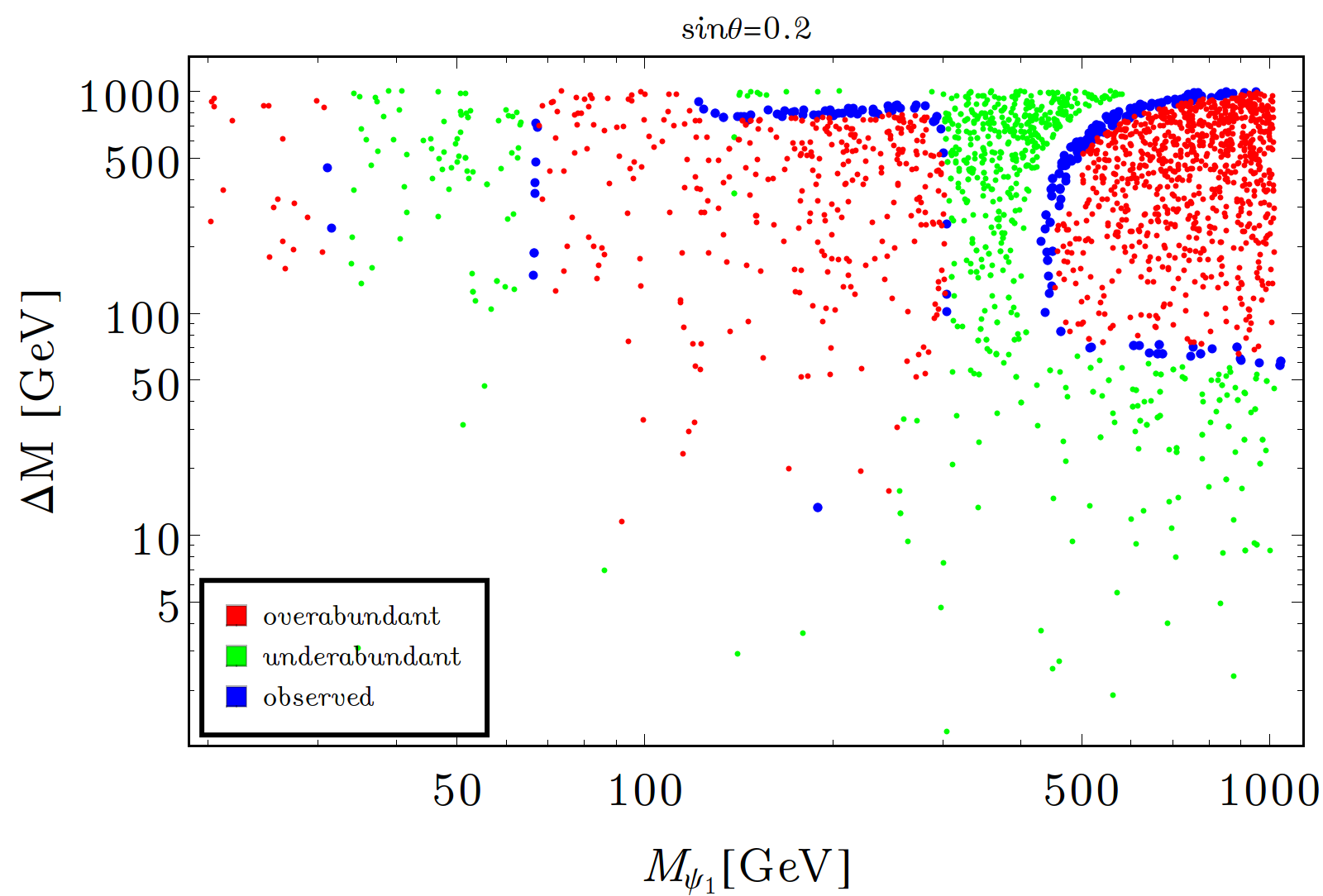}
 $$
 $$
   \includegraphics[scale=0.42]{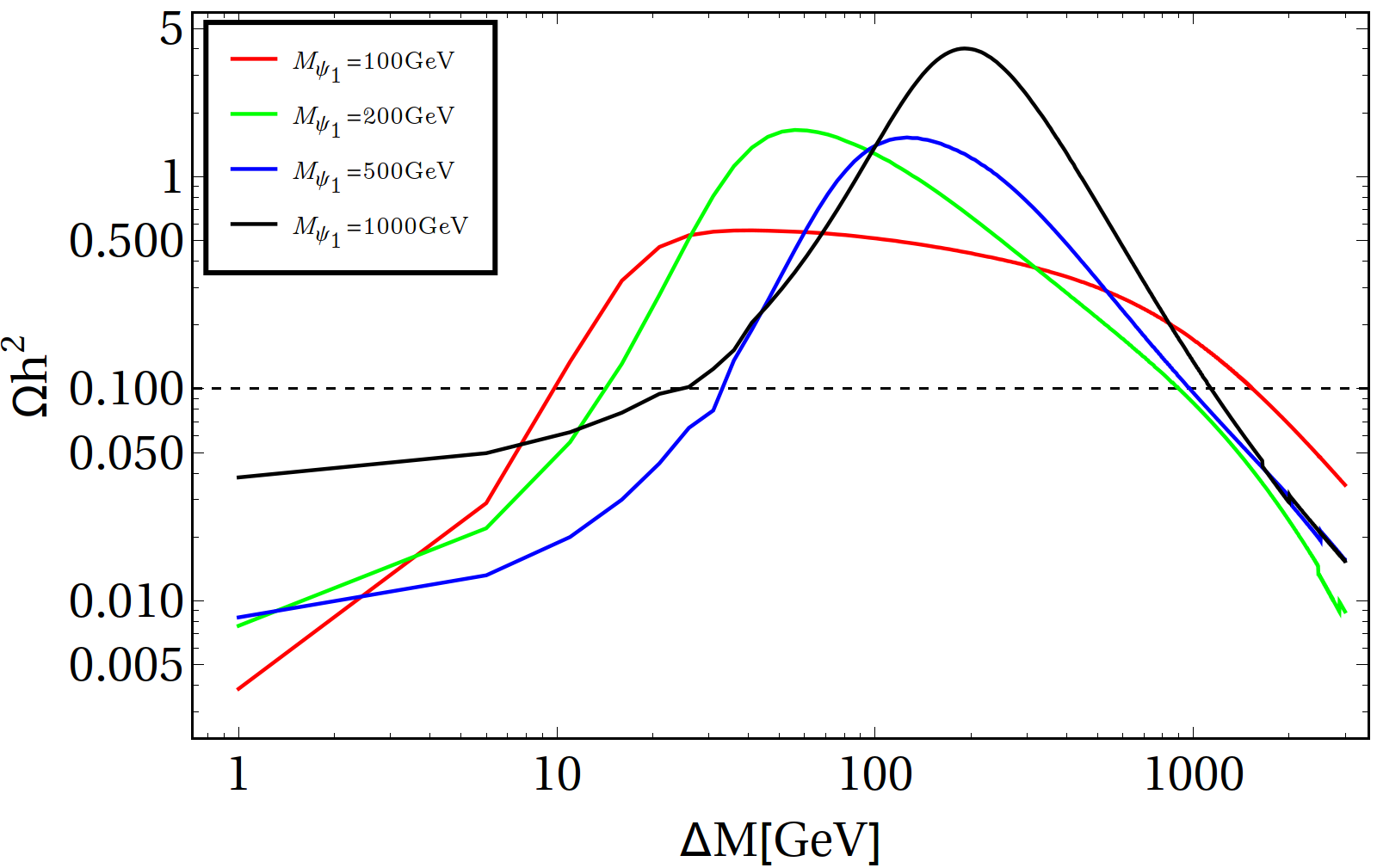}
 $$
 \caption{Top Left: Parameter space allowed by relic density in $M_{\psi_1}$-$\Delta M$ plane for different choices of the singlet-doublet VLF mixing: $\sin\theta:\{0.01-0.1\}$ (red), $\sin\theta:\{0.1-0.2\}$ (green), $\sin\theta:\{0.2-0.3\}$ (blue) and $\sin\theta:\{0.4-0.5\}$ (magenta). Top Right: In the same plane the underabundant (green) and overabundant (red) regions are shown together with observed relic density (blue) region for $\sin\theta=0.2$. Bottom: Variation of relic abundance with $\Delta M$ for different choices of DM mass $M_{\psi_1}$ for $\sin\theta=0.2$.}
 \label{fig:mDelmrelic}
\end{figure}

LHS of Fig.~\ref{fig:mDelmrelic} in the top panel shows the relic density allowed parameter space of the model in $M_{\psi_1}-\Delta M$ plane for a range of $\sin\theta$ varying within: \{0.01-0.5\} (shown in different colours). 
Both DM mass and $\Delta M$ have been varied upto 1 TeV for the scan. Now, the plot shows several interesting features. The most important effect is observed in the vicinity of $M_{\psi_1} \sim 300$ GeV, which is the value 
chosen for the triplet scalars in the analysis. Therefore for $M_{\psi_1} \gsim 300$ GeV, the annihilation to triplet channels open up (see the Feynman graphs in section \ref{sec:tripdiagram}). Annihilation to triplet is guided by 
gauge mediation and Higgs mediation, where the former dominates over the latter. For low $\sin\theta \le 0.1$ however, annihilation to triplet is not substantial through gauge mediation due to very small doublet component present in DM. 
Therefore, for such points ($\sin\theta \le 0.1$ shown by red dots), the additional annihilation channel to triplets can be accounted by taming the co-annihilation processes with larger $\Delta M$ and yields just a step in the vicinity of
 $M_{\psi_1} \sim 300$ GeV, for $\Delta M \sim $ 10 GeV to 50 GeV.  When we choose a larger range of $\sin\theta \sim \{0.1-0.2\}$, the annihilation to triplet final states become much more effective both through gauge mediation 
 (which is solely dictated by $\sin\theta$) and through Higgs mediation (where the Yukawa is proportional to both $\sin\theta$ and $\Delta M$). Therefore points with $\sin\theta \sim \{0.1-0.2\}$ requires a sharp increase in $\Delta M$ to 
 reduce co-annihilation for $M_{\psi_1} \sim 300$ GeV and ends up with the vertical column with green dots in this region. For even larger $\sin\theta \sim \{0.2-0.3\}$ (blue dots in the top LHS plot) shows two half circles as allowed relic 
 density points in the vicinity of $M_{\psi_1} \sim 300$ GeV. In order to understand this feature, we explore the exact relic density allowed points (in blue) with under abundant points (in green) and over abundant points (in red) 
 for a fixed $\sin\theta=0.2$ in top RHS panel of Fig.~\ref{fig:mDelmrelic}. 
 
 In top RHS panel of Fig.~\ref{fig:mDelmrelic}, first of all, we see that the relic density allowed half circles span either $M_{\psi_1} <300$ GeV or $M_{\psi_1} > 300$ GeV. With larger $\sin\theta \sim 0.2$ as we have here,  
 the annihilation to triplet is quite  large and therefore it always ends up with under-abundant (geen) points for $M_{\psi_1} \sim 300$ GeV. Hence with smaller $M_{\psi_1}$, when the triplet channel is not open, 
 or with larger $M_{\psi_1}$, where the annihilation to triplet is further subdued by $\frac{1}{m_{DM}^2}$ suppression, one can achieve correct relic. Now, the lower arc of the allowed half 
 circle comes from the existence of annihilation plus co-annihilation with co-annihilation taking a larger share with small $\Delta M$. As 
 we increase $\Delta M$, the co-annihilation effect gets subdued, however the annihilation through Higgs becomes important with larger Yukawa (proportional to $\Delta M$). Therefore, for a fixed $M_{\psi_1}$, there are two different 
 $\Delta M$ where one can observe correct density:  (i) a small $\Delta M$ region, where co-annihilation plays crucial role with annihilation, (ii) a larger $\Delta M$, where co-annihilation gets suppressed but larger annihilation through Higgs mediation 
 provides correct relic. This is explicitly demonstrated in the bottom panel of  Fig.~\ref{fig:mDelmrelic}, where we plot relic density versus $\Delta M$ for different fixed values of DM masses (with $\sin\theta=0.2$) and 
 the above feature is clearly observed.  We would also  like to explain the over-abundance of DM (red dots) within the allowed half circle in the top RHS plot. This is simply, because in this region, the co-annihilation effect is reduced with 
 large $\Delta M$, while the increase in Higgs mediated annihilation is not able to cope up. Apart, we also see three resonance allowed relic density regions at $M_{\psi_1}=\frac{m_Z}{2}$, $M_{\psi_1}=\frac{M_{H_1}}{2}$ and 
 $M_{\psi_1}=\frac{M_{H_2}}{2}$ corresponding to Z-boson, SM Higgs and the triplet Higgs mediation.

\begin{figure}[htb!]
$$
 \includegraphics[scale=0.45]{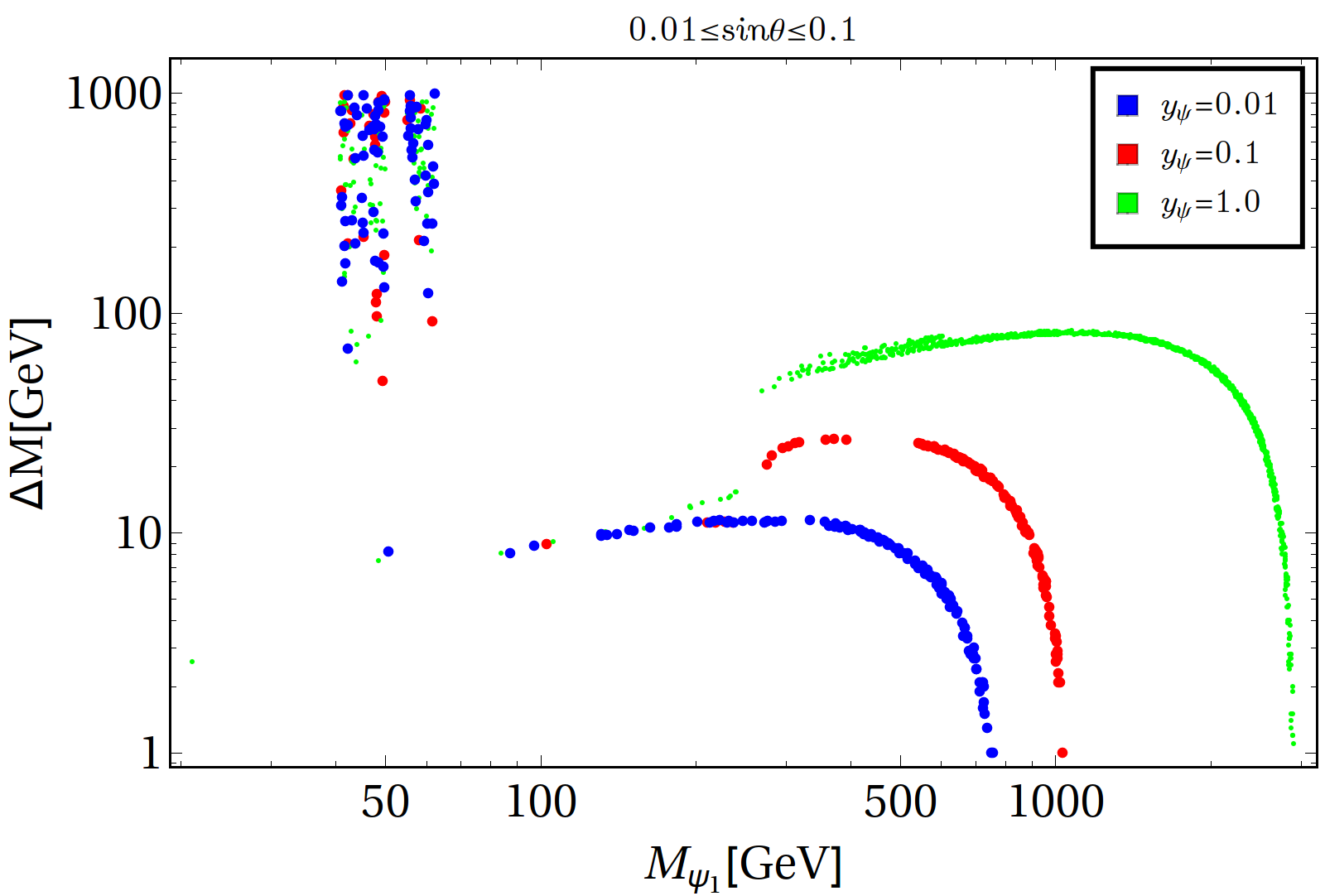}
$$
 \caption{The figure shows the effect of triplet-VLF Yukawa coupling in $M_{\psi_1}-\Delta M$ plane satisfying relic density constraint for three different choices of $y_{\psi}$:\{0.01,0.1,1.0\} shown in blue, red and green respectively. The triplet scalar VEV is fixed at $v_t=1~\rm GeV$.}
 \label{fig:vtfN}
\end{figure}

Another noteworthy feature is in Fig.~\ref{fig:vtfN}, where have shown how the relic density allowed parameter space changes pattern for different choices of the VLF-triplet Yukawa coupling $y_{\psi}$ (in Eq.~\ref{eq:lyuk}) for $0.01\le\sin\theta\le0.1$. For $y_{\psi}=0.01$, there is almost no contribution from the triplet scalar. In that case, co-annihilation plays vital role in producing the correct relic abundance and hence one needs to resort to smaller $\Delta M$, as shown by the blue curve. For larger DM mass the curve bends down due to $1/M_{\psi_1}^2$ suppression coming from the cross-section (unitarity). As $y_{\psi}$ is increased to 0.1, the triplet starts playing role. This can be understood by the rise of the red and green curves at $M_{\psi_1}\sim 300~\rm GeV$. Now, as the triplet gets into the picture, it provides enough annihilation channels and as a result co-annihilation plays a sub-dominant role here. This is again evident from the larger values of $\Delta M$ for both $y_{\psi}=0.1$ and $y_{\psi}=1.0$ curves. The drop in the high DM mass region is again due to unitarity.

\subsection{Direct search of DM}
\label{sec:dd}

In this section we shall investigate the effect of spin-independent direct search constraints on the DM parameter space. Our goal is to find how much of the parameter space, satisfied by PLANCK-observed relic density, is left after imposing the upper  limit from XENON1T. The pivotal role in this regard is played by the triplet scalar. As we shall see in the following subsection, due to the presence of the triplet, the $Z$-mediated inelastic direct search is forbidden for $\sin\theta\lsim 0.1$ for DM mass upto 1 TeV.  

\subsubsection{Emergence of pseudo-Dirac states and its effect on direct search}
\label{sec:velo}

The presence of the triplet scalar plays a decisive role in determining the fate of this model in direct search experiment as discussed in~\cite{Bhattacharya:2017sml}. Since the VEV of the neutral component of the triplet scalar induces a Majorana mass term (as seen from Eq.~\ref{eq:lyuk}), it splits the Dirac spinor $\psi_1$ into two {\it pseudo-Dirac} states $\psi_1^{\alpha,\beta}$ with mass difference proportional to the VLF-mixing angle and VEV of $\Delta^0$ (already mentioned in~\ref{sec:numass}):

\bea
\delta=2m=y_{\psi} \sin^2\theta \langle\Delta^0\rangle.
\label{eq:pseudodirac}
\eea

Now, the $Z$-mediated direct detection interaction of the DM is given as:

\bea
\mathcal{L}\supset i\bar{\psi_1}\left(\slashed{\partial}-i g_z \gamma_{\mu}Z^{\mu}\right)\psi_1,
\label{eq:zinteraction1}
\eea

where $g_z=\frac{g}{2\cos\theta_w}\sin^2\theta$, $\theta_w$ being the Weinberg angle. In presence of the pseudo-Dirac states, this interaction takes the form:

\bea
\mathcal{L}\supset \bar{\psi_1^{\alpha}}i\slashed{\partial}\psi_1^{\alpha}+\bar{\psi_1^{\beta}}i\slashed{\partial}\psi_1^{\beta}+g_z \bar{\psi_1^{\alpha}}\gamma_{\mu}\psi_1^{\beta}Z^{\mu}
\label{eq:zinteraction2}.
\eea

As one can notice, the $Z$-interaction is off-diagonal, {\it i.e,} $Z$ is coupled to $\psi_1^{\alpha}$ and $\psi_1^{\beta}$, unlike the diagonal kinetic terms. This therefore induces inelastic $Z$ mediated scattering for the fermion DM in presence of triplet. Such an inelastic scattering is kinematically allowed if~\cite{TuckerSmith:2001hy}:

\bea
\delta_{max} < \frac{\beta^2}{2}\frac{M_{\psi_1} M_N}{M_{\psi_1}+M_N},
\label{eq:deltmax}
\eea

\begin{figure}[htb!]
$$
 \includegraphics[scale=0.246]{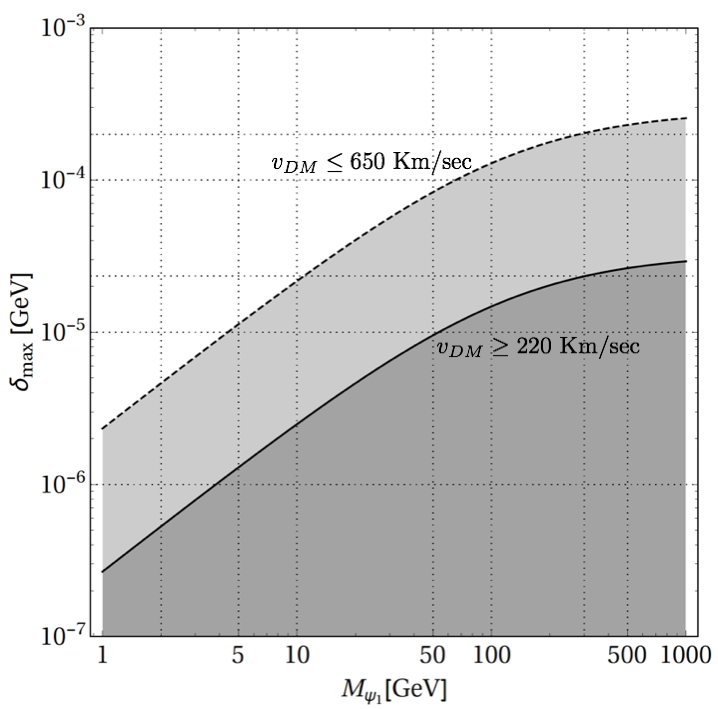} 
 \includegraphics[scale=0.30]{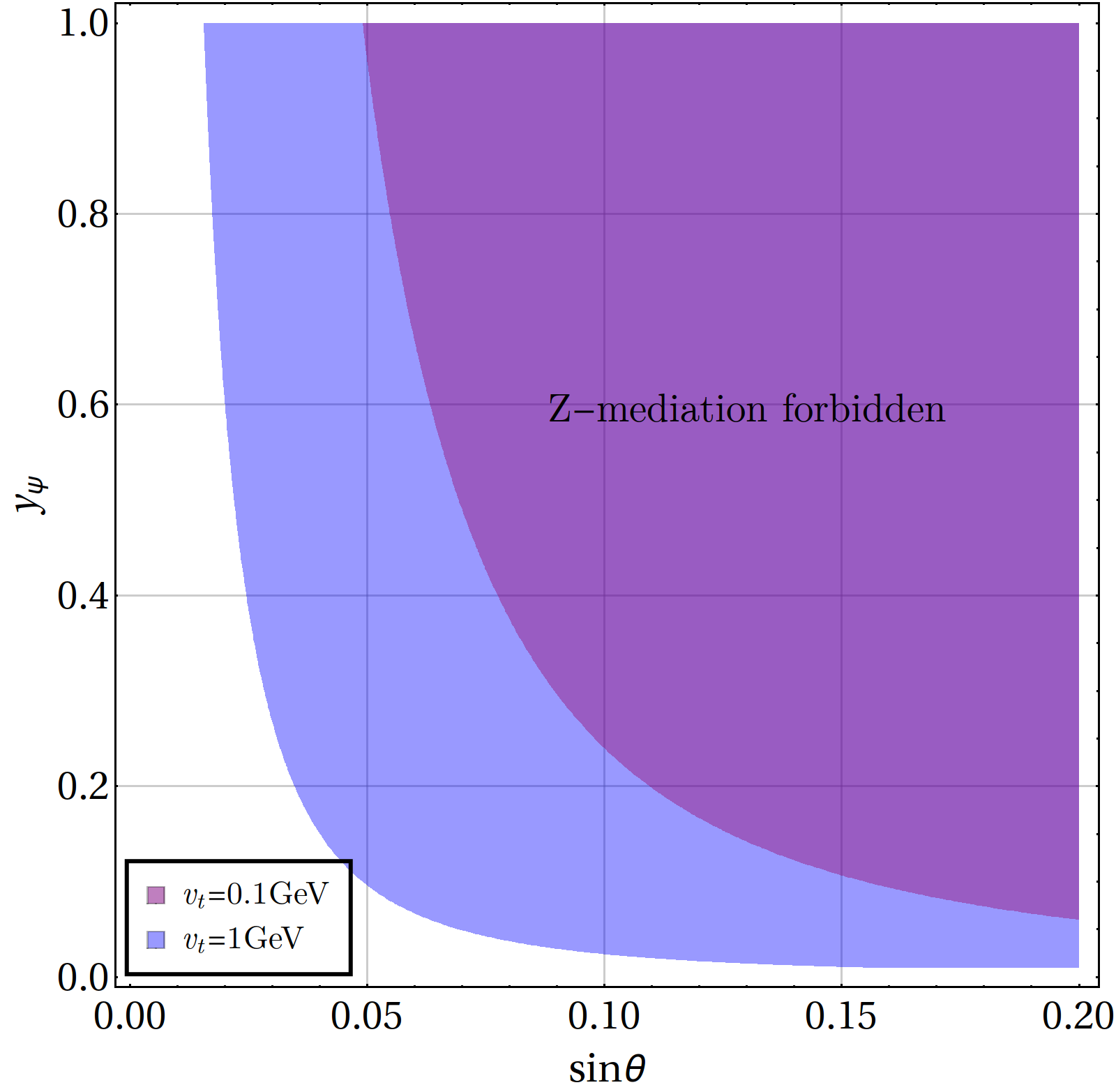}
$$
$$
 \includegraphics[scale=0.30]{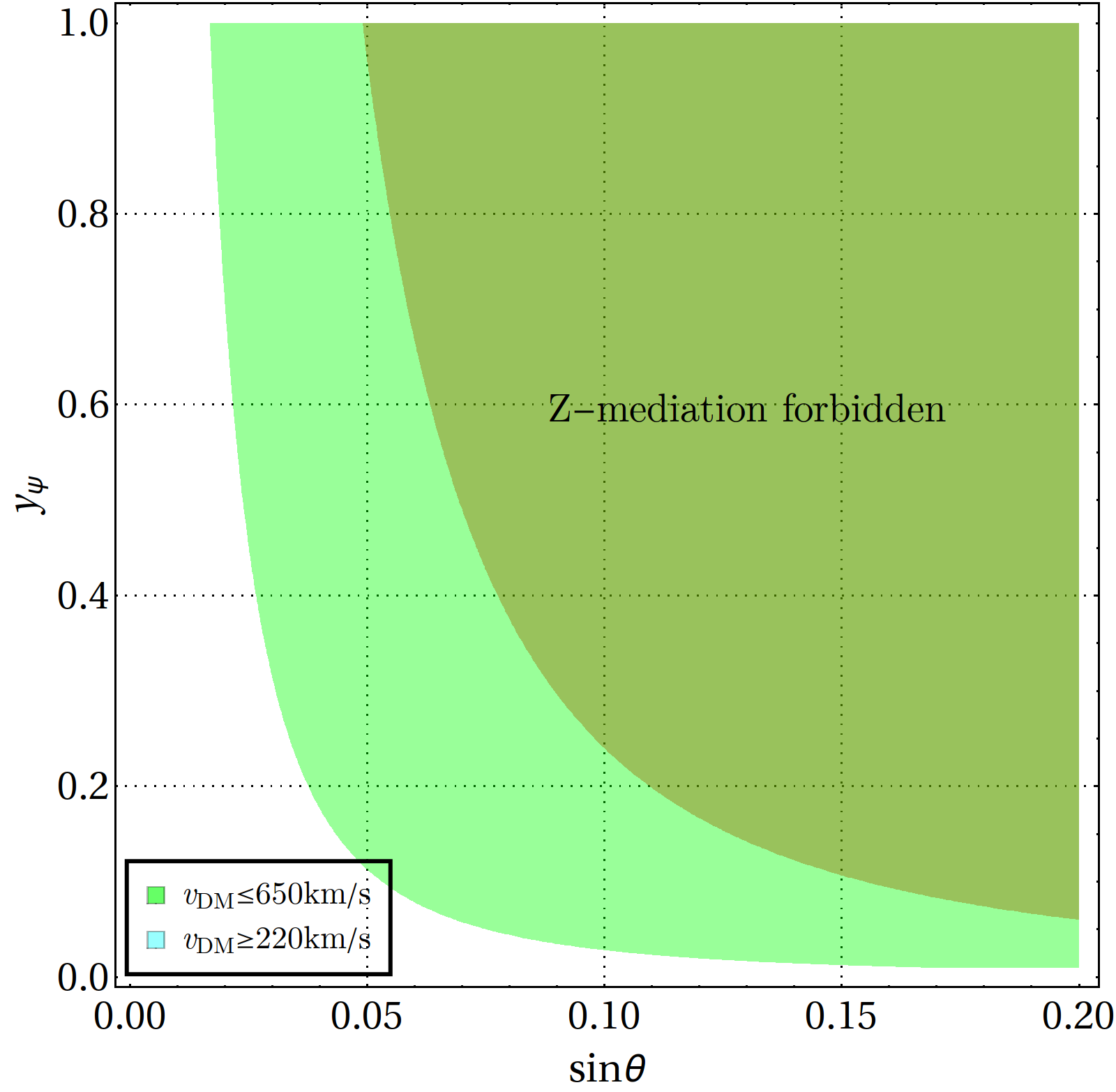}
 \includegraphics[scale=0.26]{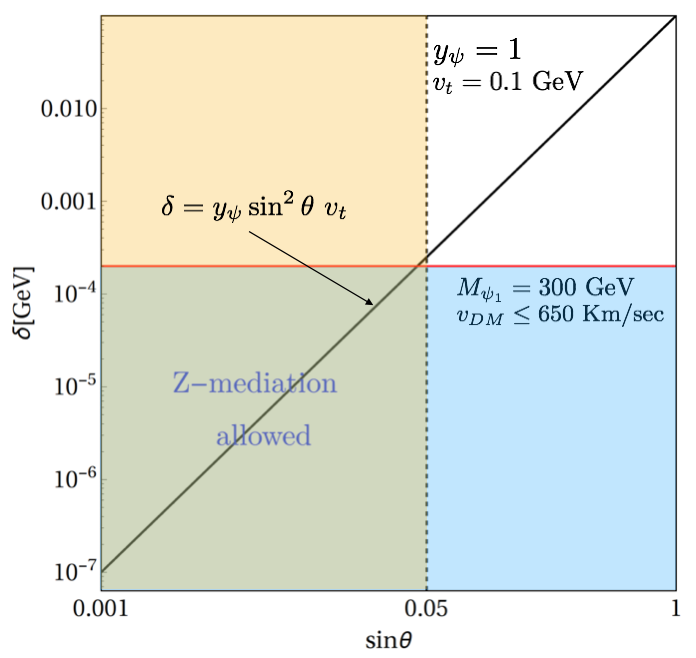}
 $$
 \caption{{\it Top left:} The grey region is where inelastic scattering of the DM via $Z$-mediation is allowed as derived from Eq.~\ref{eq:deltmax}. 
 The solid black line corresponds to DM velocity $\beta=220~\rm km/s$, while dashed black line corresponds to $\beta=v_{esc}\simeq 650~\rm km/s$. 
 {\it Top right:} The purple region shows the choices of $\sin\theta$ and $y_{\psi}$ which shall forbid the $Z$-mediated direct detection for $v_t=1~\rm GeV$  as obtained from Eq.~\ref{eq:pseudodirac}, the blue region underneath is the same for $v_t=1~\rm GeV$ assuming $\beta=v_{esc}$. 
 {\it Bottom left:} Same as top right but for a particular $v_t=1~\rm GeV$ with two regions corresponding to the lower (light green region) and upper (green region) limit on  DM velocity. {\it Bottom right:} $Z$-mediation allowed region in $\delta$ vs. $\sin\theta$ plane satisfying Eq.~\ref{eq:pseudodirac}, for DM mass of 300 GeV,  $v_t=1~\rm GeV$ and $y_{\psi}=1$. We use $v_{DM} \le 650~\rm {Km/sec}$.}
 \label{fig:minvelo}
\end{figure}

where $\beta c=v_{DM}$ can be within: $220~\rm km/s\lsim\beta.c<650~\rm km/s$, where the lower limit corresponds to the DM velocity in the local DM halo and the upper limit refers to the escape velocity ($v_{esc}$) of DM particles in the Milky Way, and $M_N$ is the {\it nucleus} mass.  Now, the present strongest bound on spin-independent direct detection cross-section comes from XENON1T, which we abide by for the available parameter space of the model. Then, using Xe nucleus mass $M_N=130~\rm amu$ and following Eq.~\ref{eq:deltmax}, we can have an upper limit on $\delta_{max}$ as a function of DM mass, below which $Z$-mediated inelastic scattering is allowed. This is shown in the upper left panel of Fig.~\ref{fig:minvelo}, where the shaded region allows such inelastic scattering. The solid and black dashed lines show the limit beyond which $Z$-mediated inelastic scattering is disallowed corresponding to the lower and upper limit of DM velocity $\beta c$. As one can see, $Z$-mediated cross-section is forbidden for $\delta\gsim 240~\rm keV$ for DM mass of $\sim$ 1 TeV 
corresponding to the upper limit on $\beta c$. This constraint can be viewed also in a different way. The minimum velocity of the DM which produces a recoil energy $E_R$ in the detector through inelastic scattering takes the form~\cite{TuckerSmith:2001hy}:

\bea
v_{min} = \sqrt{\frac{1}{2 M_N E_R}} \left(\frac{M_N E_R}{\mu_r}+\delta\right),
\label{eq:relicvelo}
\eea

where $\mu_r$ is the reduced mass of the DM-{\it nucleus} system. Eq. \ref{eq:relicvelo} will also yield a similar constraint on $\delta$ (as obtained in top left figure of Fig.~\ref{fig:minvelo}) but for a given recoil energy ($E_R$) specific to a detector used for the DM direct search. For $E_R \sim 30 \rm{keV}$, the conclusions are roughly the same. 

If this constraint on $\delta$ (derived from Eq.~\ref{eq:deltmax}) is implemented  in our model, we can have a relation between the mixing $\sin\theta$ and the triplet Yukawa $y_{\psi}$ from Eq.~\ref{eq:pseudodirac}. This is depicted in the top right panel of Fig.~\ref{fig:minvelo}, where we have shown the $Z$-mediation forbidden region of the parameter space in $\sin\theta$-$y_{\psi}$ plane for two different choices of the triplet VEVs: $v_t= \{0.1,1\}$ GeVs shown in purple and pale blue respectively. 
As the splitting is proportional to $v_t$, larger the $v_t$, larger is the $Z$-forbidden region. For this plot we have used a liberal limit of maximum possible DM velocity of 650 km/sec to avail the maximum splitting $\delta$. We can see from the top right figure that with $y_\psi<1$, in order to avoid $Z$-mediated direct search, one has to choose $\sin\theta\gsim 0.05$ for DM mass of 1 TeV. The bound on $\sin\theta$ is even more conservative to allow $Z$ mediation ($\sin\theta\gsim 0.02$) for $v_t=1~\rm GeV$ (shown by the pale blue region). A similar plot as in top right panel, is plotted in the bottom left panel to show the $Z$ forbidden region for the minimum and maximum permissible DM velocities for $v_t=1~\rm GeV$. Lastly, in the bottom right panel of Fig.~\ref{fig:minvelo}, we have illustrated a situation (following Eq.~\ref{eq:pseudodirac}) where $Z$-mediated inelastic scattering is possible for a fixed DM mass of $M_\psi=$ 300 GeV. If we choose $v_{DM}\le 650~\rm km/s$, this yields a bound on $\delta$ (following top left figure) and is shown by the red solid line below which $Z$-mediation is possible. Once we choose a specific $y_\psi=1$ and $v_t=0.1$ GeV, a bound on $\sin\theta$ is also obtained, and is shown by black dashed line. On the left side of this line $Z$-mediation is possible. If we now consider the splitting that the model can generate following Eq.~\ref{eq:pseudodirac}, for the chosen values of $y_\psi=1$ and $v_t=0.1$ GeV, we obtain a specific relation between the splitting $\delta$ to $\sin\theta$, shown by the diagonal solid black line. To summarize, the olive coloured region allows $Z$ mediated interaction, and the part of the black line within this can be realized in our model framework. We would however be interested to work in the parameter space where $Z$ mediation is forbidden, which crucially alters the direct search allowed parameter space of the model in presence of scalar triplet.

\subsubsection{Spin-independent direct detection constraint}
\label{sec:sidd}

\begin{figure}[htb!]
$$
 \includegraphics[scale=0.38]{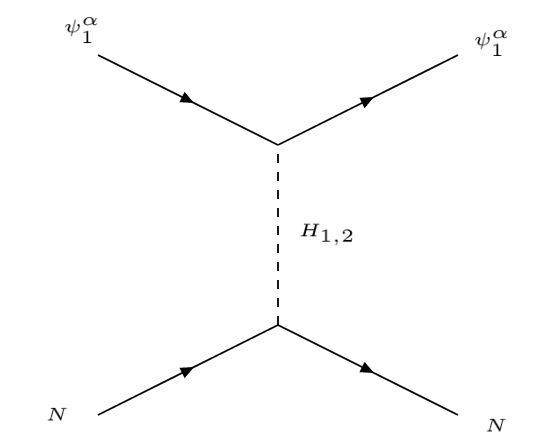}
 $$
 \caption{Feynman graph showing scattering of DM particle against the nucleus. This can be mediated both by the SM Higgs $H_1$ and the triplet Higgs $H_2$.}
 \label{fig:dd}
\end{figure}

From the previous section, we see that for a moderate choice of $y_\psi \simeq 1$, the $Z$ mediated inelastic scattering for the DM will have no contribution if we choose $\sin\theta \gsim 0.05$ limit (as seen from Fig.~\ref{fig:minvelo}). Therefore the DM particles can recoil against the nucleus, giving rise to direct search signature as shown in Fig.~\ref{fig:dd} only through Higgs ($H_{1,2}$) mediation. The spin-independent (SI) direct detection cross section per nucleon is given by
~\cite{Duerr:2015aka}:

\bea
\sigma^{SI} = \frac{1}{\pi A^2} \mu^2 \left|\mathcal{M}\right|^2,
\label{eq:sidd}
\eea

where $A$ is the mass number of the target nucleus, $\mu=\frac{M_{\psi_1}M_N}{M{\psi_1}+M_N}$ is the DM-nucleus reduced mass and $\left|\mathcal{M}\right|$ is the DM-nucleus amplitude, which reads:

\bea
\mathcal{M} = \sum_{i=1,2}\left[Z f_p^i+\left(A-Z\right)f_n^i\right].
\label{eq:amp}
\eea

The effective couplings in Eq.~\ref{eq:amp} are:

\bea
f_{p,n}^i = \sum_{q=u,d,s} f_{T_q}^{p,n} \alpha_q^{i} \frac{m_{p,n}}{m_q}+\frac{2}{27} f_{T_G}^{p,n}\sum_{Q=c,t,b} \alpha_Q^{i}\frac{m_{p,n}}{m_Q},
\label{eq:coupling}
\eea

with

\bea
\alpha_q^{1} = \frac{Y \sin 2\theta \cos\theta_0^{2}}{m_{H_1}^2}\frac{m_q}{v}\\
\alpha_q^{2} = -\frac{Y \sin 2\theta \sin\theta_0^{2}}{m_{H_2}^2}\frac{m_q}{v}. 
\label{eq: alpha}
\eea

Different coupling strengths between the DM and the light quarks are given by~\cite{Durr:2015dna}: $f_{T_u}^{p}=0.020\pm 0.004$, $f_{T_d}^{p}=0.026\pm 0.005$, $f_{T_s}^{p}=0.118\pm 0.062$, $f_{T_u}^{n}=0.014\pm 0.004$, $f_{T_d}^{n}=0.036\pm 0.008$, $f_{T_s}^{n}=0.118\pm 0.062$. The coupling of the DM with the gluons (through one loop graphs) in the target nuclei is taken into account by the effective form factor:

\bea
f_{T_G}^{p,n} = 1-\sum_{q=u,d,s} f_{T_q}^{p,n}.
\eea

\begin{figure}[htb!]
$$
 \includegraphics[scale=0.45]{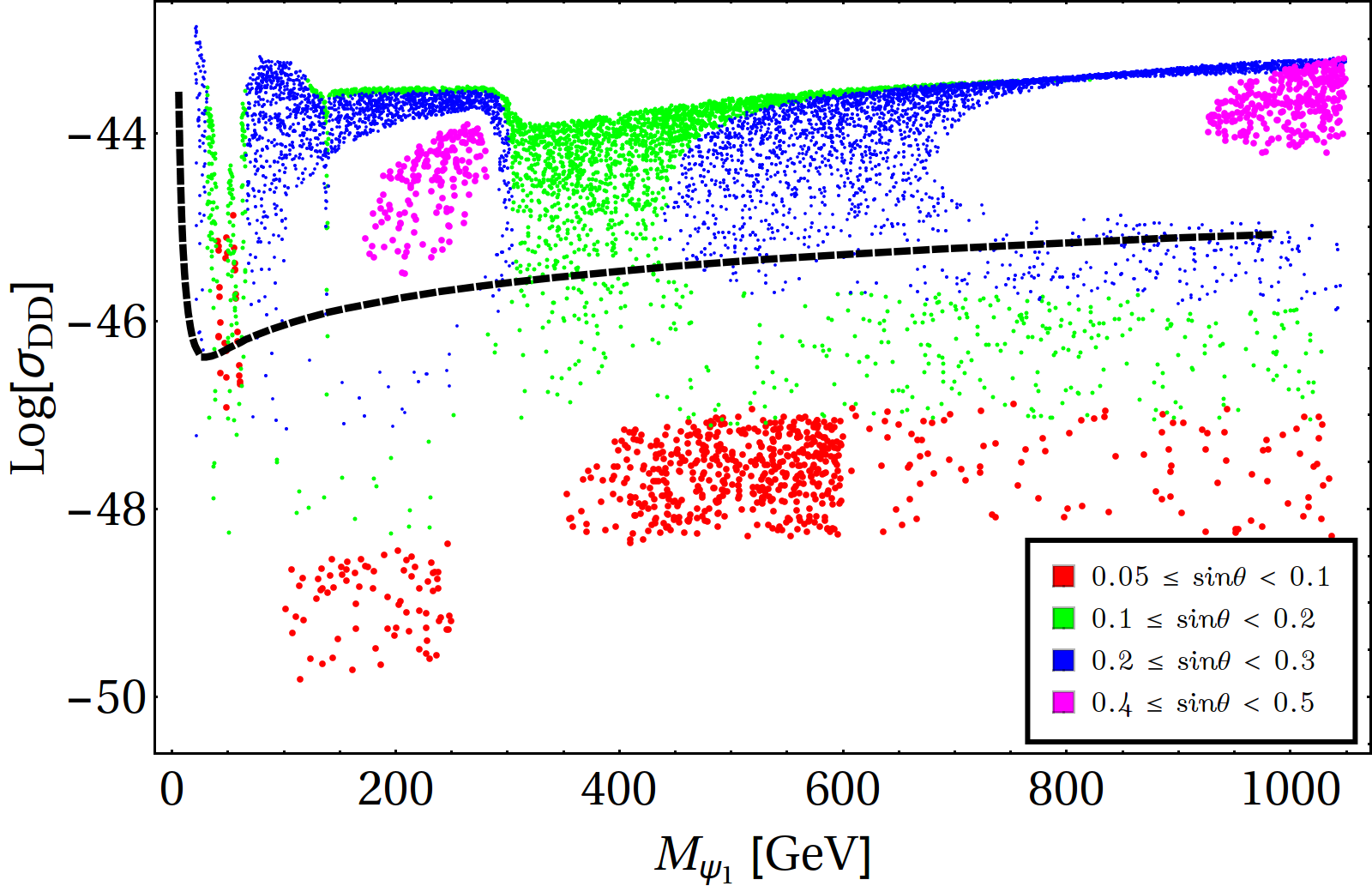}
 $$
 $$
 \includegraphics[scale=0.4]{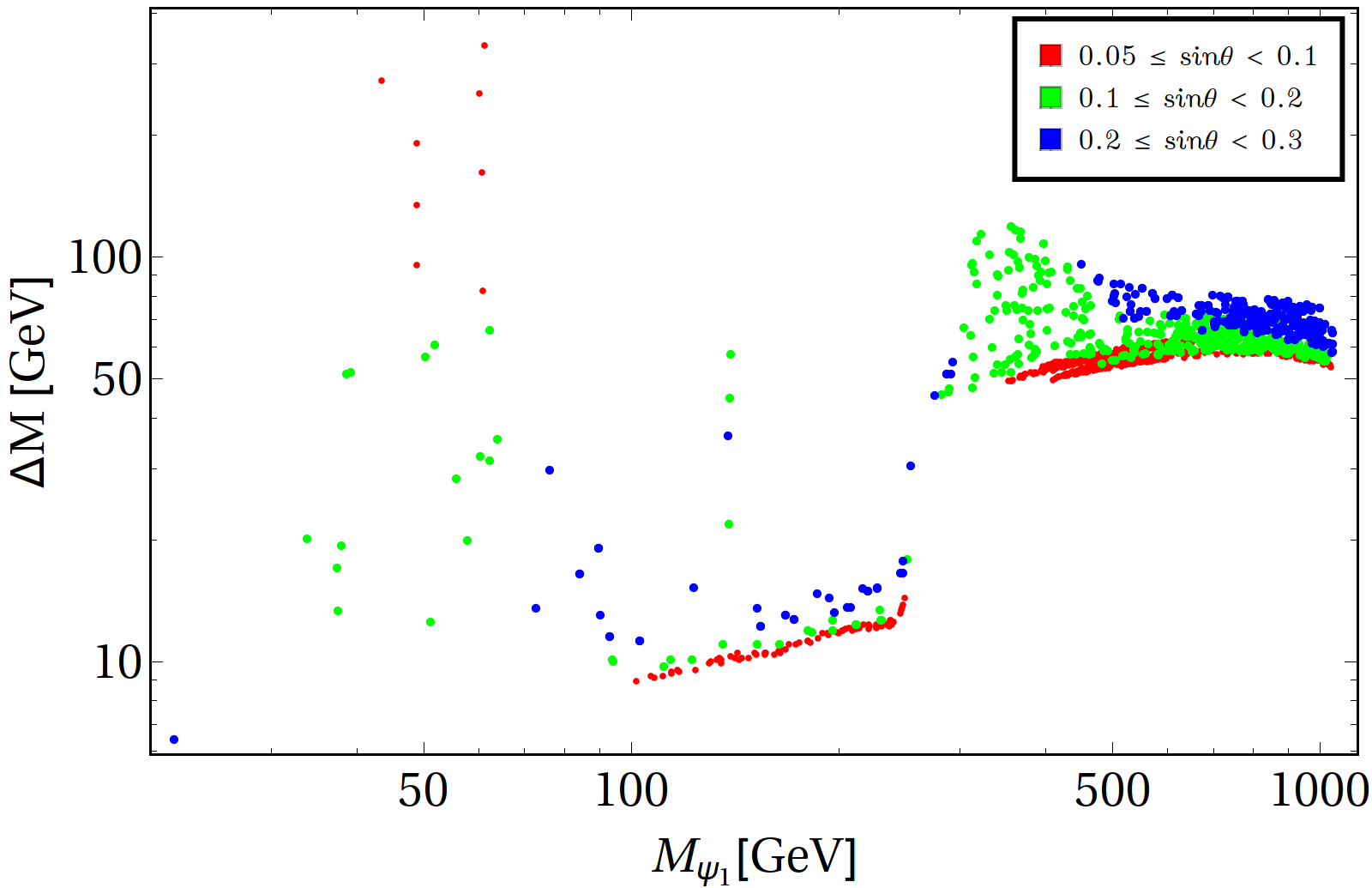}
    \includegraphics[scale=0.42]{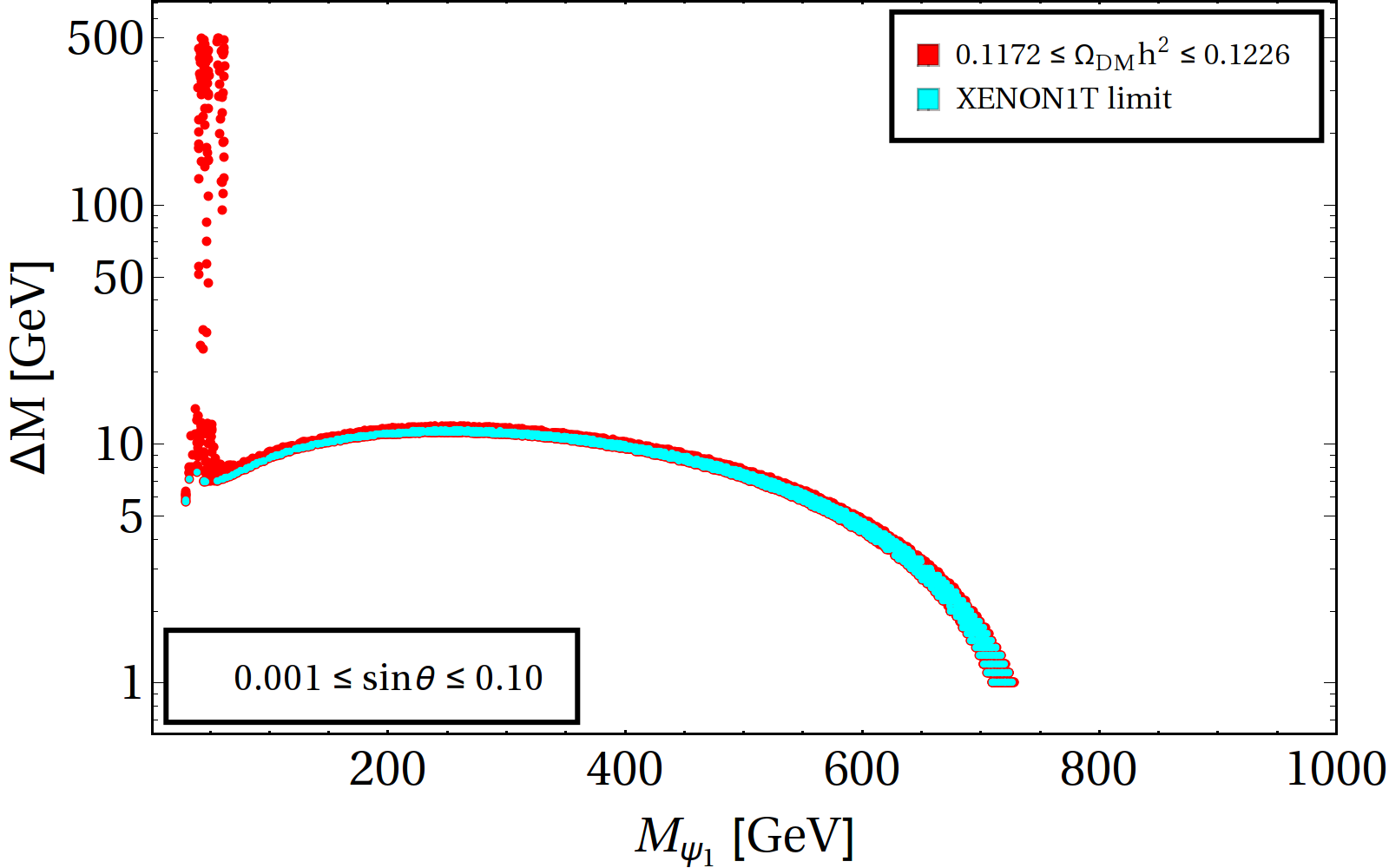}
$$
 \caption{Top: Relic density allowed parameter space satisfied by spin-independent direct detection in direct search plane. Different coloured regions correspond to different singlet-doublet VLF mixings: $\sin\theta=\{0.01-0.1\}$ in red, $\sin\theta=\{0.1-0.2\}$ in green, $\sin\theta=\{0.2-0.3\}$ in blue and $\sin\theta=\{0.4-0.5\}$ in magenta. The black dashed line corresponds to exclusion limit from XENON1T. {\it Bottom left:} Net parameter space left after satisfying relic density and direct detection constraints (color codes are same as that of left figure). {\it Bottom right:} Parameter space allowed by relic abundance and XENON1T exclusion limit but {\it without} the triplet scalar included.}
 \label{fig:direct}
\end{figure}

\begin{table}[htb!]
\begin{center}
\begin{tabular}{|c|c|c|c|c|c|c|c|c|c|}
\hline
Benchmark & $\sin\theta$ & $\Delta M$ & $M_{\psi_1}$  & $\sigma_{DD}$ & $\Omega h^2$ \\ [0.5ex] 
Point     &              &  (GeV)     & (GeV)         & $(cm^2)$      & \\ [0.5ex] 
\hline\hline

BP1 & 0.08 & 161 & 60 & $10^{-46}$ & 0.117 \\
\hline
BP2 & 0.07 & 252 & 60 & $10^{-46}$ & 0.118 \\
\hline
BP3 & 0.06 & 272 & 44 & $10^{-46}$ & 0.117 \\
\hline
BP4 & 0.05 & 332 & 61 & $10^{-46}$ & 0.119 \\
\hline\hline
BP5 & 0.07 & 64 & 47 & $10^{-48}$ & 0.117 \\
\hline
BP6 & 0.24 & 13 & 90 & $10^{-47}$ & 0.117 \\
\hline
BP7 & 0.10 & 47 & 312 & $10^{-47}$ & 0.122 \\
\hline
BP8 & 0.20 & 57 & 140 & $10^{-46}$ & 0.117 \\
\hline
\end{tabular}
\end{center}
\caption {Choices of the benchmark points used for collider analysis. Masses, mixings, relic density and direct search cross-sections for the DM candidate are tabulated.} 
\label{tab:bp}
\end{table}

Upper panel of Fig.~\ref{fig:direct} shows the parameter space allowed by the spin-independent (SI) direct detection cross section in $M_{\psi_1}$-$\sigma_{DD}$ plane. As one can see, the allowed region of parameter space that 
lies below the exclusion limit of present XENON1T data corresponds to $\sin\theta: \{0.015-0.2\}$ (shown in red and green). In the bottom left panel we have shown the net parameter space satisfied by both relic abundance and direct search. 
One should note here, large $\Delta M\gsim 100~\rm GeV$ can be achieved for small $\sin\theta:\{0.015-0.1\}$ near Higgs and $Z$ resonance region and also for $M_{\psi_1}\sim 300~\rm GeV$. Why such regions are available from relic 
density constraint, has already been elaborated before. The reason, that they are not forbidden by direct search can be attributed to forbidden $Z$ mediation, which is possible only when the triplet scalar is present in the model. 
The case of relic density and direct search allowed parameter space for the DM model without the scalar triplet is shown in the right side of bottom panel of Fig.~\ref{fig:direct}. Here we can see, the maximum splitting one may 
achieve is $\Delta M\sim 10~\rm GeV$ with small $\sin\theta$ satisfying both relic density and direct search bounds. This serves as a crucial ingredient to discover such a model at the upcoming Large Hadron Collider (LHC).

Before moving on to the collider section, we shall choose a few benchmark points (BP) which satisfy relic density, direct detection exclusion bound and all the constraints mentioned in Sec.~\ref{sec:constraint}. These are tabulated in 
Tab.~\ref{tab:bp}, where the input parameters and also relic density and direct search outcomes have been mentioned. The BPs are chosen based on different choices of $\Delta M$, where large $\Delta M$ can be probed at the LHC, 
while small $\Delta M$ are better suited for ILC search as we demonstrate. BP1-BP4 can therefore be probed at the LHC because of large $\Delta M$. Due to small $\Delta M$, BP5 and BP6 can be seen at very early run of ILC, while 
BP7 and BP8 can only be probed at ILC with $\sqrt{s}=1~\rm TeV$. Note that, a lower limit on pair-produced charged heavy vector-like leptons have been set by LEP: $m_L>101.2~\rm GeV$ at 95 \%\ C.L. for $L^{\pm}\to \nu W$ 
final states~\cite{Achard:2001qw}. So, our benchmark points are safe from LEP bounds.

We also note here that some of the above benchmark points are subject to the choice of the scalar triplet mass, which is taken as $\sim 300$ GeV in this analysis. Such a choice helps us getting interesting DM phenomenology 
for DM mass in the vicinity of the scalar triplet mass and allows a larger available parameter space. However scalar triplet mass of this order is little fine tuned when neutrino mass is concerned, where the Yukawa coupling required 
turns out to be exceedingly small. While, one may choose a heavier scalar triplet and achieve a larger singlet doublet mixing ($\sin\theta$), a large $\Delta M$ (as in BP7) would have to be shifted to a higher DM mass accordingly. 
We will also then be deprived of collider signature of the scalar triplet.

\section{Collider phenomenology}
\label{sec:collider}

In this section we shall discuss the possibility of probing the model at the ongoing and future collider experiments. As we have already seen, due to the presence of the scalar triplet, large $\Delta M$ is allowed by relic abundance and direct detection bounds in the vicinity of the $Z$ and Higgs resonance. Apart from that, moderate $\Delta M$ can also be achieved near the triplet resonance and at $M_{\psi_1}\simeq M_{H_2}$. We shall see, in the following sections, large $\Delta M$ (and hence larger missing energy) is always favorable at the LHC, au contraire, ILC search is more favoured for smaller $\Delta M$ regions (missing energy peaks at small values). Thus, due to the presence of the triplet scalar, this model provides a scope of being probed both at LHC and ILC searches, which correspond to complementary $\Delta M$ regions of the parameter space. As we have examined, in order to unveil this model at the collider experiments, a high luminosity is required for LHC, while the model may show up in the early runs of ILC at a much lower luminosity. In subsection.~\ref{sec:lhcsearch} we have elaborated the LHC analysis with important kinematic distributions and event rates for both signal (BP1-BP4) and dominant SM backgrounds for $\sqrt{s}=14~\rm TeV$. In subsection.~\ref{sec:ilcsearch} the same is done from ILC perspective for both $\sqrt{s}=350~\rm GeV$, corresponding to an early ILC run and $\sqrt{s}=1~\rm TeV$, corresponding to future prediction.

\subsection{Sensitivity of the signal at the LHC}
\label{sec:lhcsearch}

The charged companion of the VLF doublets can be produced at the LHC via $Z,\gamma$ mediation. These charged particles can decay to DM ($\psi_1$) via $W^{\pm}$, producing missing energy in the final state. Note here, for the BPs chosen for LHC ({\it i.e,} BP(1-4)), the decay happens on-shell as $\Delta M>m_W$ .The charged $W$-bosons further decays into leptons and jets, which are registered in the detector, and also to neutrinos which escape the detector and adds to missing energy. The model, in general, can give rise to three different final states:

\begin{itemize}
 \item Hadronically quiet Opposite sign dilepton (OSD) with missing energy $\left(\ell^{+}\ell^{-}+\slashed{E_T}\right)$.
 \item Single lepton, with two jets plus missing energy $\left(\ell^{\pm}+jj+\slashed{E_T}\right)$.
 \item Four jets plus missing energy $\left(jjjj+\slashed{E_T}\right)$.
\end{itemize}

\begin{figure}[htb!]
$$
 \includegraphics[scale=0.25]{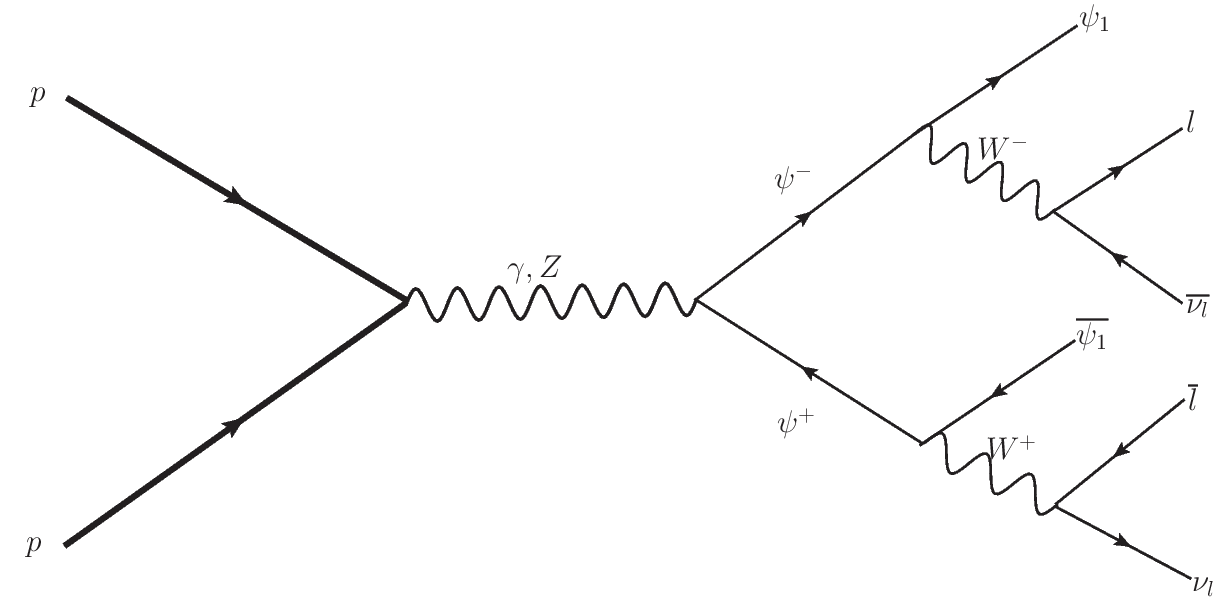}
 $$
 \caption{OSD+$\slashed{E_T}$ final state at the LHC.}
 \label{fig:osdsig}
\end{figure}

As the hadronic final states are infested with SM background, particularly at LHC, while leptonic channels are cleaner, we shall only analyze the OSD final states with missing energy (Fig.~\ref{fig:osdsig}). 

\subsubsection{Object reconstruction and simulation strategy at the LHC}
\label{sec:simul}

We have used {\tt LanHep}~\cite{Semenov:2008jy} to implement the model framework and used {\tt CalcHEP}~\cite{Belyaev:2012qa} in order to generate the parton level events. These events then showered through {\tt PYTHIA}~\cite{Sjostrand:2006za} for hadronization. All events have been simulated at a center of mass energy of $\sqrt{s}=14~\rm TeV$, using {\tt CTEQ6l}~\cite{Placakyte:2011az} as the parton distribution function. To mimic the collider environment, the leptons and jets are re-constructed using the following criteria:

\begin{itemize}
 \item {\it Lepton ($l=e,\mu$):} Leptons are identified with a minimum transverse momentum $p_T>20$ GeV and pseudorapidity $|\eta|<2.5$. Two leptons are isolated objects if their mutual distance 
 in the $\eta-\phi$ plane is $\Delta R=\sqrt{\left(\Delta\eta\right)^2+\left(\Delta\phi\right)^2}\ge 0.2$, while the separation between a lepton and a jet has to satisfy $\Delta R\ge 0.4$.
 
 \item {\it Jets ($j$):} All the partons within $\Delta R=0.4$ from the jet initiator cell are included to form the jets using the cone jet algorithm {\tt PYCELL} built in {\tt PYTHIA}. We demand $p_T>20$ GeV for a clustered object to be considered as jet. Jets are isolated from unclustered objects if $\Delta R>0.4$. Although our signal events (hadronically quiet OSD) do not carry jets, the definition of jet turns out to be important in order for  the signal to be identified with zero jet veto.  
 
 \item {\it Unclustered Objects:}  All the final state objects which are neither clustered to form jets, nor identified as leptons, belong to this category. All particles with $0.5<p_T<20$ GeV and $|\eta|<5$, are considered as unclustered. Again, unclustered objects do not enter into our signal definition, but is important in identifying missing energy of the event.
 
 \item {\it Missing Energy ($\slashed{E}_T$):} The transverse momentum of all the missing particles (those are not registered in the detector) can be estimated from the momentum imbalance in the transverse direction associated to the visible particles. Missing energy (MET) is thus defined as:
 \bea
 \slashed{E}_T = -\sqrt{(\sum_{\ell,j} p_x)^2+(\sum_{\ell,j} p_y)^2},
 \eea
 where the sum runs over all visible objects that include the leptons, jets and the unclustered components. 
 
 \item {\it Invariant dilepton mass $\left(m_{\ell\ell}\right)$}: We can construct the invariant dilepton mass variable for two opposite sign leptons by defining:
 \bea
 m_{\ell\ell}^2 = \left(p_{\ell^{+}}+p_{\ell^{-}}\right)^2.
 \eea
 Invariant mass of OSD events, if created from a single parent, peak at the parent mass, for example, $Z$ boson. As the signal events (Fig.~\ref{fig:osdsig}) do not arise from a single parent particle, invariant mass cut plays a crucial role in eliminating the $Z$ mediated SM background. 
 
 \item $H_T$: $H_T$ is defined as the scalar sum of all isolated jets and lepton $p_T$'s:
 \bea
 H_T = \sum_{\ell,j} p_T. 	
 \eea
 Of course, for our signal, the sum only includes the two leptons that are present in the final state. 
 \end{itemize}
 
It is very important for collider analysis to estimate the SM background that mimic the signal. All the dominant SM backgrounds have been generated in {\tt MadGraph}~\cite{Alwall:2014hca} and then showered through {\tt PYTHIA}. 

\subsubsection{Event rates and signal significance at the LHC}
\label{sec:events}

\begin{figure}[htb!]
$$
 \includegraphics[scale=0.4]{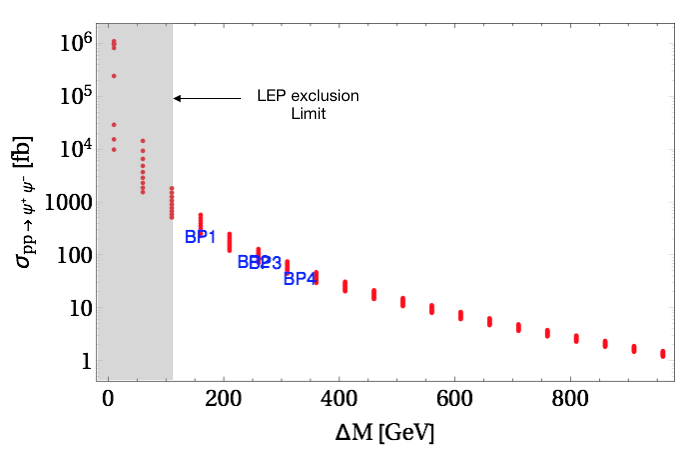}
$$
  \caption{Variation of production cross section $\sigma_{pp\to\psi^{+}\psi^{-}}$ at LHC with $\Delta M$ for $\sqrt s=14$ TeV. DM mass is varied between $M_{\psi_1}:\{1-65\}$ GeV. 
  Different benchmark points (BP1-BP4, see Tab.~\ref{tab:sigevent}) are also indicated in blue.  BP2 and BP3 are superimposed on each other as they have almost the same production cross-section. 
  LEP limit on charged fermion mass is also shown by the shaded region.}
 \label{fig:csvar}
\end{figure}

\begin{figure}[htb!]
$$
 \includegraphics[scale=0.5]{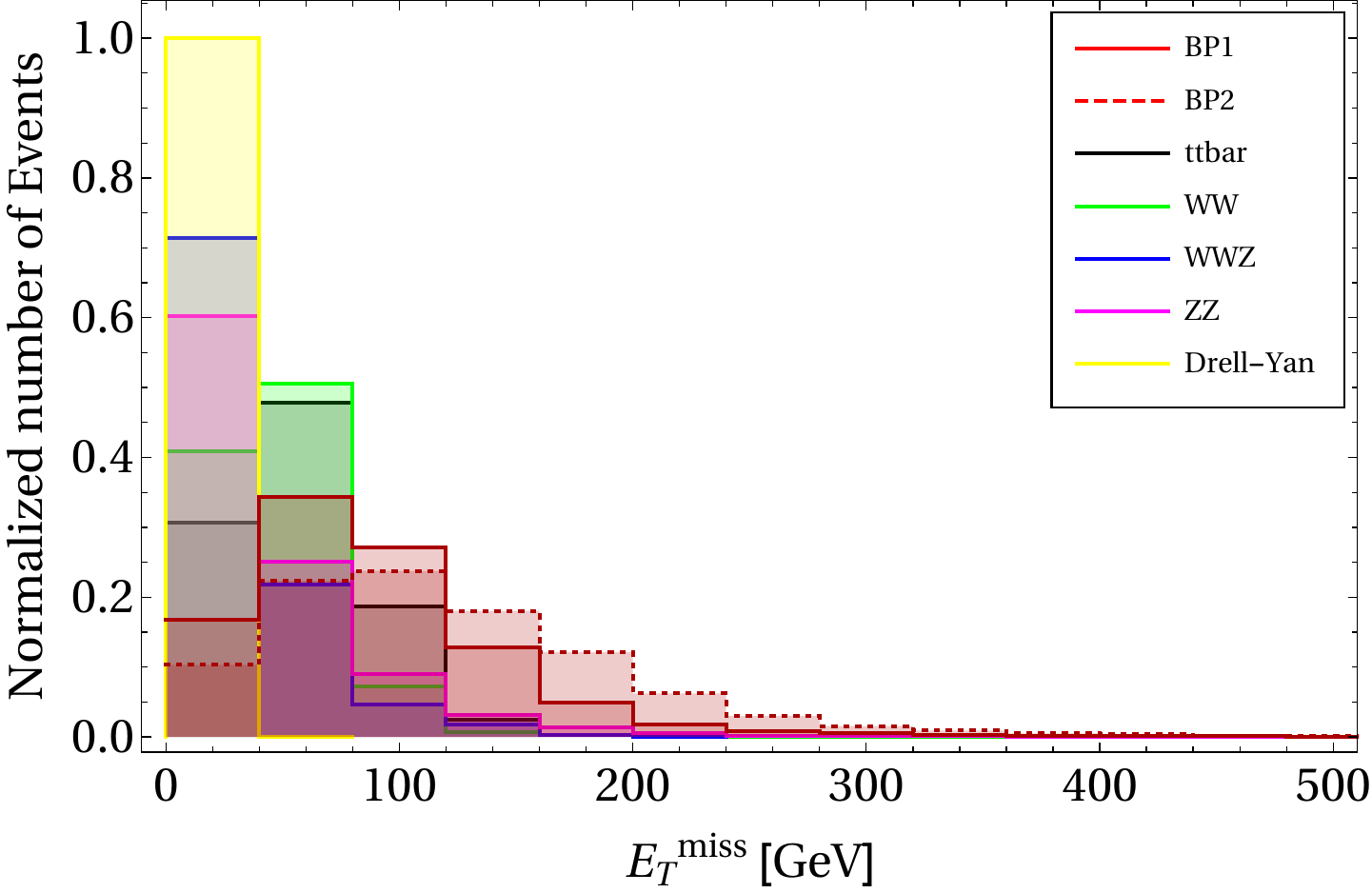}
  \includegraphics[scale=0.5]{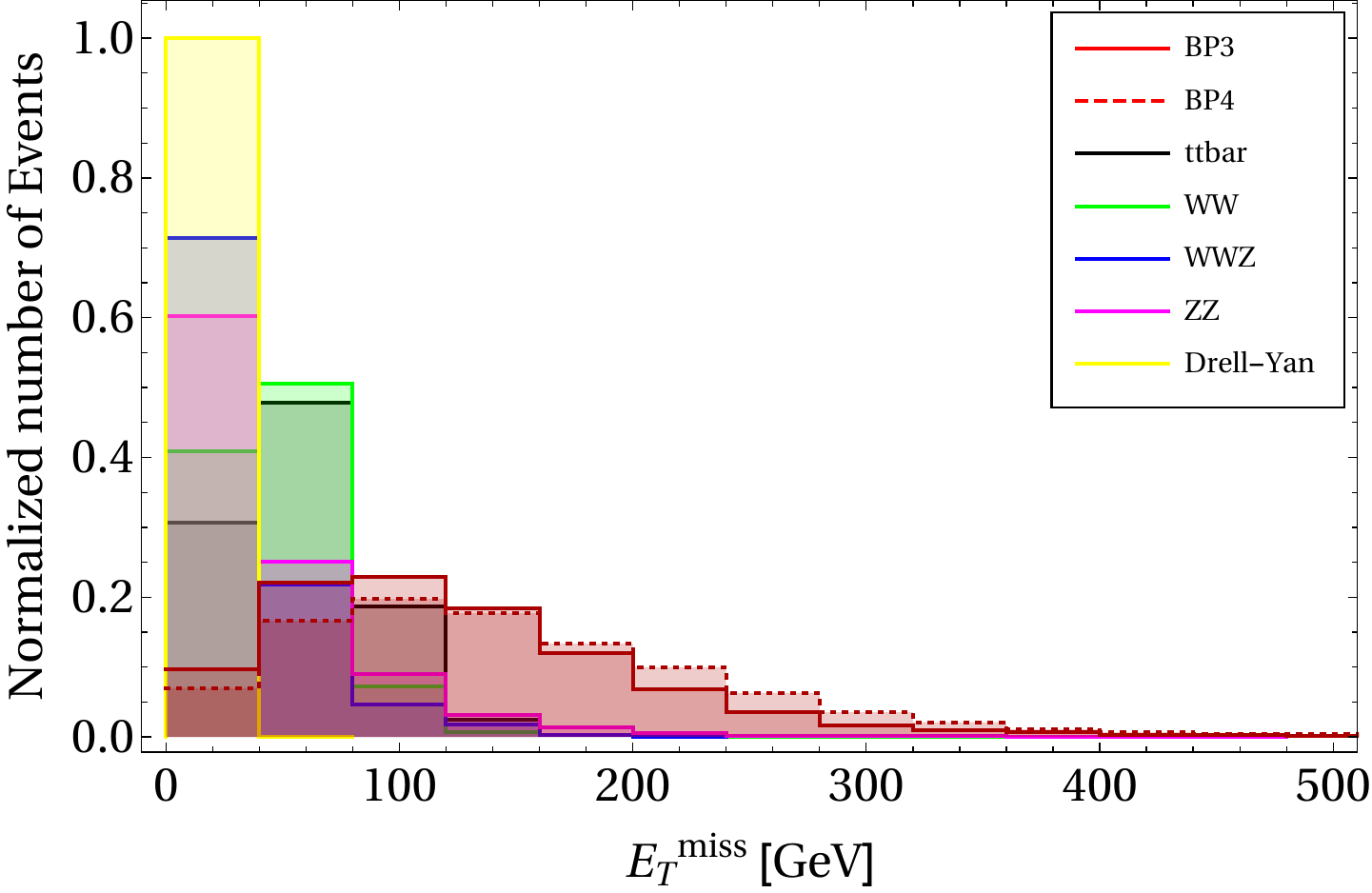}
 $$
 $$
 \includegraphics[scale=0.5]{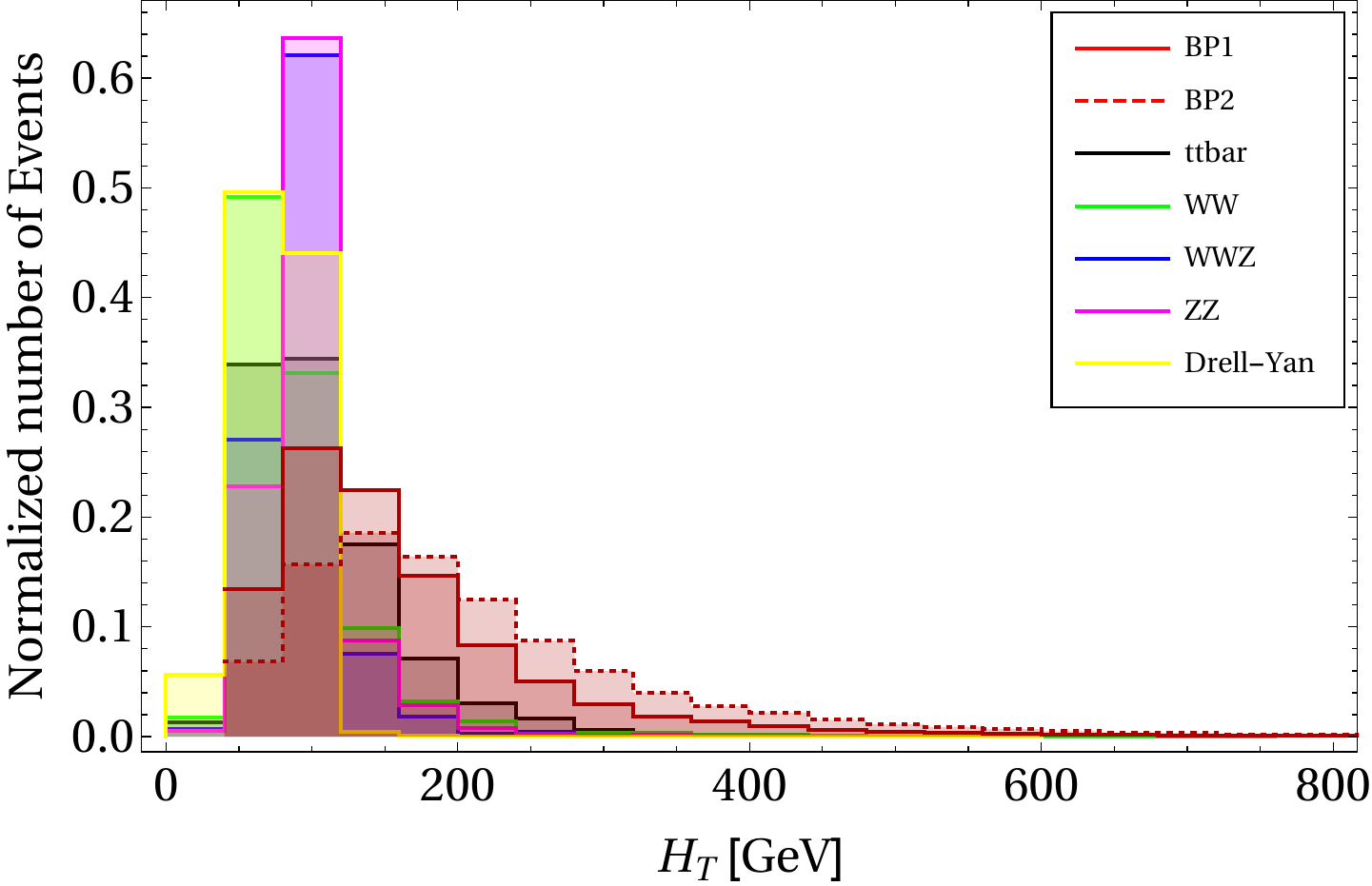}
  \includegraphics[scale=0.5]{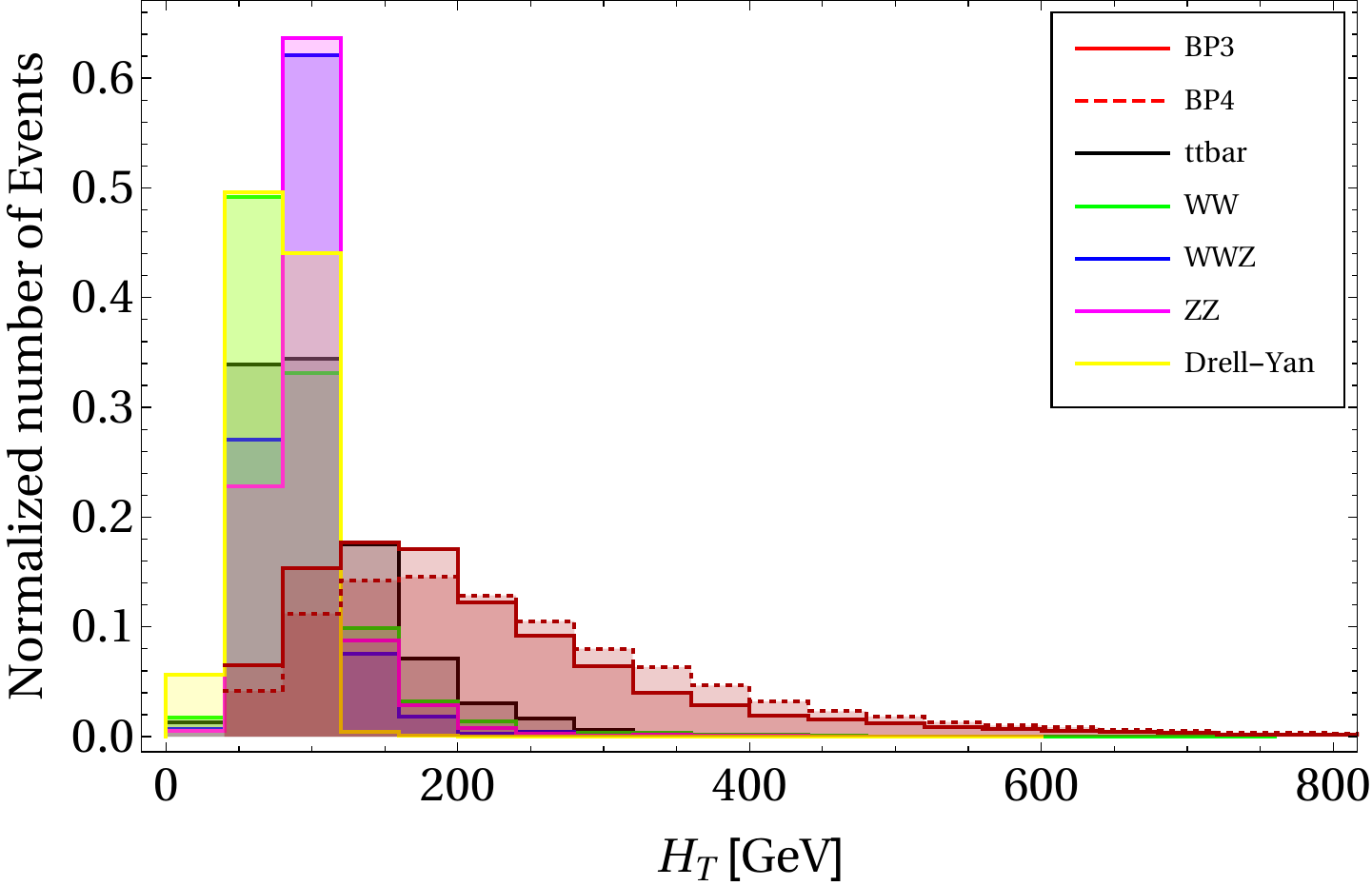}
 $$
 \caption{Top: Missing energy distribution for OSD+$\slashed{E}_T$ final state for the benchmark points are shown in red. Those of the dominant SM backgrounds are also shown with 
 different colours. Bottom: $H_T$ distribution for the same. The simulation is done assuming LHC with $\sqrt s=14$ TeV.}
 \label{fig:metht}
\end{figure}

We have shown the variation of production cross-section $\sigma_{pp\to\psi^{+}\psi^{-}}$ at LHC for $\sqrt s=14$ TeV with $\Delta M$ for different DM masses ranging between $M_{\psi_1}:\{1-65\}$ GeV in Fig.~\ref{fig:csvar}. As expected, with larger $\Delta M$ the cross-section for $\psi^{+}\psi^{-}$ falls due to phase space suppression with $M_{\psi^\pm}=M_{\psi_1}+\Delta M$. The production cross-section for the benchmark points (BP1-BP4), relevant for the LHC search are also indicated in the same plot. We see that, BP2 and BP3 fall on each other as they have almost equal production cross-section. LEP exclusion for the charged fermion is also shown by the shaded grey region ($M_{\psi^\pm}>101.2$ GeV).

In Fig.~\ref{fig:metht}, the MET and $H_T$ distribution for the BPs (along with the SM dominant backgrounds) are shown in top and bottom panels respectively. The cross-section for all the SM backgrounds have been calculated upto next-to-leading order using appropriate $K$-factors~\cite{Alwall:2014hca}. Since the background dominates over the signal, we have employed $\slashed{E_T}$ and $H_T$ cuts to distinguish the signal region from the background. For the background the only source of missing energy is the SM neutrinos, while for the signal, along with the SM neutrinos, MET also comes from the DM produced during the decay of the charged VLFs.  With larger $\Delta M$ the MET distribution gets flattened as more $p_T$ is being carried away by the DM. This is what is seen from the MET distributions, particularly we see that BP1 with least $\Delta M$ is almost falling on top of SM background. So, the efficiency of using an MET cut to select the signal is also the least here. It is therefore obvious that the other BPs like BP5-BP8 (not shown in this distribution) will not be able to survive any large MET cut. $H_T$ distributions are almost similar to that of MET. We have finally employed following cuts (with zero jet veto) in order to separate the signal from the background:

\begin{itemize}
 \item $\slashed{E_T}> 300~\rm GeV$ is employed to kill all the backgrounds. Although, as it can be seen from Fig.~\ref{fig:metht}, $\slashed{E_T}>150~\rm GeV$ is good enough to separate the siganl from the background, but the $W^{+}W^{-}$ background will still persist, hence we chose a hard cut on MET.
 \item $H_T>100~\rm GeV$ is used to reduce the background further, without harming the signal events.
 \item Invariant mass cut over the $Z$-window $|m_z-15|<m_{ll}<|m_Z+15|$ is required to get rid-off the $ZZ$ background to a significant extent.
\end{itemize}

Next, we would like to see the number of signal and corresponding background events using the cuts mentioned above. In Tab.~\ref{tab:sigevent}, we have tabulated the number of events for the signal at a future luminosity of $\mathcal{L}=100~\rm fb^{-1}$ with all the cuts incorporated. The cross-sections are also quoted in each case and a set of two different MET cuts have been illustrated to demonstrate the cut-flow. With larger MET cut the number of final state signal events get diminished as expected. The effective number of events at a particular luminosity ($\mathcal{L}$) as has been mentioned in Tab.~\ref{tab:sigevent} is obtained from the simulated events in the following way:

\bea
N_{\text{eff}} = \frac{\sigma_{\text{p}}\times n}{N}\times\mathcal{L},
\label{eq:neff}
\eea
where $\sigma_p$ is production cross-section as shown in Fig.~\ref{fig:csvar}, $n$ is the number of events generated out of $N$ simulated events (after putting all the cuts and showering through {\tt PYTHIA}) 
and $\mathcal{L}$ is the luminosity, which we have considered to be $100~\rm fb^{-1}$.

\begin{table}[htb!]
\begin{center}
\scalebox{1.0}{
\begin{tabular}{|c|c|c|c|c|c|c|c|c|}
\hline

Benchmark Point & $\sigma^{\psi^{+}\psi^{-}}$ (fb) & $\slashed{E}_T$ (GeV) & $\sigma^{\text{OSD}} (fb)$ & $N^{\text{OSD}}_{\text{eff}}@\mathcal{L}=100~\rm fb^{-1}$ \\
\hline\hline

&   & $>$ 200 & 0.13  & 13 \\
\cline{3-5}
BP1 & 218.19  & $>$ 300 & 0.04  & 4 \\
\hline 

&   & $>$ 200 & 0.15 & 15 \\
\cline{3-5}
BP2 & 74.80 & $>$ 300 & 0.04  & 4 \\
\hline 

&   & $>$ 200 & 0.17 & 17 \\
\cline{3-5}
BP3 & 71.80  & $>$ 300  &  0.04 & 4 \\
\hline

&   & $>$ 200 & 0.13 & 13 \\
\cline{3-5}
BP4 &  35.93   & $>$ 300 & 0.03  & 3 \\
\hline
\end{tabular}
}
\end{center}
\caption {Signal events with $\sqrt{s}$ = 14 TeV at the LHC for luminosity $\mathcal{L} = 100~fb^{-1}$ for the benchmark points (BP1-BP4) in Tab.~\ref{tab:bp}.}
\label{tab:sigevent}
\end{table}

\begin{table}[htb!]
\begin{center}
\scalebox{1.0}{
\begin{tabular}{|c|c|c|c|c|c|c|c|c|}
\hline

Backgrounds & $\sigma_{production}$ (pb) & $\slashed{E}_T$ (GeV) & $\sigma_{OSD}$ & $N^{\text{OSD}}_{\text{eff}}@\mathcal{L}=100~\rm fb^{-1}$ \\
\hline\hline

&   & $>$ 200 & $<$0.81  & 0 \\
\cline{3-5}
$t\bar{t}$ & 814.64    & $>$ 300 & $<$0.81  & 0 \\
\hline 

&   & $>$ 200 & 1.99 & 199 \\
\cline{3-5}
$W^{+}W^{-}$ & 99.98 & $>$ 300 & $<$0.49  & $<$1 \\
\hline 

&   & $>$ 200 & 0.04 & 4 \\
\cline{3-5}
$W^{+}W^{-}Z$ & 0.15   & $>$ 300  &  0.01 & 1 \\
\hline

&   & $>$ 200 & $<$0.07 & 0 \\
\cline{3-5}
$ZZ$ &  14.01   & $>$ 300 & $<$0.07  & 0 \\
\hline
\end{tabular}
}
\end{center}
\caption {Events for dominant SM backgrounds with $\sqrt{s}$ = 14 TeV at the LHC for luminosity $\mathcal{L} = 100~fb^{-1}$. The cross-sections are quoted in NLO order by multiplying with appropriate $K$-factors (see text).}
\label{tab:bckevnt}
\end{table} 

\begin{figure}[htb!]
 $$
 \includegraphics[scale=0.6]{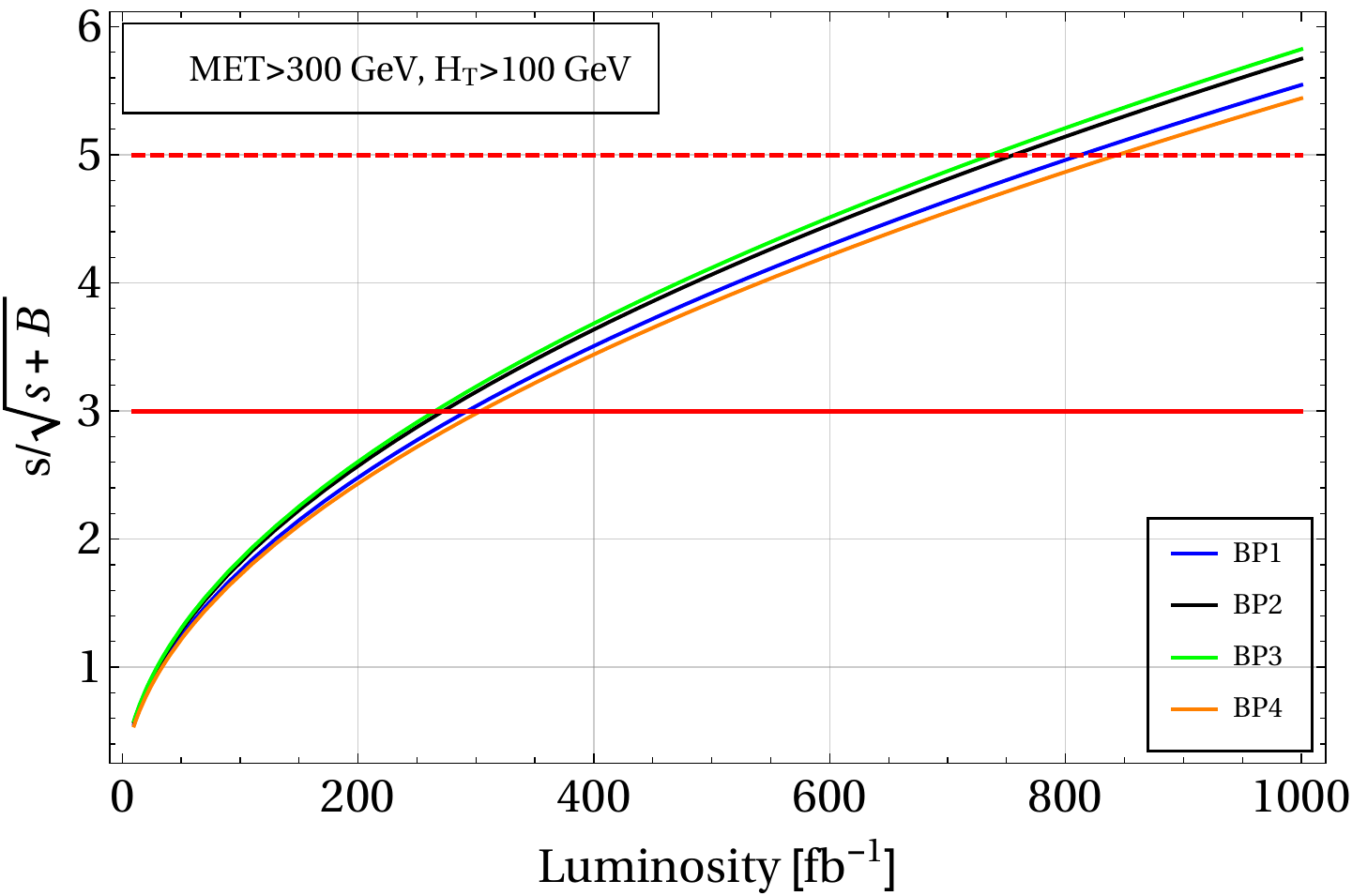}
 $$
 \caption{Signal significance for different BPs, where we have used $\slashed{E_T}>300~\rm GeV$ and $H_T>100~\rm GeV$. The solid red and dashed red lines correspond to 3$\sigma$ and 5$\sigma$ discovery limits respectively.}
 \label{fig:signi}
\end{figure}

Tab.~\ref{tab:bckevnt} enlists the number of events coming from dominant SM backgrounds after using the same set of cuts mentioned before. Events from $t\bar{t}$ and $ZZ$ can be eliminated to a significant extent by demanding zero jet veto and putting a high MET cut (along with the $m_{\ell\ell}$ cut for $ZZ$ events in particular). The hard MET cut also helps to get rid off the $W^{+}W^{-}$ background. The only background that remains (although with only one event) is that from $W^{+}W^{-}Z$. But the cuts employed also eliminate some of the signal events, making the significance is low. 

The discovery potential of hadronically quiet OSD signal for different BPs are shown in Fig.~\ref{fig:signi}, as a function of luminosity. We have chosen $\slashed{E_T}>300~\rm GeV$ and $H_T>100~\rm GeV$ to compute the signal significance so that the SM background is minimum. As one can see from Tab.~\ref{tab:sigevent}, the number of signal events left after imposing the cuts are more or less the same for all the benchmark points. This is also reflected in Fig.~\ref{fig:signi}, where we can see all the BPs reach a 
5$\sigma$ discovery at a luminosity $\mathcal{L}\sim 800~\rm fb^{-1}$. Here we would like to remind once more, the possibility of getting a signal excess in hadronically quiet OSD channel is due to the presence of the scalar triplet, without which the model would have failed to produce any such signature at the LHC. We will later discuss the possibility of seeing a displaced vertex signal and this adds to the Complementarity of the search strategy of this model.

\subsection{Sensitivity of the signal at the ILC}
\label{sec:ilcsearch}

\begin{figure}[htb!]
$$
 \includegraphics[scale=0.25]{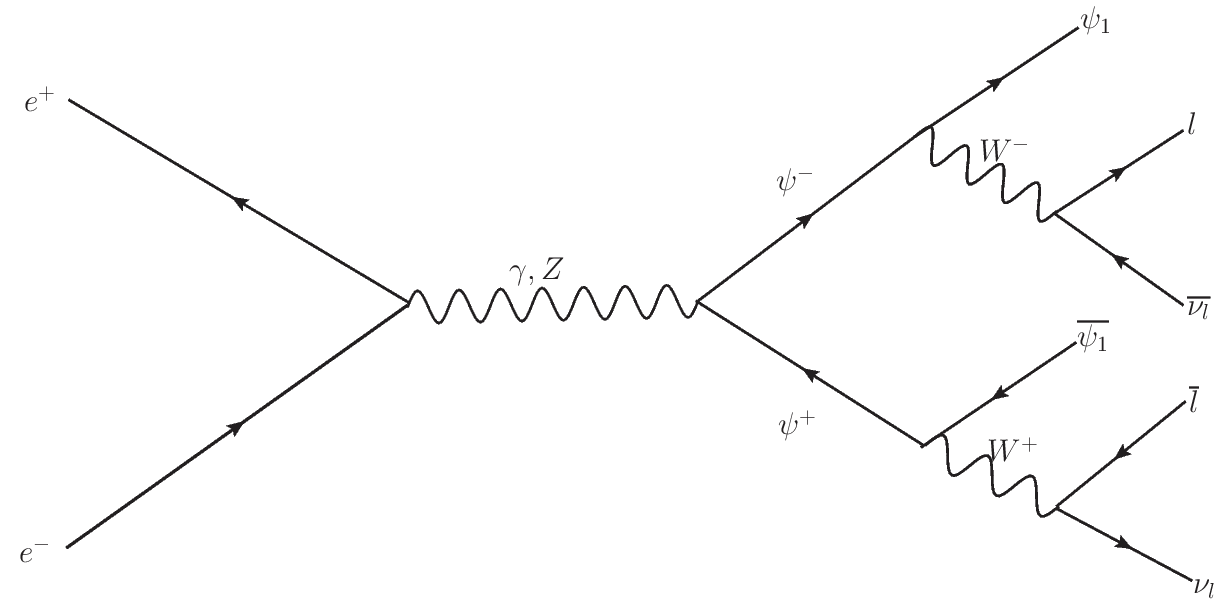}
$$
  \caption{OSD+$\slashed{E_T}$ signal at the ILC.}
 \label{fig:ilcprod}
\end{figure}

The VLFs can also be produced at the ILC via gauge mediation as shown in Fig.~\ref{fig:ilcprod}. The model thus can be probed at the ILC in the same $\ell^+\ell^-+\slashed{E_T}$ final state as that of the LHC. However, one may note that unlike LHC, jet rich final state signal at the ILC is not disfavored due to smaller SM background contribution due to absence of QCD processes like $t\bar{t}$. Therefore, we are still left with the SM gauge boson productions to potentially mimic our signal. One can still analyse the single lepton plus jet channel or dijet channel at the ILC, but to show the complementarity of the hadronically quiet dilepton final state signature at the LHC and the ILC, we analyze this particular channel in details here. The main goal is to show  sensitivity of the signal for different choices of $\Delta M$ that can be probed at the ILC, which can not be probed at the LHC. We shall demonstrate, because of smaller $\Delta M$, BP5-BP8 are suitable for ILC searches. Of the four BPs, BP5 and BP6 can be probed at the early run of ILC with $\sqrt{s}=350~\rm GeV$, while BP7 and BP8 need higher $\sqrt{s}$.

\subsubsection{Object reconstruction and simulation strategy at the ILC}
\label{sec:ilcobj}

As before, we have generated the parton-level signal events in {\tt CalcHEP} and showered them through {\tt PYTHIA}, while the relevant background events are generated via {\tt MadGraph}. Now, for event reconstruction, we have used the following criteria~\cite{Kamon:2017yfx}:

\begin{itemize}
\item Leptons are required to have $p_T(l)>10~\rm GeV$ where $l=\mu,e$ with pseudorapidity $|\eta|<2.4$. Two leptons are said to be isolated if $\Delta R\ge 0.2$, while a lepton and a jet can be identified as separate objects if $\Delta R\ge 0.4$.
 
 \item Jets are reconstructed using the cone jet algorithm in-built in {\tt PYTHIA}. Objects with $p_T(j)>20~\rm GeV$ and $|\eta|<3.0$ are considered as jets. Again, this is required so that we select events for the desired signal with 
 zero jet veto.
\end{itemize}

Now, ILC will be providing highly polarized electron beam ($P_{e^-}:$~80 \%) and moderately polarized positron beam ($P_{e^+}:$~20 \%)~\cite{Behnke:2013xla}. We have used $+$ sign for right polarization and $-$ for left polarization. In order to minimize the SM background, we have looked into three different polarizations of the incoming beam:

\begin{itemize}
 \item 80 \%\ left polarized $e^-$ and 20\%\ right polarized $e^+$ beam ($\left[P_{e^-},P_{e^+}\right]:$~[-80 \%,+20 \%]).
 \item 80 \%\ right polarized $e^-$ and 20\%\ left polarized $e^+$ beam ($\left[P_{e^-},P_{e^+}\right]:$~[+80 \%,-20 \%]).
 \item Unpolarized incoming beams ($\left[P_{e^-},P_{e^+}\right]:$~[0 \%,0 \%]).
\end{itemize}

\subsubsection{Event rates and signal significance at the ILC}
\label{sec:events}

Production cross-section of the dominant SM backgrounds with different beam polarizations are tabulated in Tab.~\ref{tab:polbcktab1} and Tab.~\ref{tab:polbcktab} for $\sqrt{s}=350~\rm GeV$ and $\sqrt{s}=1~\rm TeV$ respectively. One can notice that all the SM background cross-sections are minimum for $(P_{e^-},P_{e^+})$=(+80\%,-20\%). This is because left handed particles form a doublet under $SU(2)$, boosting the SM gauge boson production for dominantly left polarised beams. On the other hand, right handed electrons are singlet under $SU(2)$ and therefore, dominantly right polarized beams will suppress the SM gauge boson production. The case of unpolarised beam falls in between the two extreme cases described here. The signal cross-section will also change similarly due to the choice of beam polarization. However, the final state fermions being vector-like, the change will only appear at the SM vertex (left vertex of Fig.~\ref{fig:ilcprod}) due to change in polarization. Therefore, the change in cross-section for the signal due to change in polarization of the electron beam will be milder. The signal $\psi^+\psi^-$ production cross-section with the polarization of the beams is tabulated in Tab.~\ref{tab:polsigtab1} and Tab.~\ref{tab:polsigtab} for $\sqrt{s}=350~\rm GeV$ and $\sqrt{s}=1~\rm TeV$ respectively. We have therefore chosen dominantly right polarized beams i.e. $(P_{e^-},P_{e^+})$=(+80\%,-20\%) 
for the maximum signal sensitivity of the model at ILC. 

\begin{table}[htb!]
\begin{center}
\scalebox{1.0}{
\begin{tabular}{|c|c|c|c|c|c|c|c|c|}
\hline
$P_{e^-}$ & $P_{e^+}$ & $\sigma\left(W^+W^-\right)$ (pb) & $\sigma\left(W^+W^-Z\right)$ (pb) & $\sigma\left(ZZ\right)$ (pb)\\
\hline\hline
&   &  & & \\
-80\%\ & +20\%\  & 24.37 & 0.026 & 1.08 \\
&   &  & & \\
\hline
&   &  & & \\
+80\%\ & -20\%\  & 1.90 & 0.002 & 0.49 \\
&   &  & & \\
\hline
&   &  & & \\
0\%\ & 0\%\  & 11.31 & 0.012 & 0.67  \\
&   &  & & \\
\hline
\end{tabular}
}
\end{center}
\caption {Dominant SM background cross-sections for different polarization of the $e^-e^+$ beams at $\sqrt{s}=350~\rm GeV$ at the ILC.}
\label{tab:polbcktab1}
\end{table}

\begin{table}[htb!]
\begin{center}
\scalebox{1.0}{
\begin{tabular}{|c|c|c|c|c|c|c|c|c|}
\hline
$P_{e^-}$ & $P_{e^+}$ & $\sigma\left(W^+W^-\right)$ (pb) & $\sigma\left(W^+W^-Z\right)$ (pb) & $\sigma\left(ZZ\right)$ (pb)\\
\hline\hline 
&   &  & & \\
-80\%\ & +20\%\  & 5.87 & 0.12 & 0.23 \\
&   &  & & \\
\hline 
&   &  & & \\
+80\%\ & -20\%\  & 0.43 & 0.009 & 0.11 \\
&   &  & & \\
\hline 
&   &  & & \\
0\%\ & 0\%\  & 2.65 & 0.05 & 0.15  \\
&   &  & & \\
\hline
\end{tabular}
}
\end{center}
\caption {Dominant SM background cross-sections for different polarization of the $e^-e^+$ beams at $\sqrt{s}=1~\rm TeV$ at the ILC.}
\label{tab:polbcktab}
\end{table}

\begin{table}[htb!]
\begin{center}
\scalebox{1.0}{
\begin{tabular}{|c|c|c|c|c|c|c|c|c|}
\hline
$P_{e^-}$ & $P_{e^+}$ & $\sigma$(BP5) (fb) & $\sigma$(BP6) (fb) \\
\hline\hline
 &   &  &  \\
-80\%\ & +20\%\  & 1225.7 & 1252.5 \\
&   &  & \\
\hline
&   &  &   \\
+80\%\ & -20\%\  & 690.32 & 705.13 \\
&   &  &   \\
\hline
&   &  &   \\
0\%\ & 0\%\  & 958.01 & 978.56 \\
&   &  &   \\
\hline
\end{tabular}
}
\end{center}
\caption {Variation of $\psi^+\psi^-$ production cross-section at $\sqrt{s}=350~\rm GeV$ with different choices of polarization of the incoming beam at the ILC for benchmark points BP5 and BP6.}
\label{tab:polsigtab1}
\end{table}

\begin{table}[htb!]
\begin{center}
\scalebox{1.0}{
\begin{tabular}{|c|c|c|c|c|c|c|c|c|}
\hline
$P_{e^-}$ & $P_{e^+}$ & $\sigma$(BP5) (fb) & $\sigma$(BP6) (fb)   & $\sigma$(BP7) (fb) & $\sigma$(BP8) (fb) \\
\hline\hline
 &   &  & & &\\
-80\%\ & +20\%\ & 156.44 & 155.48 & 136.9 & 155.17 \\
&   & && & \\
\hline
&   &  &  &&\\
+80\%\ & -20\%\ & 90.46 & 90.41 & 79.16 & 89.55 \\
&   &  & & &\\
\hline
&   &  &  &&\\
0\%\ & 0\%\ & 123.45 & 123.48 & 108.27 & 122.44 \\
&   &  & &&\\
\hline
\end{tabular}
}
\end{center}
\caption {Variation of $\psi^+\psi^-$ production cross-section at $\sqrt{s}=1~\rm TeV$ with different choices of polarization of the incoming beam at the ILC for 
benchmark points BP5, BP6, BP7 and BP8.}
\label{tab:polsigtab}
\end{table} 

\begin{figure}[htb!]
$$
 \includegraphics[scale=0.45]{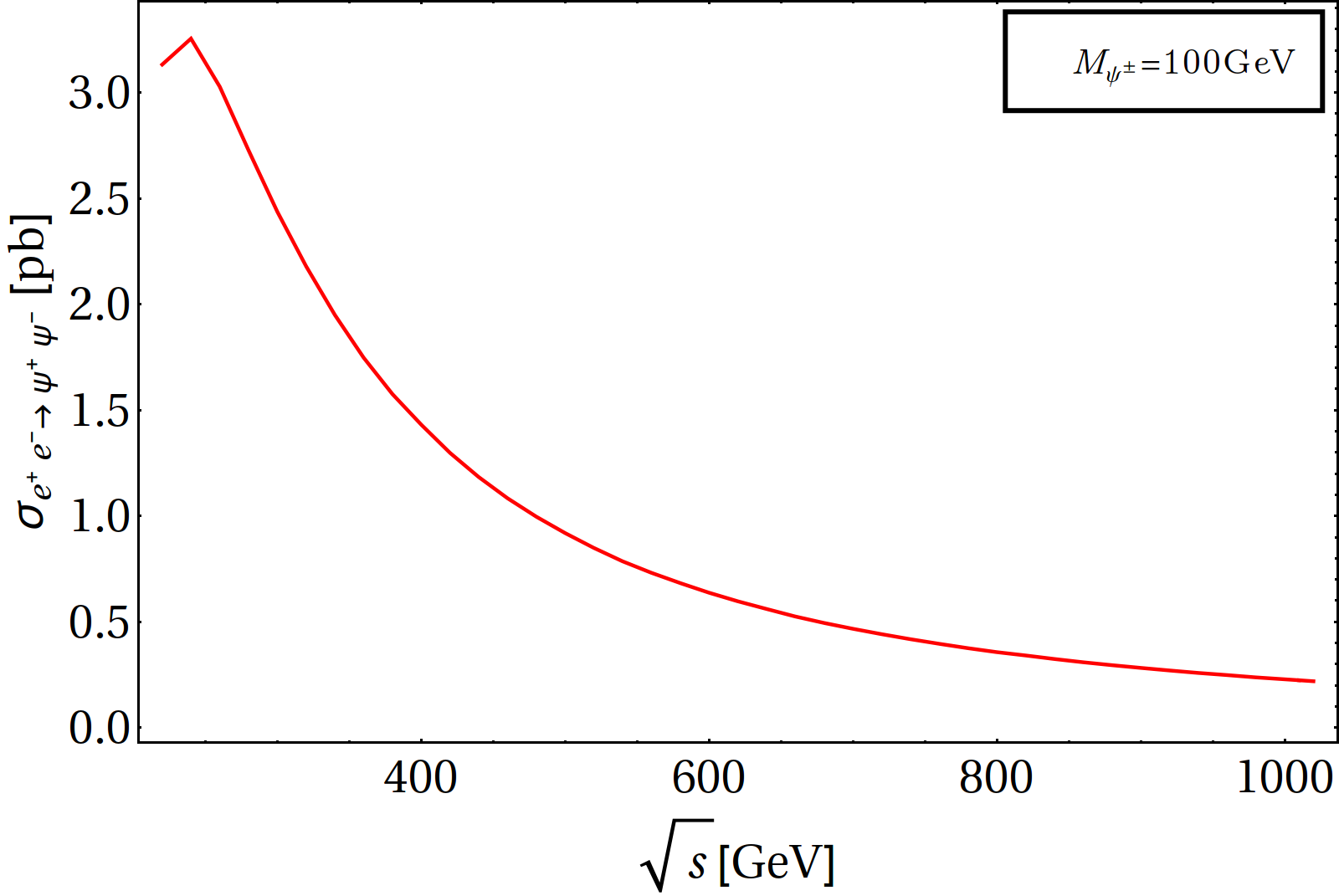}~~
$$
  \caption{Variation of production cross section for the signal $\psi^+\psi^-$ with $\sqrt{s}$ at ILC. $M_{\psi^{\pm}}=100$ GeV is chosen as an illustration.}
 \label{fig:csvarPlt_ilc}
\end{figure}

It is important to note here, all the cross-sections, irrespective of the signal or the SM background, diminish significantly at higher center-of-mass energy with $\sqrt{s}=1~\rm TeV$. This is simply due to the fact that cross-section diminishes as $\frac{1}{s}$. This is shown for $\psi^+\psi^-$ production cross-section with $M_\psi^{\pm}=100$ GeV in Fig.~\ref{fig:csvarPlt_ilc}. 

\begin{figure}[htb!]
$$
 \includegraphics[scale=0.5]{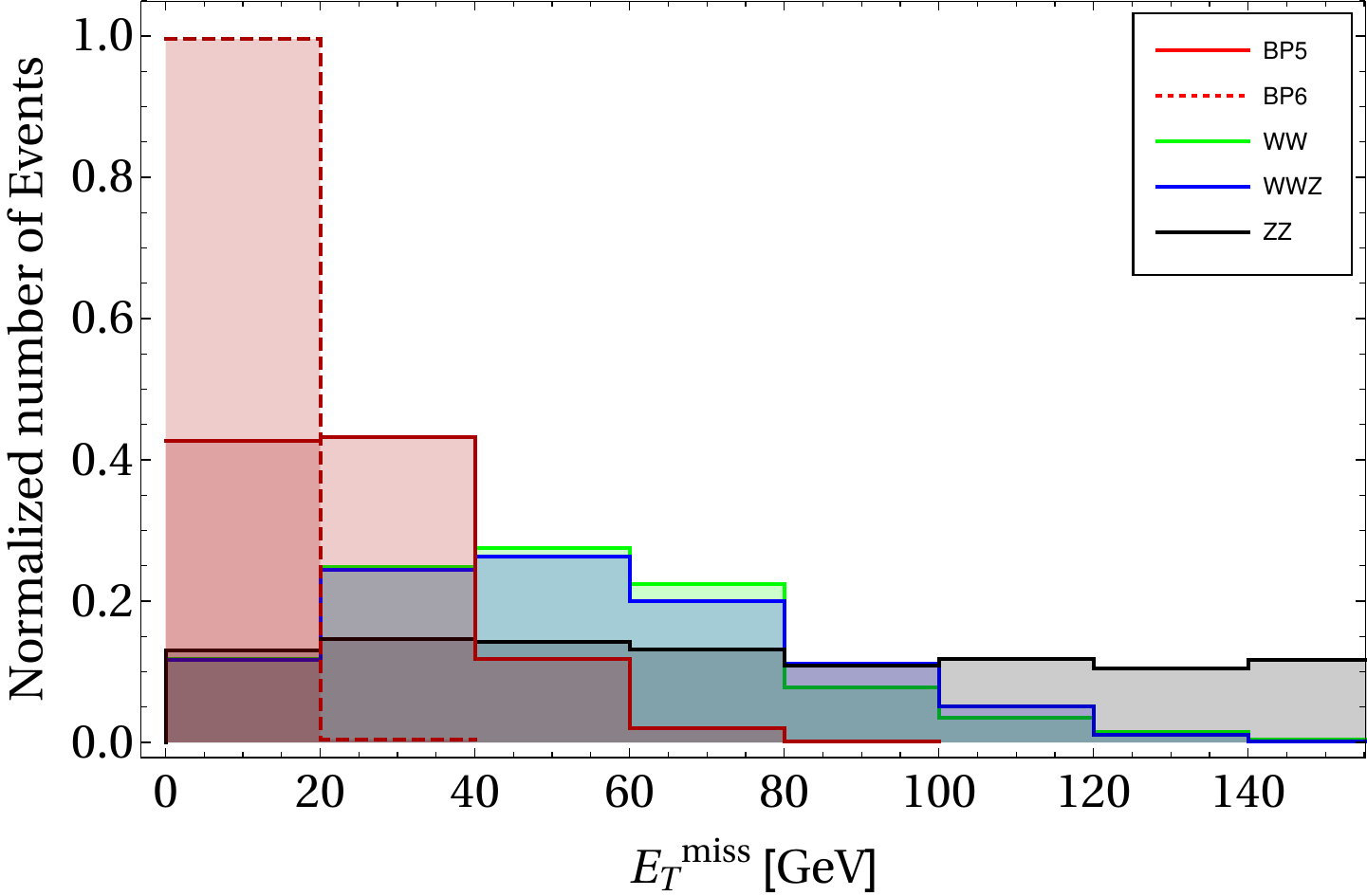}
  \includegraphics[scale=0.5]{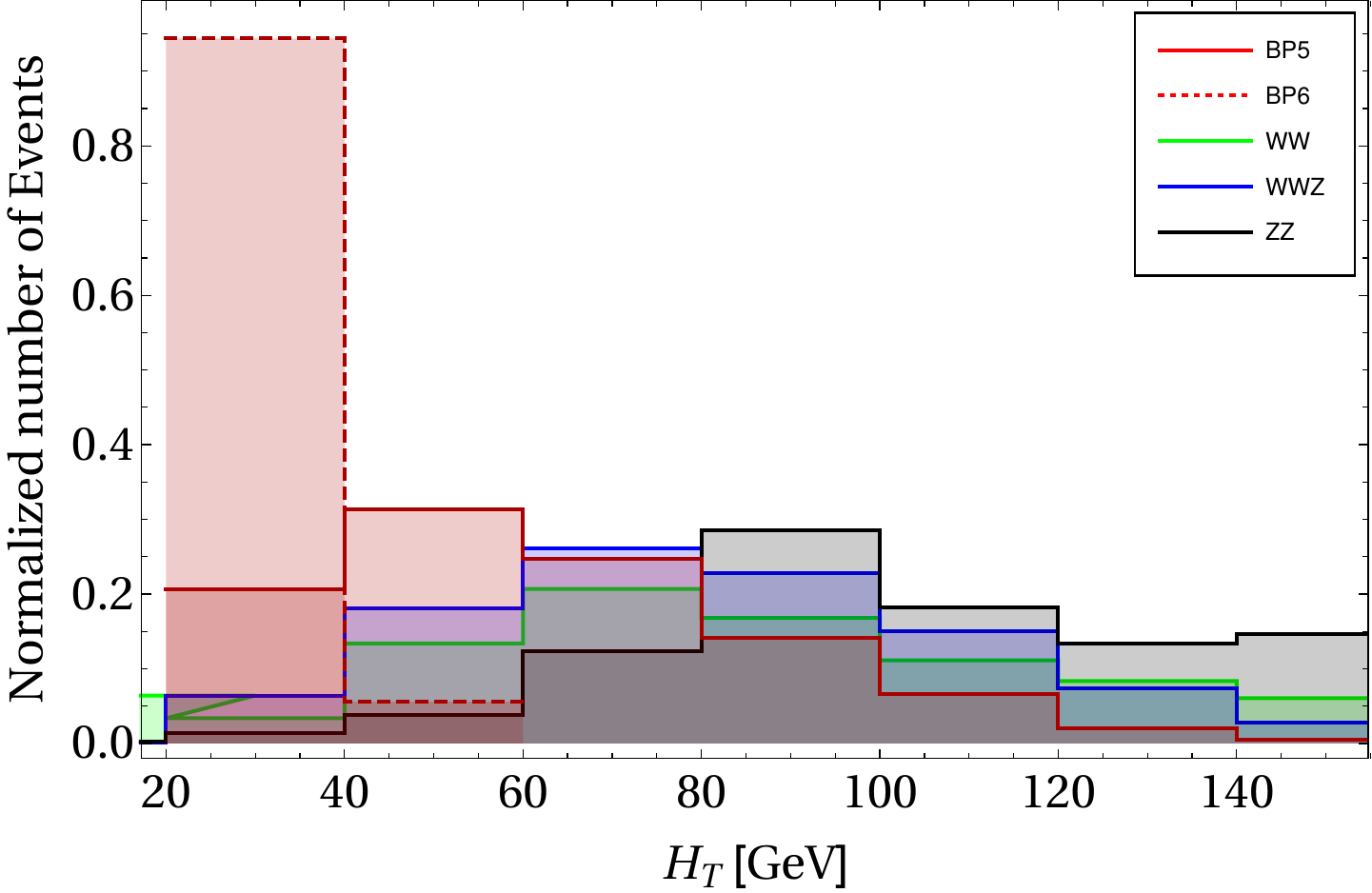}
 $$
$$
 \includegraphics[scale=0.5]{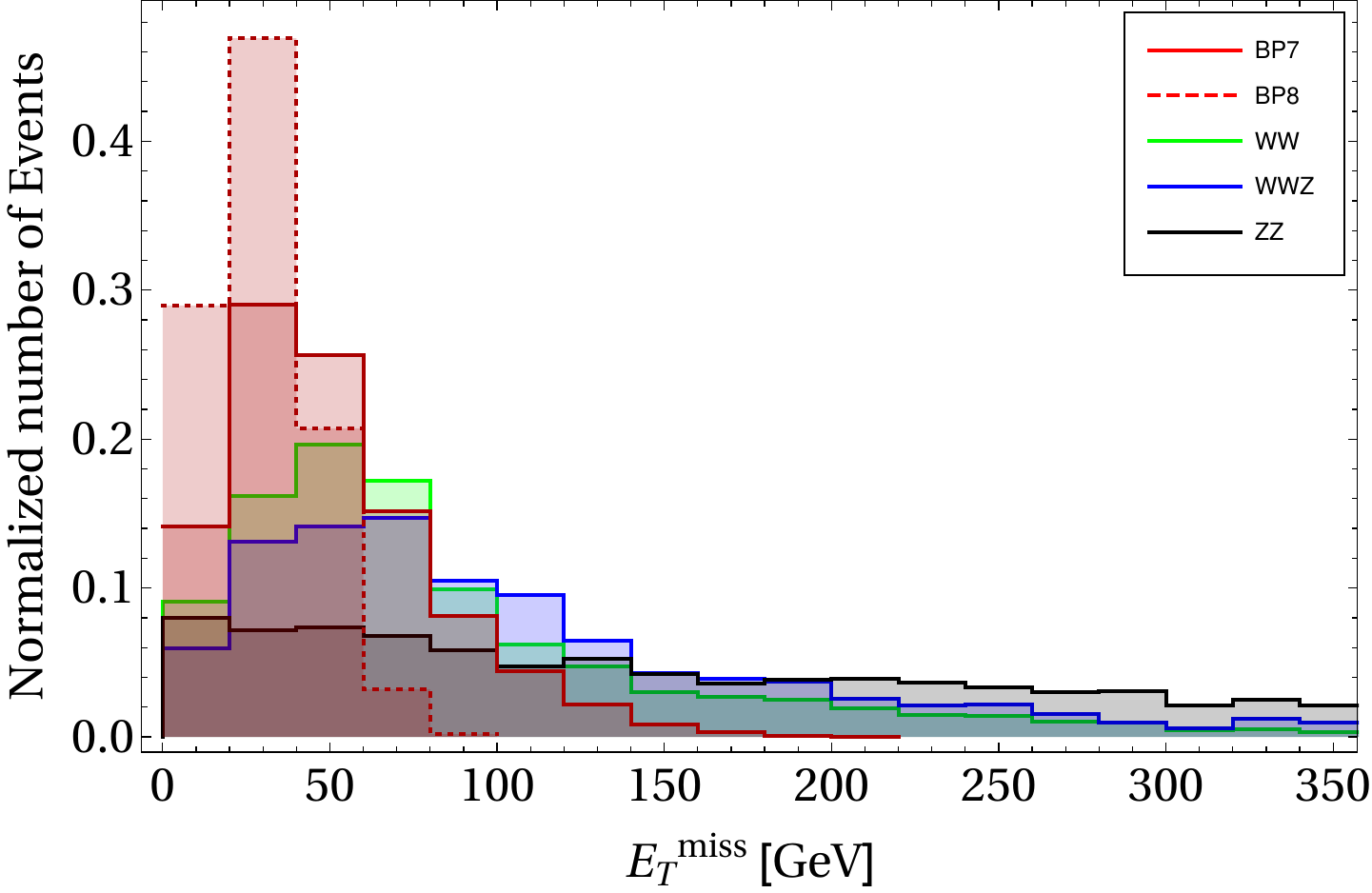}
  \includegraphics[scale=0.5]{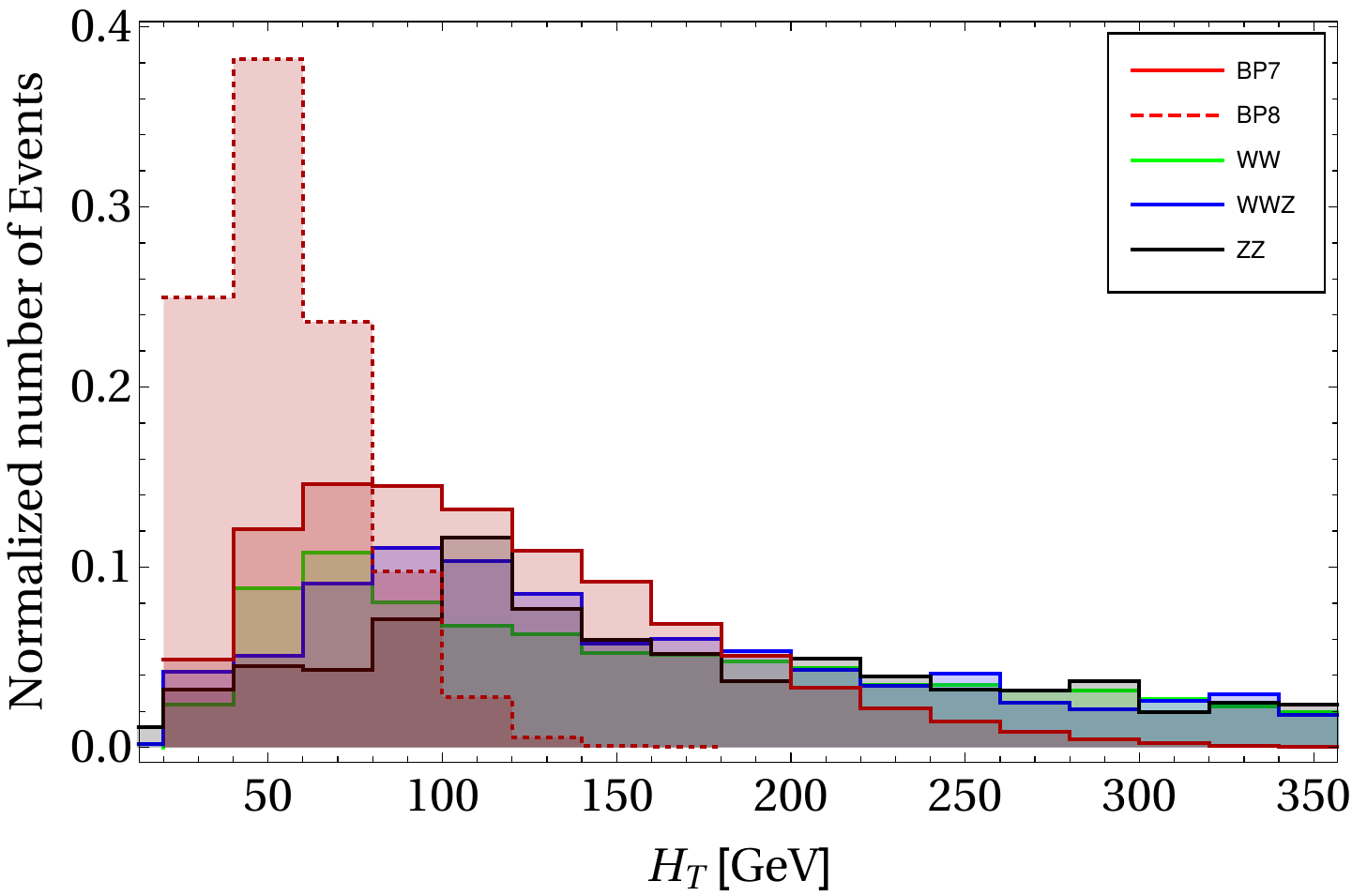}
 $$
 \caption{{\it Top Left:} MET distribution for OSD+$\slashed{E}_T$ final state at $\sqrt{s}=350~\rm GeV$ for BP5 and BP6 (shown in red). Corresponding dominant SM backgrounds are also shown with 
 different colors. {\it Top Right:} $H_T$ distribution for the same. {\it Bottom Left:} MET distribution for OSD+$\slashed{E}_T$ final state at $\sqrt{s}=1~\rm TeV$ for BP7 and BP8 (in red). Corresponding dominant SM backgrounds are also shown with different colors. {\it Bottom Right:} $H_T$ distribution for the same.}
 \label{fig:methtilc}
\end{figure}

\begin{figure}[htb!]
$$
 \includegraphics[scale=0.45]{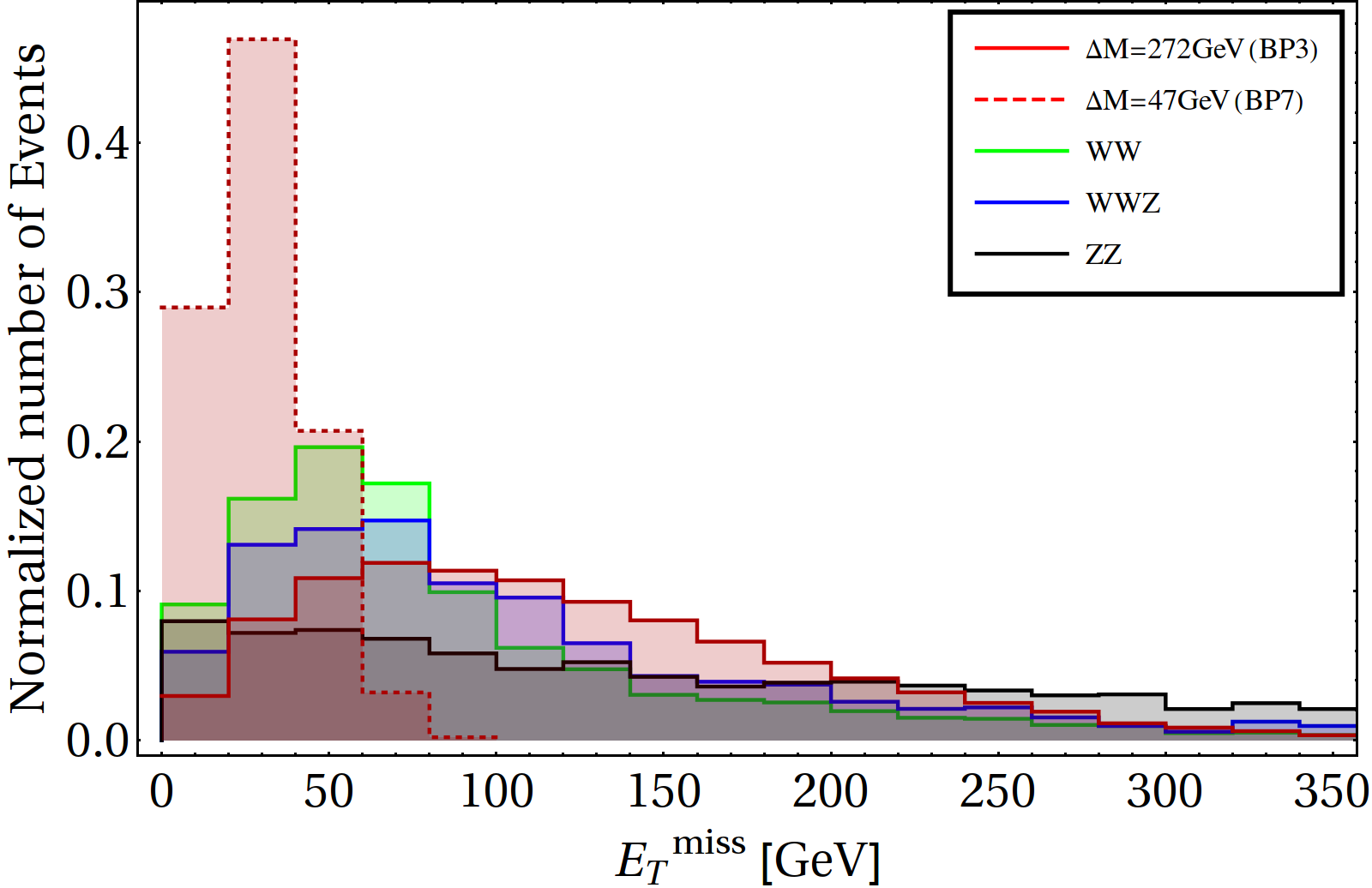}
  $$
 \caption{Comparison of MET distribution of BP3 and BP7 at the ILC with $\sqrt{s}=1~\rm TeV$.}
 \label{fig:metcmprilc}
\end{figure}

Now, once we have chosen the right combination of the beam polarisation to suppress SM background, we are in a position to analyse a favourable cut flow for the signal events. We plot, the main kinematic variables: MET and $H_T$ distribution for all the BPs, along with the SM backgrounds in Fig.~\ref{fig:methtilc}. This is done for both $\sqrt{s}=350~\rm GeV$ in the upper panel and for $\sqrt{s}=1~\rm TeV$ in the lower panel of Fig.~\ref{fig:methtilc}. We see that our benchmark points (BP5-BP8) produce a sharp peak in MET and $H_T$ at lower values, while the SM background distribution is flatter. This is because, in signal events, the mass difference ($\Delta M$) between the charged fermions to that of the DM is small. This essentially dictates that momentum available for the DM or for those of the SM leptons are on the smaller side. On the other hand, due to large mass difference between the produced SM gauge boson and the SM leptons, the available momentum for the leptons can be much larger. Therefore, we can safely choose a judicious upper cut on MET and $H_T$ to retain such signals and diminish SM backgrounds further. In order to show this dependence of MET on $\Delta M$ explicitly, we have compared the MET distribution for BP3 (with $\Delta M=272~\rm GeV$) and BP7 (with $\Delta M=47~\rm GeV$) in Fig.~\ref{fig:metcmprilc}. As already pointed out, due to larger $\Delta M$, BP3 produces larger missing energy and MET distribution becomes flatter and gets submerged into the SM background. BP7, with smaller $\Delta M$, peaks at lower end of the distribution. We choose therefore the following selection cuts for selecting signal events: 

\begin{itemize}
 \item MET cut of $\slashed{E_T}< \{100,50\}~\rm GeV$, which retains most of the signals while killing majority of the background for $\sqrt{s}=1~\rm TeV$, while for $\sqrt{s}=350~\rm GeV$ the MET cut is even milder: $\slashed{E_T}<\{30,20\}~\rm GeV$.
 \item A $H_T$ cut of $H_T<150~\rm GeV$ to reduce the background further for $\sqrt{s}=1~\rm TeV$. For $\sqrt{s}=350~\rm GeV$ we employed: $H_T<50~\rm GeV$.
 \item An invariant mass cut around $Z$-window: $|m_z-15|<m_{ll}<|m_Z+15|$ helps to get rid off the $Z$-dominated background in both cases. 
\end{itemize}

\begin{table}[htb!]
\begin{center}
\scalebox{1.0}{
\begin{tabular}{|c|c|c|c|c|c|c|c|c|}
\hline
Benchmark Point & $\sigma^{\psi^{+}\psi^{-}}$ (fb) & $\slashed{E}_T$ (GeV) & $\sigma^{\text{OSD}} (fb)$\\
\hline\hline 
BP5 & 690.32 & $<$ 30 & 6.27\\
\cline{3-4}
&   & $<$ 20 & 3.64\\
\hline 
BP6 & 705.13 & $<$ 30 & 3.11\\
\cline{3-4}
&   & $<$ 20 & 3.09 \\
\hline
\end{tabular}
}
\end{center}
\caption {Signal events with $\sqrt{s}$ = 350 GeV at the ILC.}
\label{tab:sigeventilc1}
\end{table}

\begin{table}[htb!]
\begin{center}
\scalebox{1.0}{
\begin{tabular}{|c|c|c|c|c|c|c|c|c|}
\hline
Benchmark Point & $\sigma^{\psi^{+}\psi^{-}}$ (fb) & $\slashed{E}_T$ (GeV) & $\sigma^{\text{OSD}} (fb)$  \\
\hline\hline
BP7 &  79.16   & $<$ 100 & 2.04 \\
\cline{3-4}
&   & $<$ 50 & 1.85 \\
\hline
BP8 &  89.55   & $<$ 100 & 1.84 \\
\cline{3-4}
&   & $<$ 50 & 1.24 \\
\hline
\end{tabular}
}
\end{center}
\caption {Signal events with $\sqrt{s}$ = 1 TeV at the ILC.}
\label{tab:sigeventilc}
\end{table}

\begin{figure}[htb!]
 $$
\includegraphics[scale=0.53]{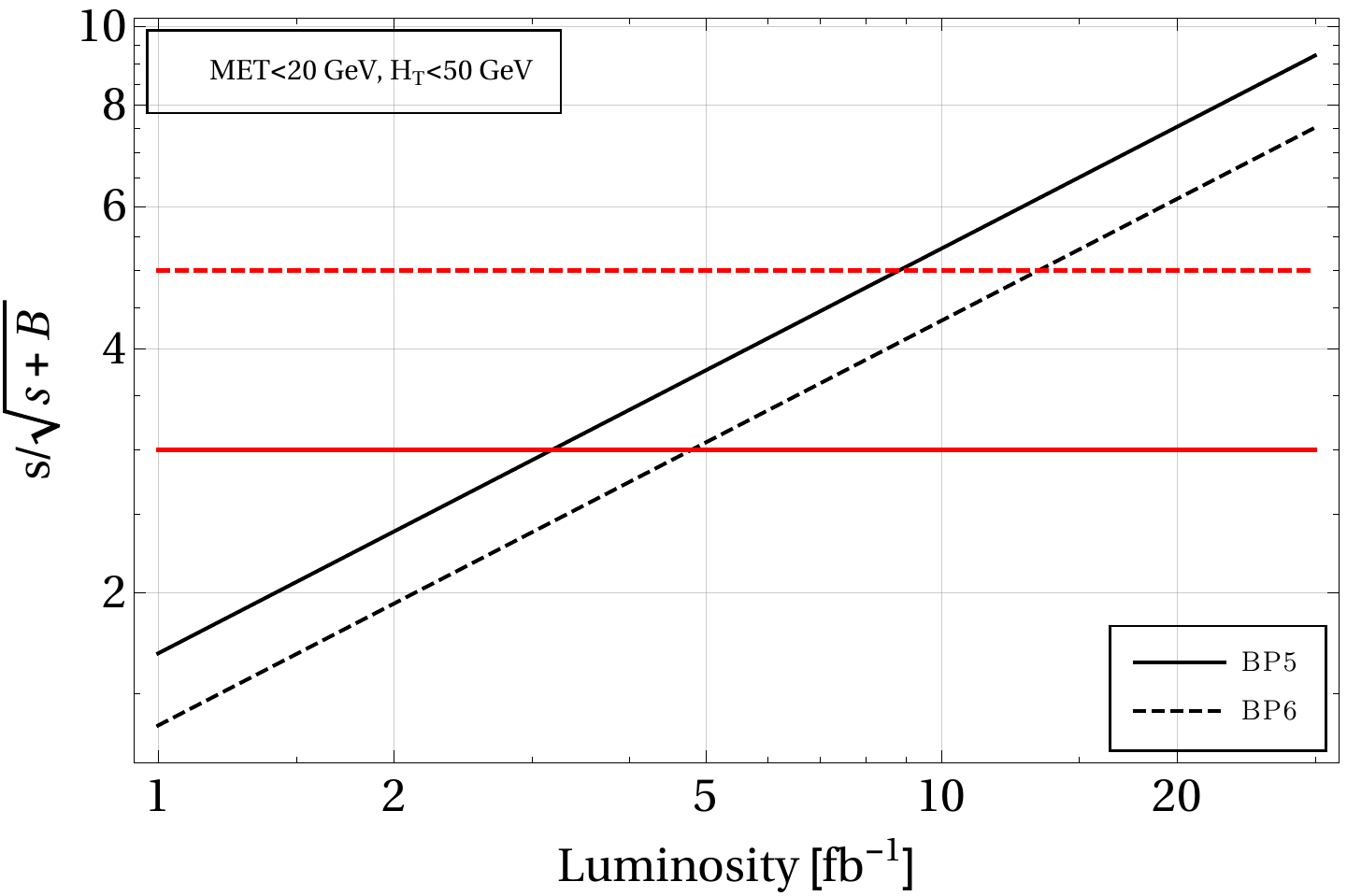}
\includegraphics[scale=0.53]{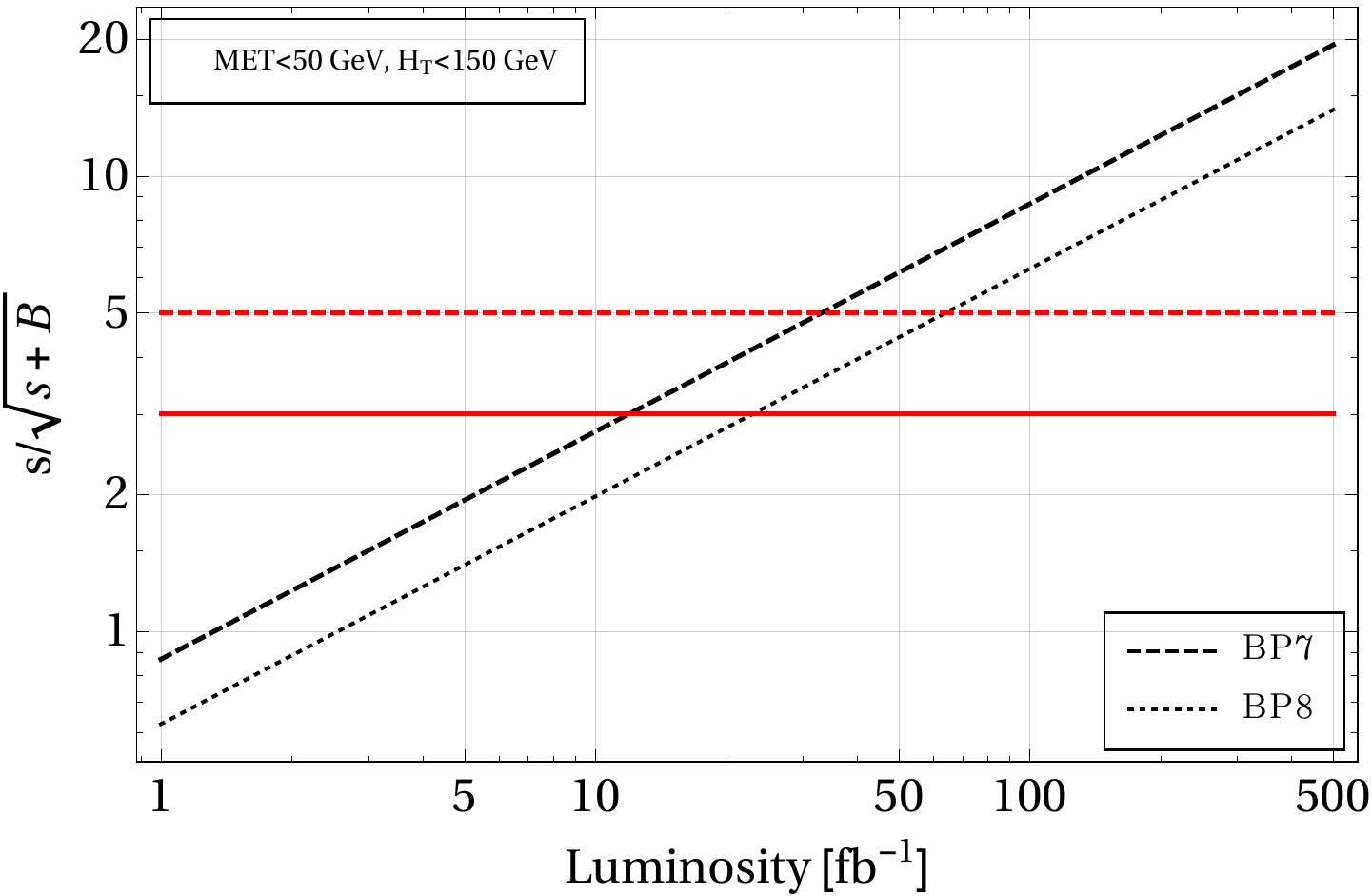}
 $$
 \caption{Left: Signal significance for BP5 and BP6 at the ILC for $\sqrt{s}=350~\rm GeV$. Right: Significance of BP7 and BP8 at $\sqrt{s}=1~\rm TeV$. In both the plots The solid red and dashed red lines correspond to 3$\sigma$ and 5$\sigma$ discovery limits respectively.}
 \label{fig:signiilc}
\end{figure}

We have finally tabulated the number of signal and background events at the ILC for both $\sqrt{s}=350~\rm GeV$ and $\sqrt{s}=1~\rm TeV$ for the chosen polarization $(P_{e^-},P_{e^+})$=(+80\%,-20\%). In Tab.~\ref{tab:sigeventilc1} and Tab.~\ref{tab:sigeventilc}, we have shown the variation in signal events with the cuts applied for $\sqrt{s}=350~\rm GeV$ and $\sqrt{s}=1~\rm TeV$ respectively. The same for the dominated SM background are also tabulated in  Tab.~\ref{tab:bckevntilc1} and Tab.~\ref{tab:bckevntilc} for $\sqrt{s}=350~\rm GeV$ and $\sqrt{s}=1~\rm GeV$ respectively. In order to find the discovery potential of such signals at the ILC we have again computed the signal significance. This is shown in Fig.~\ref{fig:signiilc}. As one can see, for $\sqrt{s}=350~\rm GeV$ a 5$\sigma$ discovery reach is possible at a very low luminosity (left panel of Fig.~\ref{fig:signiilc}): $\mathcal{L}\sim 8~\rm fb^{-1}$ for BP5. For $\sqrt{s}=1~\rm TeV$ the same can be reached for BP7 at a luminosity $\mathcal{L}=30~\rm fb^{-1}$ as shown in the right panel of 
Fig.~\ref{fig:signiilc}. This tells us, there is a chance that this model might show up at a very early run of the ILC, compared to that of LHC which demands a much higher luminosity to be probed.

\begin{table}[htb!]
\begin{center}
\scalebox{1.0}{
\begin{tabular}{|c|c|c|c|c|c|c|c|c|}
\hline
Background & $\sigma_{production}$ (pb) & $\slashed{E}_T$ (GeV) & $\sigma_{OSD}(fb)$  \\
\hline\hline
$W^+W^-$ & 1.90 & $<$ 30 & 3.80  \\
\cline{3-4}
&   & $<$ 20 & 1.88\\
\hline
$W^+W^-Z$ & 0.002 & $<$ 30 & 0.001 \\
\cline{3-4}
&   & $<$ 20 & 0.009\\
\hline 
$ZZ$ & 0.49 & $<$ 30 & 0.18 \\
\cline{3-4}
&   & $<$ 20 & 0.11 \\
\hline 
\end{tabular}
}
\end{center}
\caption {Events for dominant SM background with $\sqrt{s}$ = 350 GeV at the ILC.}
\label{tab:bckevntilc1}
\end{table}

\begin{table}[htb!]
\begin{center}
\scalebox{1.0}{
\begin{tabular}{|c|c|c|c|c|c|c|c|c|}
\hline
Background & $\sigma_{production}$ (pb) & $\slashed{E}_T$ (GeV) & $\sigma_{OSD}(fb)$ \\
\hline\hline 
$W^+W^-$ & 0.43 & $<$ 100 & 4.97 \\
\cline{3-4}
&   & $<$ 50 & 2.61  \\
\hline
$W^+W^-Z$ & 0.009 & $<$ 100 & 0.03 \\
\cline{3-4}
&   & $<$ 50 & 0.01  \\
\hline
$ZZ$ & 0.11 & $<$ 100 & 0.13  \\
\cline{3-4}
&   & $<$ 50 & 0.08 \\
\hline 
\end{tabular}
}
\end{center}
\caption {Events for dominant SM background with $\sqrt{s}$ = 1 TeV at the ILC.}
\label{tab:bckevntilc}
\end{table}

We  conclude this section, by again pointing out that small $\Delta M$, i.e. small mass difference between the charged fermions and DM, can only be probed at the ILC through hadronically quiet OSD events, thanks to the absence of $t\bar{t}$ and Drell Yan type background events. While this has been established with some benchmark points (BP5-BP8) in presence of scalar triplet, the feature can also be captured for the same fermion DM model~\cite{Bhattacharya:2015qpa} in absence of scalar triplet. The scalar triplet rather paves the way for probing the model at higher $\Delta M$ region at the LHC. The complementarity of the LHC and the ILC searches for the model is an interesting noteworthy feature of this analysis.  

\section{Role of vector like lepton in collider signature of scalar triplet}
\label{sec:tripcollider}

\begin{figure}[htb!]
 $$
\includegraphics[scale=0.5]{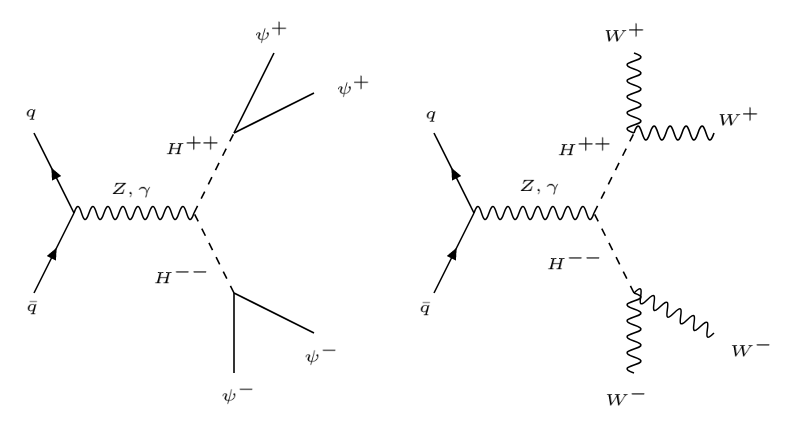}
$$
$$
\includegraphics[scale=0.5]{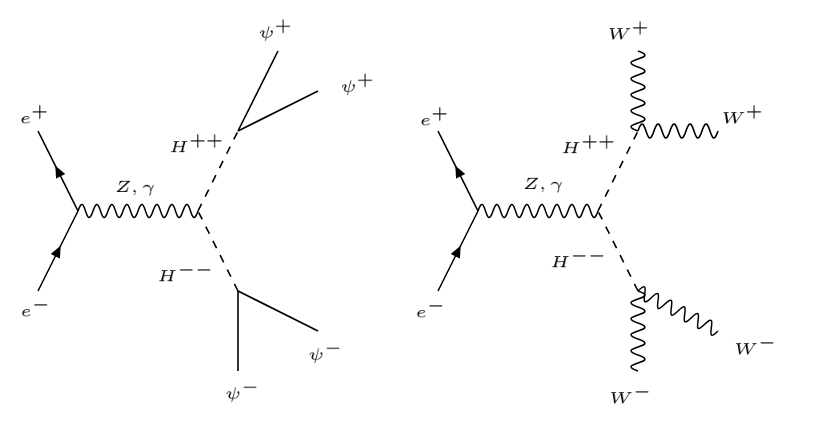}
 $$
 \caption{Top left: Production and subsequent decay of the doubly charged scalars at the LHC in presence of VLF. Top right: Same for models without the VLFs (eg. type-II seesaw model). Bottom left: Production and subsequent decay of the doubly charged scalars at the ILC in presence of VLF. Bottom right:  Same for models without the VLFs (eg. type-II seesaw model).}
 \label{fig:dblychrgd}
\end{figure}

The scalar triplet sector itself can produce rich phenomenology at collider, being equipped with single ($H^\pm$) and doubly charged ($H^{\pm\pm}$) 
scalars as we have demonstrated in Sec.~\ref{sec:model}. Collider phenomenology of scalar triplet has already been elaborated in many references 
before~\cite{Ghosh:2017pxl,Agrawal:2018pci}. Our aim is definitely not to repeat the same exercise here. We would however like to point out to an interesting feature 
of this model, where the signature of the doubly charged scalars get affected by the presence of VLFs as we have in our model. The Feynman graphs for producing 
doubly charged scalars and their subsequent decay to produce hadronically quiet four lepton (HQ4l) signature is shown in Fig.~\ref{fig:dblychrgd} at LHC (top panel) 
and ILC (bottom panel). In the left panel we show the decay branching of doubly charged scalars through charged vector like lepton ($\psi^\pm \to \psi^0+\ell^\pm+\nu_\ell$) 
and in the right panel, we show the branching through usual $W$ mediation to produce HQ4l. In absence of vector like fermion, the diagrams on the right 
panel only contribute to HQ4l signature. We will be interested in exploring the distinction of such a situation from the usual case of scalar triplet for example, 
Type II seesaw, if any.  

\begin{figure}[htb!]
 $$
\includegraphics[scale=0.4]{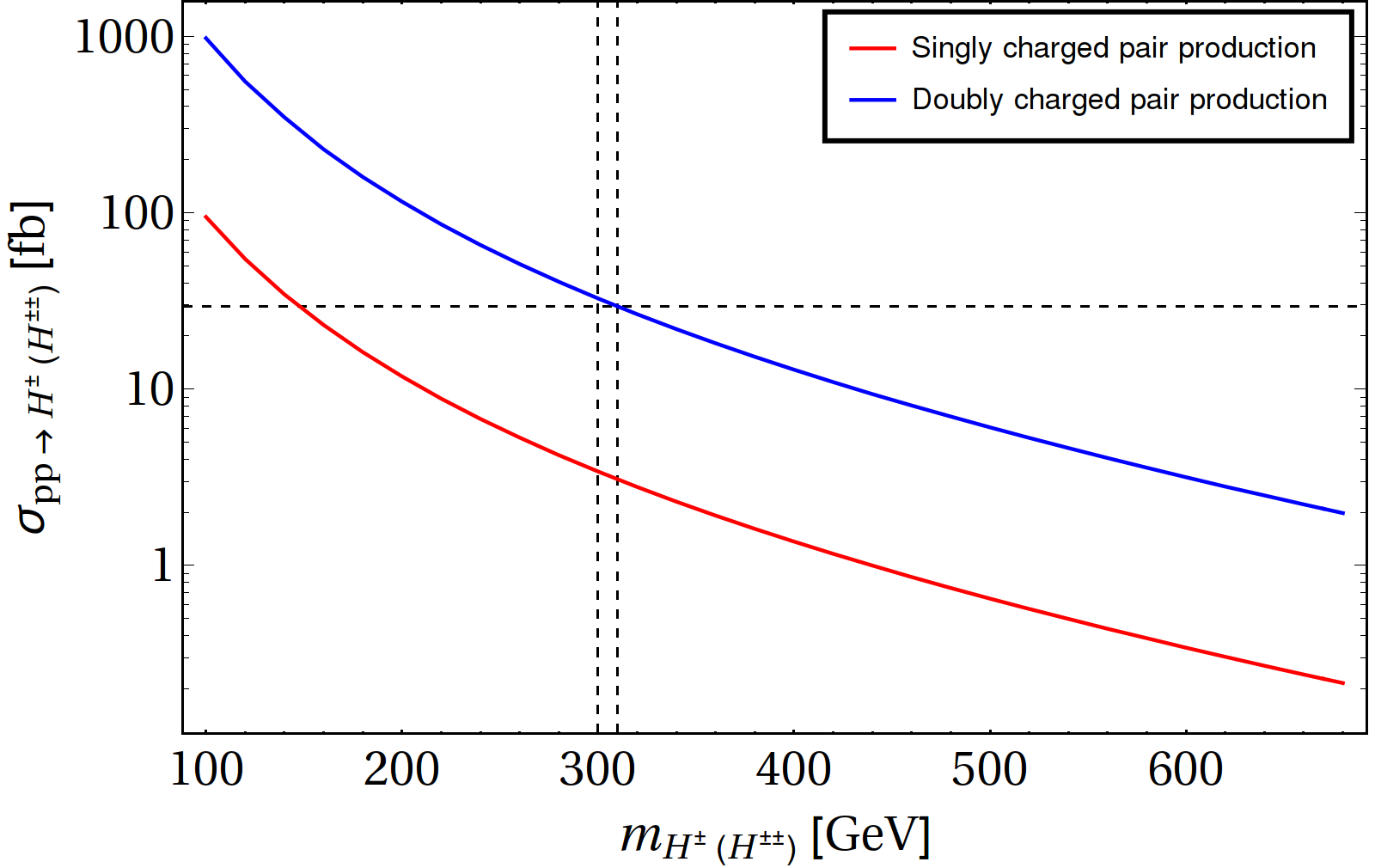}
\includegraphics[scale=0.4]{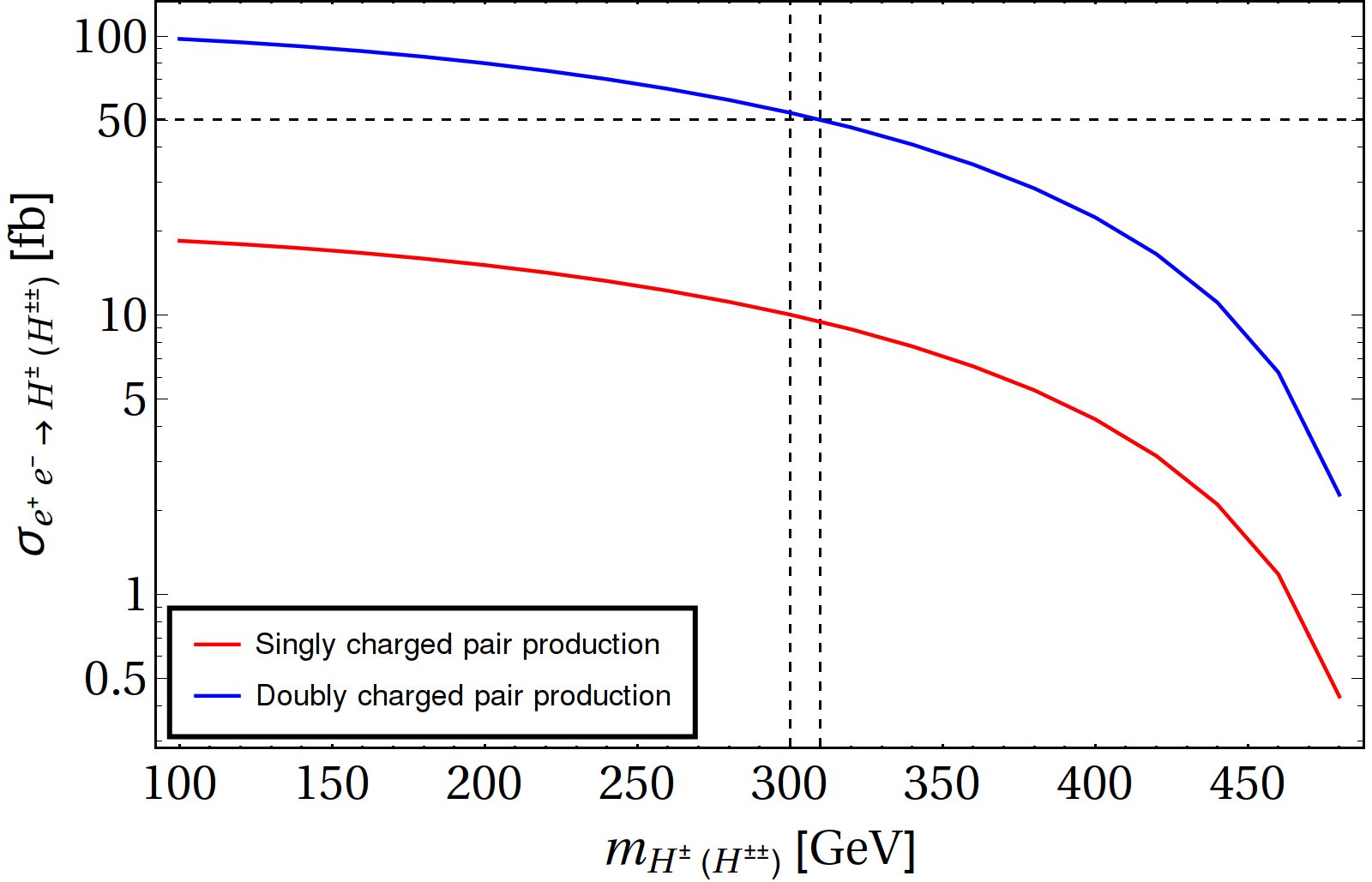}
 $$
 \caption{Production cross-section of single and double charged scalars belonging to scalar triplet, as a function of mass in Left: LHC at $\sqrt{s}=14~\rm TeV$ and on 
 Right : ILC at $\sqrt{s}=1~\rm TeV$. Our choice of benchmark with $m_{H^0}=300$ GeV and $m_{H^{\pm, \pm\pm}}=310$ GeV and corresponding production 
 cross section is also indicated.}
 \label{fig:prodtrip}
\end{figure}

In Fig.~\ref{fig:prodtrip}, we first show the production cross-section of singly charged and doubly charged scalar at LHC (left) and at ILC (right). We see that, 
naturally the production cross-section of doubly charged scalar is larger than the singly charged scalar. We also point out that the production of doubly charged scalar at LHC 
for the chosen benchmark point of our analysis with $m_{H^{\pm}}=300$ GeV and $m_{H^{\pm\pm}}=310$ GeV is quite high $\sim$ 30 fb. We will analyse the final state 
signal of HQ4l produced by these processes, and highlight only on the cases where the vector like lepton can enter into the decay chain. Now, with our chosen 
BPs (Tab.~\ref{tab:bp}), $\psi^\pm$ can be produced from the decay of $H^{\pm\pm}$ only for BP5 and BP6 as for other BPs: $m_{H^{\pm\pm}}<2 m_{\psi^\pm}$.  
For these two benchmark points, the branching ratios of the doubly charged scalar is tabulated in Table~\ref{tab:br-dble}, which shows that $H^{\pm\pm}$ 
dominantly decays to charged vector like lepton pair over $WW$.

\begin{table}[htb!]
\begin{center}
\scalebox{1.0}{
\begin{tabular}{|c|c|c|c|c|c|c|c|c|}
\hline
Benchmark Point & $\mathcal{B}\left(H^{++}\to\psi^+\psi^+\right)$ & $\mathcal{B}\left(H^{++}\to W^+W^+\right)$ \\
\hline\hline 
BP5 & 0.989 & 0.011    \\
BP6 & 0.992 & 0.008   \\
\hline
\end{tabular}
}
\end{center}
\caption {Branching fraction of $H^{\pm\pm}$ to $\psi^{+}\psi^+$ and $W^{+}W^{+}$ for the chosen benchmark points BP5 and BP6.}
\label{tab:br-dble}
\end{table} 
\begin{figure}[htb!]
 $$
\includegraphics[scale=0.43]{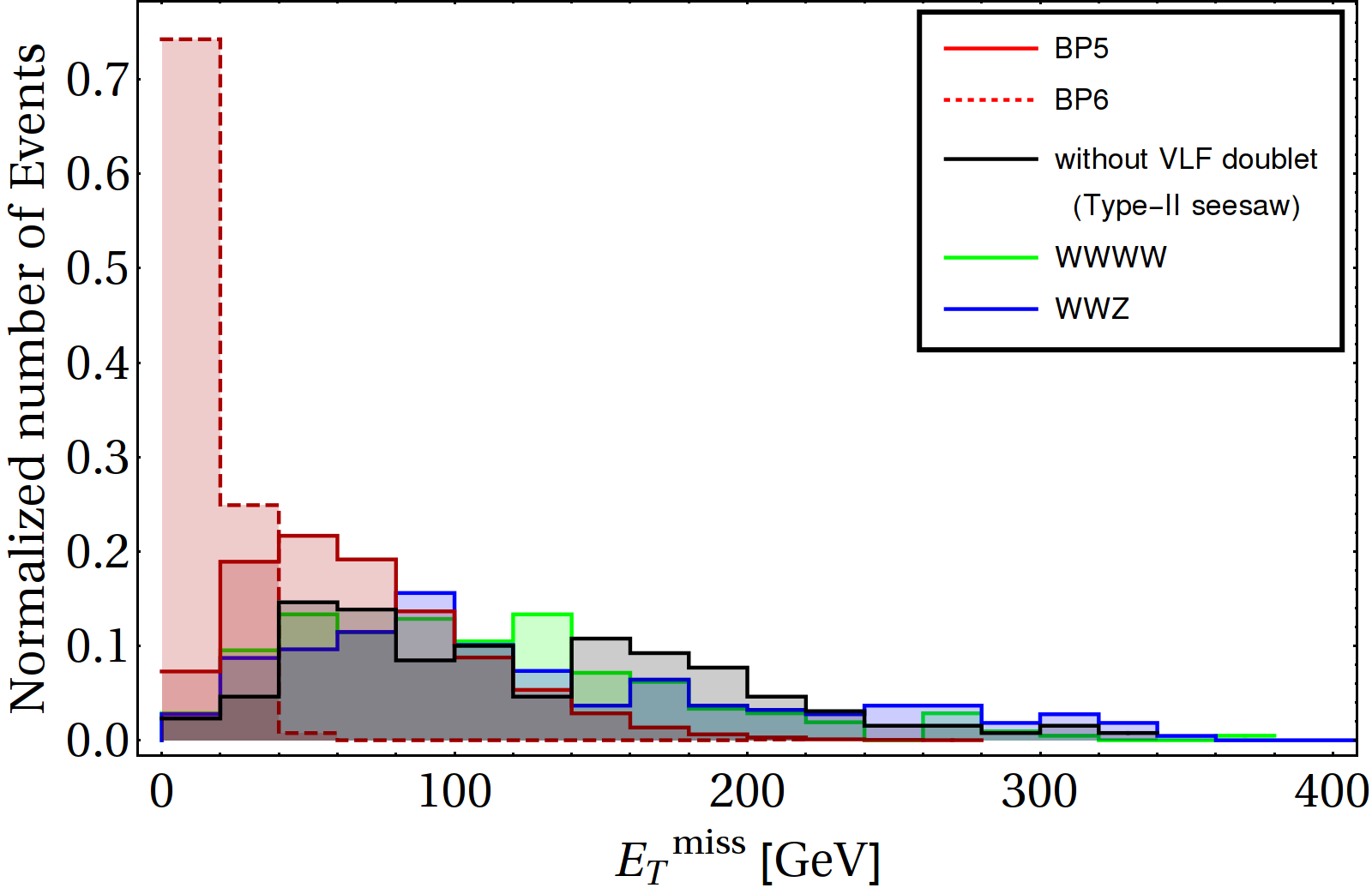}
\includegraphics[scale=0.43]{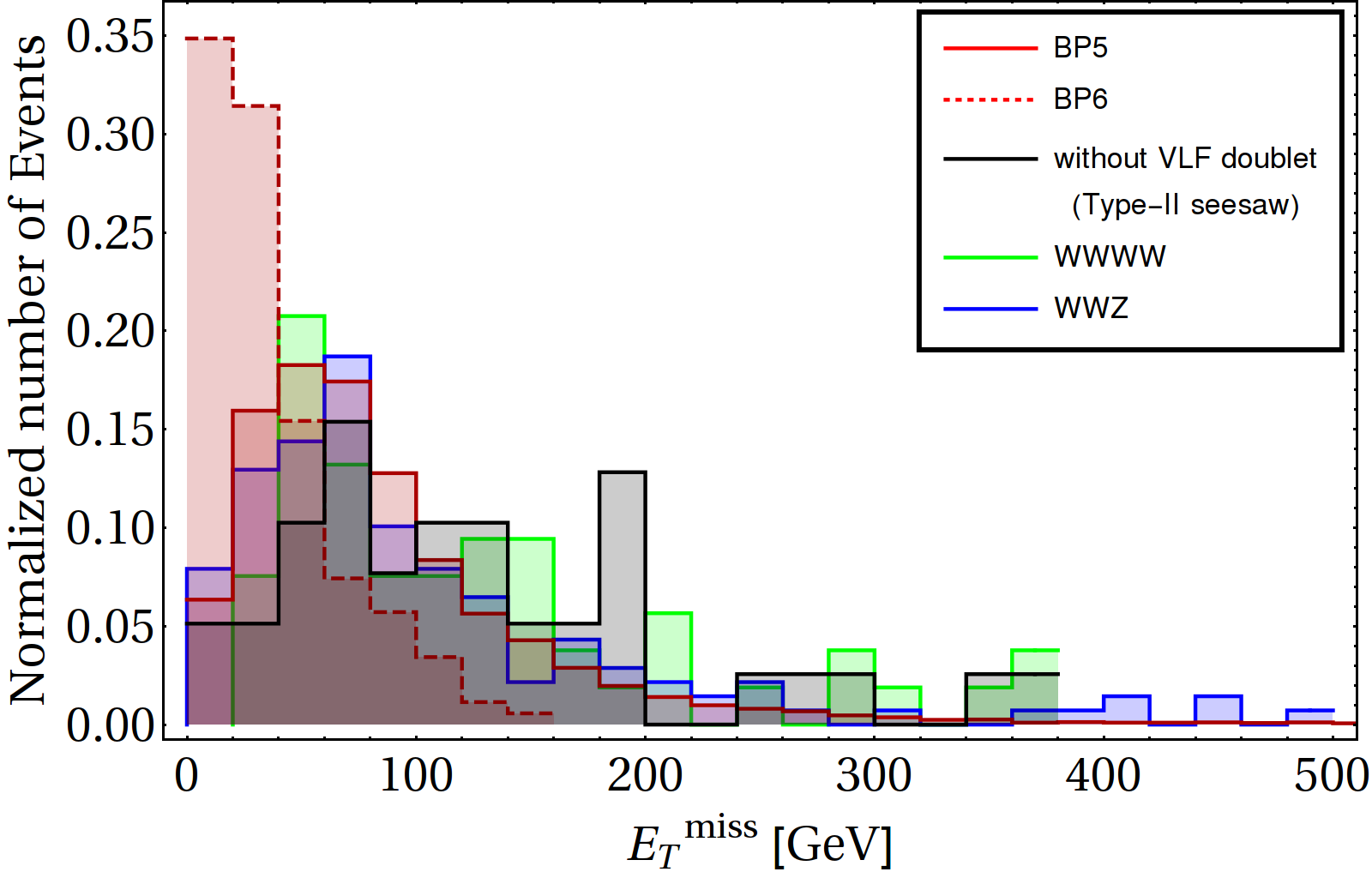}
 $$
 \caption{Missing energy distribution for HQ4l channel for benchmark points BP5, BP6 and type-II seesaw model (the case without VLF) 
 along with dominant SM background. On Left: ILC search at $\sqrt{s}=1~\rm TeV$ and on Right: LHC search at $\sqrt{s}=14~\rm TeV$ are shown.}
 \label{fig:mettrip}
\end{figure}

We now follow the same strategy for collider signal analysis as elaborated in Sec.~\ref{sec:lhcsearch} and Sec.~\ref{sec:ilcsearch}  
for doubly charged scalar production and their subsequent decay to HQ4l.

In Fig.~\ref{fig:mettrip}, we have shown the missing energy (MET) distribution for the benchmark points BP5 and BP6, along with dominant SM background for ILC 
with $\sqrt{s}=1~\rm TeV$ (left) and LHC with $\sqrt{s}=14~\rm TeV$ (right). In case of ILC (left), we see a similar behaviour of MET distribution as obtained for 
hadronically quiet dilepton channel in Sec.~\ref{sec:ilcsearch}, where the distribution is essentially guided by $\Delta M$. Therefore smaller $\Delta M$ (BP6) 
peaks at a lower value of MET compared to BP5 and both can easily be distinguished at low MET from SM background. However, the distribution is more flattened 
for the case without VLF (i.e. usual type-II seesaw model), where the missing energy distribution becomes almost identical to that of SM background, naturally as the 
decay of doubly charged scalar occurs through $WW$ channel. Thus, in our model, where doubly charged scalar can decay through VLF and 
the charged VLF has small mass splitting with DM, (i.e. $\Delta M \lsim m_W$, which occurs in a large DM allowed parameter space), can be 
distinguished from the usual signature of doubly charged scalar belonging to a triplet (eg: type-II seesaw models) and from SM backgrounds 
if we use an upper-cut on $\slashed{E_T}:\slashed{E_T}\lsim100~\rm GeV$. 

The MET distribution for the signals at LHC (shown in the RHS of Fig.~\ref{fig:mettrip}) have a similar behaviour as that of ILC. The SM background, however, 
is a bit different from what we have seen before for hadronically quiet dilepton channel in Sec.~\ref{sec:lhcsearch} due to the absence of $t\bar{t}$ and 
Drell-Yan background. Therefore, it is important to note here that HQ4l channel through scalar triplet production can probe smaller $\Delta M$ regions (like BP5 and BP6) 
unlike the case of dilepton channel with an upper-cut on MET : $\slashed{E_T}\lsim 40~\rm GeV$ at LHC. A similar cut will also help us to disentangle the 
presence of VLF from the usual type-II seesaw models.  


\begin{table}[htb!]
\begin{center}
\scalebox{1.0}{
\begin{tabular}{|c|c|c|c|c|c|c|c|c|}
\hline
Benchmark Point & $\sigma^{H^{\pm\pm}}$ (fb) & $\sigma^{4\ell^{\pm}} (fb)$ (no $\slashed{E_T}$) & $\slashed{E}_T$ (GeV) & $\sigma^{4\ell^{\pm}} (fb)$\\
\hline\hline 
BP5 &  & 3.64  & & $5.80\times 10^{-4}$ \\
& 29.38 & & 20-40 &\\
BP6 &  & 0.10 & & $3.22\times 10^{-5}$  \\
\hline
\end{tabular}
}
\end{center}
\caption {Cross-section for signal from doubly charged Higgs production and its decays to HQ4l final state for $\sqrt{s}$ = 14 TeV at the LHC for the benchmarks BP5 and BP6.}
\label{tab:tripsigevntlhc}
\end{table}

\begin{table}[htb!]
\begin{center}
\scalebox{1.0}{
\begin{tabular}{|c|c|c|c|c|c|c|c|c|}
\hline
SM Backgrounds & Production cross-section (fb) & $\slashed{E}_T$ (GeV) & $\sigma^{4\ell^{\pm}} (fb)$
\\
\hline\hline 
$WWWW$ & 0.6 &  & $2.4\times 10^{-8}$ \\
&  &  &\\
&& 20-40  &\\
$WWZ$ & 150 &   & $1.047\times 10^{-5}$ \\
&&&\\
\hline\hline
type-II seesaw & 29.38 &  & $1.17\times 10^{-6}$ \\
\hline
\end{tabular}
}
\end{center}
\caption {Cross-section for SM Background and type-II seesaw model for doubly charged Higgs production and its decays to HQ4l final state for $\sqrt{s}$ = 14 TeV at the LHC.}
\label{tab:tripbckevntlhc}
\end{table} 



\begin{table}[htb!]
\begin{center}
\scalebox{1.0}{
\begin{tabular}{|c|c|c|c|c|c|c|c|c|}
\hline
Benchmark Point & $\sigma^{H^{\pm\pm}}$ (fb) & $4\ell^{\pm}$ events (no $\slashed{E_T}$) & $\slashed{E}_T$ (GeV) & $\sigma^{4\ell^{\pm\pm}} (fb)$\\
\hline\hline 
BP5 &  & 27.36 & & 0.0221 \\
& 50.13 & &$<$ 100 &\\
BP6 &  & 1.31 & & 0.0013 \\
\hline
\end{tabular}
}
\end{center}
\caption {Hadronically quiet four lepton signal from doubly charged Higgs production and its decays for $\sqrt{s}$ = 1 TeV at ILC for the benchmark points BP5 and BP6.}
\label{tab:tripsigevntilc}
\end{table}

\begin{table}[htb!]
\begin{center}
\scalebox{1.0}{
\begin{tabular}{|c|c|c|c|c|c|c|c|c|}
\hline
SM Backgrounds & Production cross-section (fb) & $\slashed{E}_T$ (GeV) & $\sigma^{4\ell^{\pm\pm}} (fb)$
\\
\hline\hline 
$WWWW$ & 1.74 &  & $1.649\times 10^{-6}$ \\
&  &  &\\
&& $<$ 100  &\\
$WWZ$ & 9.98 &  & $1.047\times 10^{-5}$ \\
&&&\\
\hline\hline
Type-II seesaw & 50.13 &  & $5.711\times 10^{-5}$ \\
\hline
\end{tabular}
}
\end{center}
\caption {SM Background cross-section for HQ4l channel and scalar triplet framework like type-II seesaw model in absence of vector like lepton for $\sqrt{s}$ = 1 TeV at ILC.}
\label{tab:tripbckevntilc}
\end{table} 

In Tab.~\ref{tab:tripsigevntlhc} and Tab.~\ref{tab:tripbckevntlhc}, we have tabulated HQ4l event cross-sections for the benchmark points (BP5 and BP6) and 
dominant SM backgrounds along with the case of scalar triplet in absence of VLF (referred as Type II Seesaw model) at the LHC environment. 
We have chosen $20<\slashed{E_T}<40~\rm GeV$ as the MET cut (guided by the MET distribution), in order to separate the signal from the 
SM background and from the usual type-II seesaw scenario. One can note from Tab.~\ref{tab:tripsigevntlhc}, without any MET cut, 
number of HQ4l events for BP5 is several times larger than that of BP6. This happens due to the fact that 
BP5 has larger $\Delta M$ and therefore the leptons emerging out of the decay has larger probability of surviving the lepton $p_T$ cut. But due to smaller $\Delta M$, 
the peak of the distribution for BP6 is more pronounced than BP5 at low MET region. As a result, with the chosen MET cut: $20<\slashed{E_T}<40~\rm GeV$, 
although BP5 loses more events than BP6, but still retains more events than BP6. This is reflected in the final state cross-section in Tab.~\ref{tab:tripsigevntlhc}. 
As a consequence, BP5 will have a larger significance over BP6. 

The numbers for HQ4l events from signal (BP5 and BP6) at ILC is shown in Tab.~\ref{tab:tripsigevntilc}. Subsequently, MET cut sensitivity of corresponding SM 
background and scalar triplet frameworks without VLF is shown in Tab.~\ref{tab:tripbckevntilc}. We have used MET cut of $\slashed{E_T}<100~\rm GeV$ to kill 
SM background to a significant extent. In this case as well, like that of LHC, BP5 retains more events than BP6 and hence possesses higher significance. 
As one can understand from both these tables, in order to obtain finite number of final state signal events, we require high integrated luminosity 
$\mathcal{L}\sim10^5~\rm fb^{-1}$ at ILC, even if the SM backgrounds are vanishingly small. 


\begin{figure}[htb!]
 $$
\includegraphics[scale=0.43]{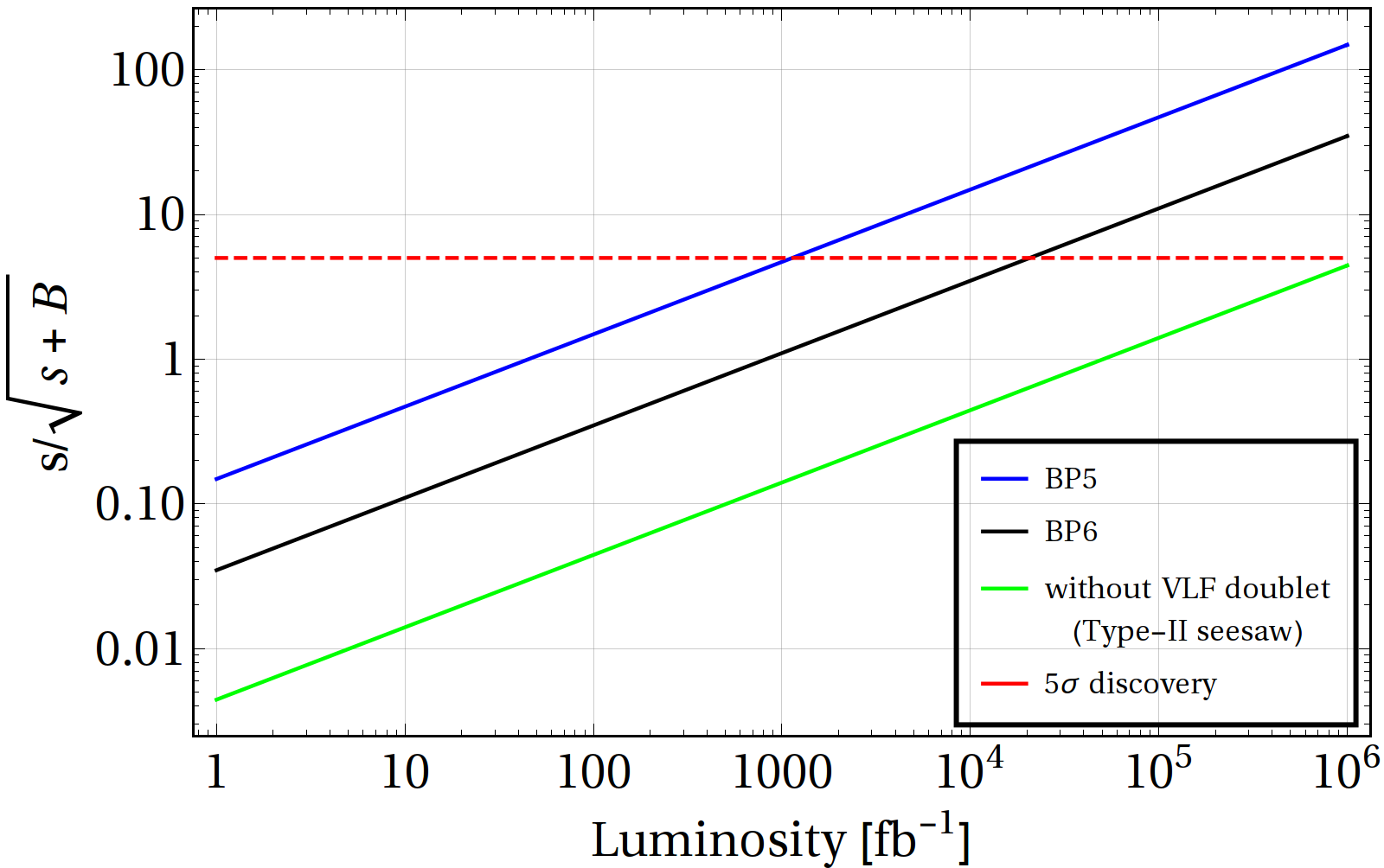}
\includegraphics[scale=0.43]{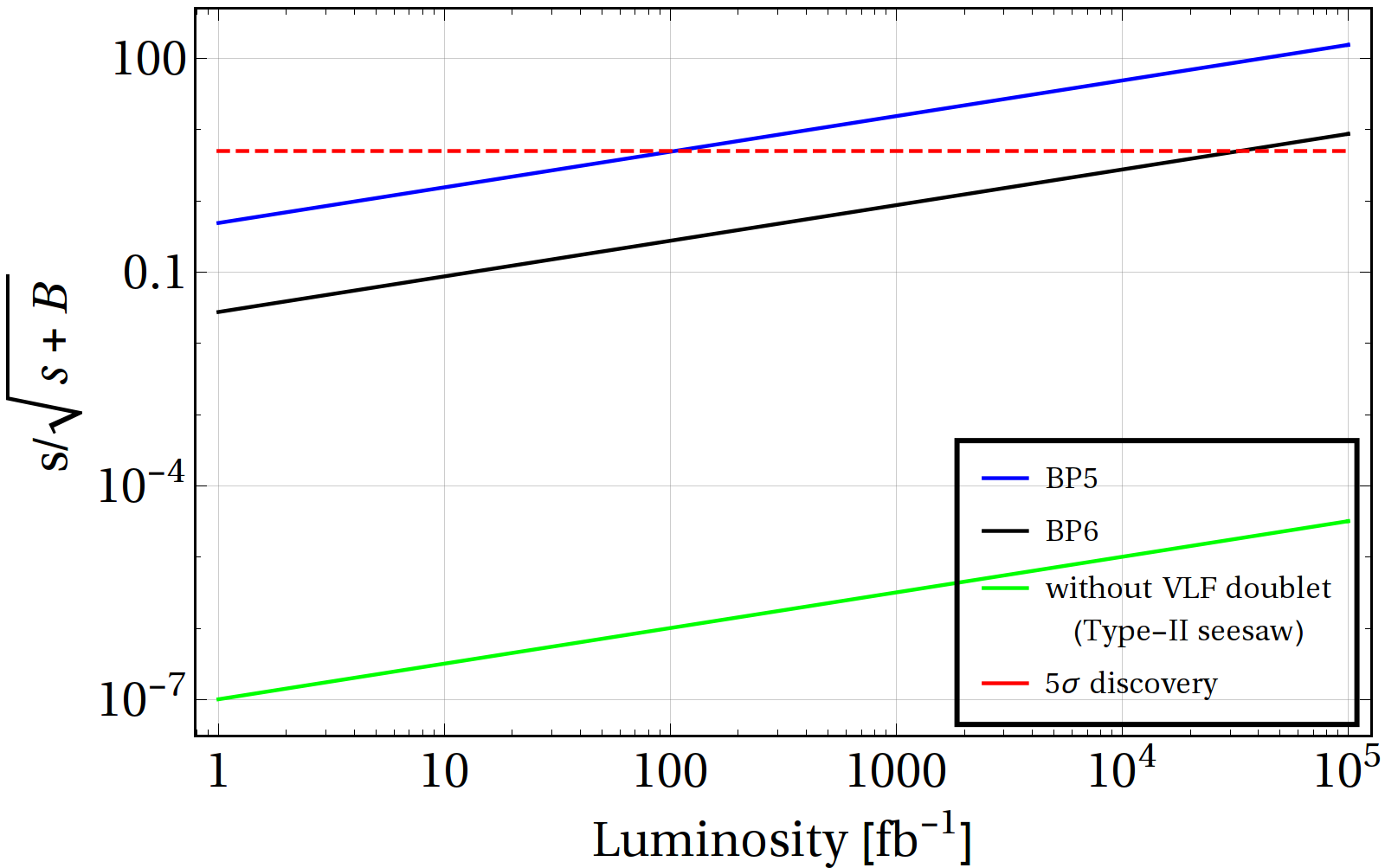}
 $$
 \caption{Left: Variation of significance with integrated luminosity at the ILC for hadronically four lepton final state where the red dashed line is the $5\sigma$ significance. Right: Same for the LHC.}
 \label{fig:tripilcsigni}
\end{figure}

Together, the discovery potential for HQ4l events through production of doubly charged scalars in our model, where the decay for such scalars dominantly occur 
through the charged VLFs, are shown in terms of Luminosity in Fig.~\ref{fig:tripilcsigni} at ILC (in the left panel) and at LHC (in the right panel). 
This clearly indicates that a judicious choice of MET cut can not only make such signal distinguished from the SM background, but can also segregate our model 
from the usual scalar triplet models as in Type II seesaw frameworks (compare the green line with blue and black thick lines in Fig.~\ref{fig:tripilcsigni}). 
In conclusion, we can say the HQ4l signature due production of the doubly charged scalars can help us to distinguish this model from that of usual type-II seesaw 
models both at the ILC and the LHC for very high integrated luminosity.


\section{Displaced vertex signature and Complementarity of different search strategies}
\label{sec:complmntr}

Finally, we would like to highlight the displaced vertex signature of this model, which is elaborated in~\cite{Bhattacharya:2017sml}. If the mass difference between $\psi^{\pm}$ and $\psi_1$ is less than that of $W$-mass, 
then the charged fermions will decay via three body process. In such cases we can see a displaced vertex signature for our model at the LHC, provided the track length (which is inverse of the 3-body decay width) 
is $\sim\mathcal{O}(1~\rm mm)$. Now, the decay width is given by~\cite{Bhattacharya:2017sml}:

\bea
\Gamma = \frac{G_F^2\sin^2\theta M_{\psi}^5}{24\pi^3} \xi,
\label{eq:decay3}
\eea
where $G_F$ is the Fermi coupling constant and the function $\xi$ is given by:

\bea
\xi = \frac{1}{4} \sqrt{\alpha}\left(x^2,y^2\right) \zeta_1\left(x,y\right)+6 \zeta_2\left(x,y\right) \ln\left(\frac{2 x}{1+x^2-y^2-\alpha^{1/2}}\right).
\eea

Here $\zeta_1$ and $\zeta_2$ are two polynomials of $x=M_1/M_{\psi}$ and $b=m_{\ell}/M_{\psi}$, where $m_{\ell}$ is the mass of charged leptons. Upto order $\mathcal{O}(y^2)$, $\zeta_{1,2}$ are given as:

\bea
\begin{split}
\zeta_1(x,y) &= \left(x^6-2 x^5-7 x^4\left(1+y^2\right)+10 x^3\left(y^2-2\right)+x^2\left(12 y^2-7\right)+3 y^2-1\right) \\
\zeta_2(x,y) &=\left(x^5+x^4+x^3\left(1-2 y^2\right)\right),
\end{split}
\eea

where $\alpha=1+x^4+y^4-2 x^2-2 y^2-2 x^2 y^2$ is the phase space. The length of the displaced vertex is given as 
$c\tau\equiv\frac{c}{\Gamma}$, where $\Gamma$ can be obtained from Eq.~\ref{eq:decay3}.

\begin{figure}[htb!]
 $$
\includegraphics[scale=0.37]{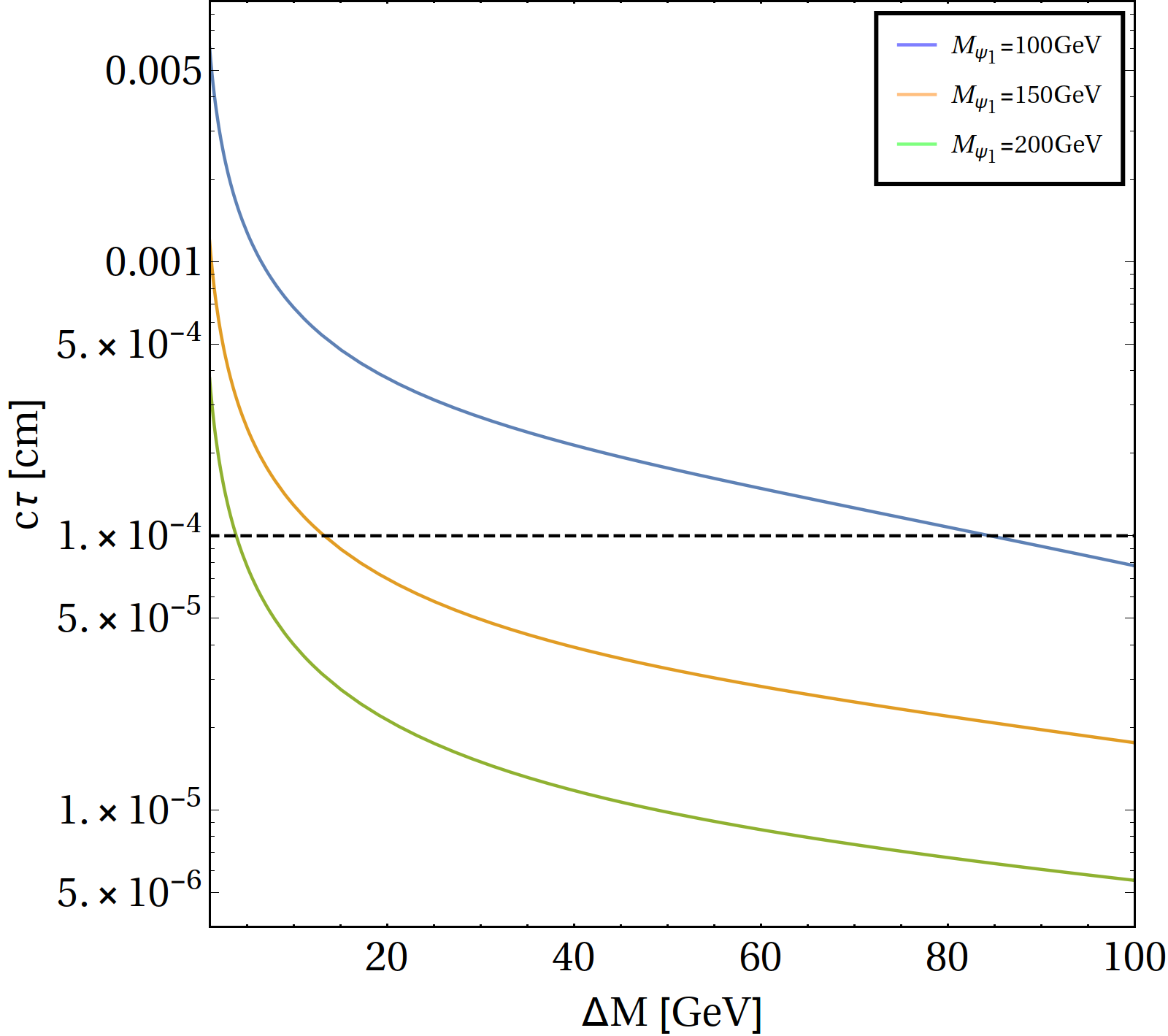}
\includegraphics[scale=0.356]{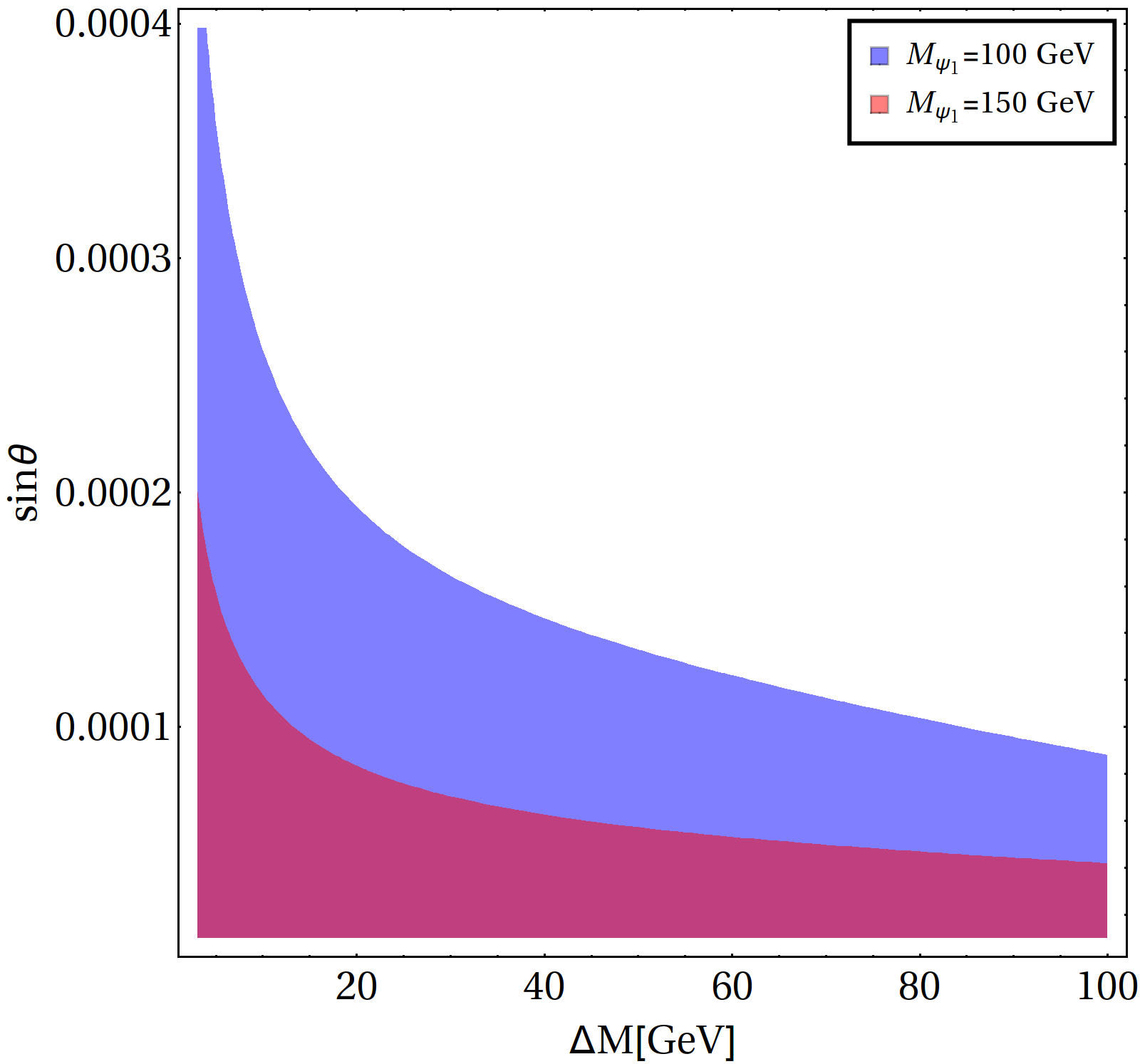}
 $$
 \caption{Left: Displaced vertex length ($c\tau$ in cm) versus $\Delta M$ for three different choices of DM mass $M_{\psi_1}=\{100,150,200\}$ GeVs for $\sin\theta =10^{-4}$. The horizontal  black dashed line corresponds to displaced vertex length of $10^{-4}$ cm. Right: Limit on $\sin\theta$ for producing displaced vertex of $c\tau \ge 10^{-4}$ cm as a function of $\Delta M$ for two specific DM masses 100 and 150 GeVs.}
 \label{fig:displaced}
\end{figure}

In Fig.~\ref{fig:displaced}, we have shown two different parametrisation for a realizable displaced vertex signature produced in this model. In the left panel, we plot the displaced vertex length $c\tau$ as a function of 
$\Delta M$ for a fixed $\sin\theta \sim 10^{-4}$. We illustrate three different choices of DM mass $M_{\psi_1}=\{100,150,200\}$ GeVs. The horizontal black dashed line corresponds to displaced vertex length of 
$10^{-4}$ cm ($i.e,1~\rm\mu m$). In the right panel, we show the limit on $\sin\theta$ for producing displaced vertex of $c\tau \ge 10^{-4}$ cm as a function of $\Delta M$ for two specific DM masses 100 and 150 GeVs. 
The upshot is, if we have to detect a measurable displaced vertex length at collider, $\sin\theta$ has to be extremely small. However, with small $\sin\theta$, the allowed parameter space behaves similar to that of 
$\sin\theta \lsim 0.1$, which has to heavily rely on co-annihilation effects to obtain correct relic density and is allowed by direct search bounds. It is also important to note that the presence of triplet scalar do not at all alter 
the displaced vertex signature discussed before for the fermion DM alone~\cite{Bhattacharya:2015qpa}.

\begin{figure}[htb!]
 $$
 \includegraphics[scale=0.4]{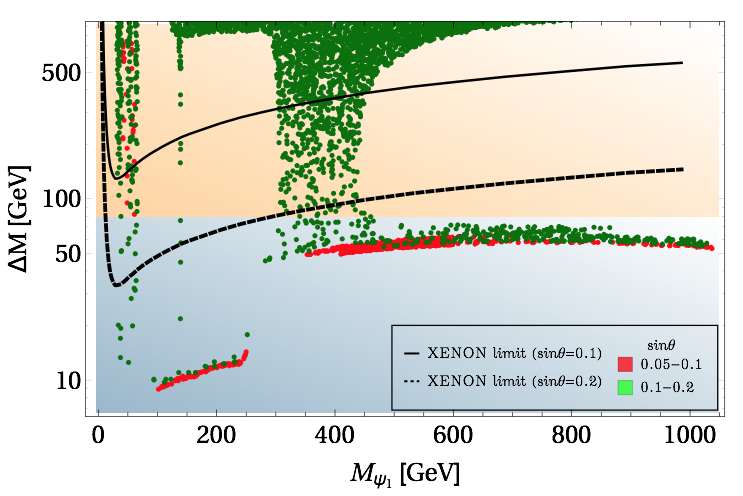}
 $$
\caption{Summary of the available parameter space in $M_{\psi_1}-\Delta M$ plane from relic density, direct search constraints and collider sensitivity. Red and green points correspond to PLANCK-observed relic abundance satisfying region (same as in LHS of Fig.~\ref{fig:mDelmrelic}); Black thick and black dashed lines correspond to XENON1T upper bound for $\sin\theta=\{0.1,0.2\}$. The blue shaded region can be probed at ILC, while the orange shaded region can potentially be probed at LHC. We have assumed the triplet scalar to have a mass $\sim$ 300 GeV. } 
\label{fig:complmnt}
\end{figure}

Finally, all the searches and constraints for this DM model put together, allow us to visualize how all these different searches are complementary to one another. Such a summary plot is shown in Fig.~\ref{fig:complmnt}. 
The green and red points correspond to observed relic abundance that are allowed by PLANCK data. XENON1T direct detection limit as shown in Fig.~\ref{fig:dd}, are indicated by the solid and dashed black lines for 
$\sin\theta=\{0.1,0.2\}$ respectively. The blue shaded region for $\Delta M< M_W$ can be probed at the ILC, while the region above with $\Delta M> M_W$ shown by orange shaded region, can be probed at the LHC. 
It is difficult to calculate the significance for the signal cross-section in this plane, and therefore the fading in the colour shades imply that the cross-section diminishes with large $M_{\psi_1}$ as well as with 
large $\Delta M$. From the figure, firstly we identify that, for small $\sin\theta \lsim 0.1$ there is a large region (red points) that fall below the direct search exclusion, which can be seen at future direct detection 
experiments for almost all of the DM mass $M_{\psi_1}$. Small $\sin\theta$ region can generally be probed as OSD signal excess at the ILC, while the small $\Delta M$ region can be probed via displaced vertex 
signature at the LHC. $Z$ and $H$-resonance regions for $\sin\theta \lsim 0.1$ can only be probed through OSD signature at the LHC. For larger $\sin\theta$, direct search allowed points again limit within 
$\Delta M \lsim 50$ GeV, and therefore favourable for the ILC search. Displaced vertex signature ($c\tau\lsim 1$ mm) requires further suppressed values of $\sin\theta \sim 10^{-4}$ (as illustrated in 
Fig.~\ref{fig:displaced}), and hence could not be shown in the plot.    

\section{Conclusions}
\label{sec:conclusion}

The paper focuses on a beyond SM (BSM) framework by introducing two vector-like fermions: a singlet $\chi$ and a doublet $\psi$, where the DM emerges as a lightest component, $\psi_1$, as an admixture of the neutral 
component of $\psi$ and $\chi$. The phenomenology of the model crucially dictates small singlet-doublet mixing ($\sin \theta$) to abide by the non observation of DM in direct search. Moreover, the observed relic density of 
DM restricts the mass splitting between DM and NLSP ($\Delta M$) to less than 10 GeV due to which the model can be probed at LHC only through displaced vertex signature of the NLSP. In this analysis we show that the ILC, 
however, can probe such model even in small mixing limit through the production and subsequent decays of the charged companion $\psi^{\pm}$ via hadronically quiet opposite sign dilepton (OSD) channels. 
This is possible due to primarily the nature of the missing energy distribution for signal and background events at the ILC,  which allows to put an upper cut on missing energy for event selection, 
while the possibility of utilising the polarization of $\{e^{-}, e^{+}\}=\{80\%,-20\%\}$ reduces SM background significantly, retaining most of the signal events due to vectorlike fermion nature.   
  

The presence of a scalar triplet of hypercharge 2 in the model can produce non-zero masses for the active neutrinos, as required by solar and atmospheric oscillation data. This also alters the DM phenomenology crucially. 
In presence of the scalar triplet, the dark fermion $\psi_1$ splits into two pseudo-Dirac states $\psi_1^\alpha$ and $\psi_1^\beta$. As a result, the $Z$-mediated DM-nucleon scattering at direct search experiments become inelastic. 
Assuming the mass splitting between the two pseudo-Dirac states to be of order 100 KeV, we showed that the DM-nucleon scattering through $Z$ mediation is forbidden. This helps to achieve larger singlet-doublet mixing in the DM state. 
In fact, we showed that the doublet component can be as large as 20\%. Moreover, the mass splitting between the DM and NLSP can be chosen to be as large as a few hundred GeVs. These are the two key factors which paved 
a path for detecting the DM model at the LHC through hadronically quiet OSD channel. However, the broadening of the mass splitting can not be obtained in all region of the parameter space, rather it is specific to the 
Higgs, $Z$ and triplet scalar resonance regions, as well as when the DM mass is equal or slightly larger than the triplet scalar. So, the LHC search for a signal excess can only be possible in such regions of the DM mass 
parameter with large $\Delta M$. On the other hand, if we embed the singlet-doublet fermion DM in presence of an additional scalar singlet DM~\cite{Bhattacharya:2018cgx}, the possibility of exploring larger $\Delta M$ regions 
enhance significantly due to DM-DM conversion and therefore signal excess of hadronically quiet dilepton channel at LHC spans a large range of fermion DM mass range.

On the other hand, the doubly charged scalar present in the scalar triplet can also be produced at the collider (both LHC and ILC) through Drell-Yan process, which yields hadronically quiet four lepton signature. It is interesting to 
note that in presence of vector like lepton as we have in this model, the doubly charged scalars attain a significant branching to the charged vector like lepton, whenever kinematically allowed. This in turn leave its imprint in the 
missing energy profile. A judicious choice of MET cut can therefore distinguish our model from the usual case of scalar triplet scenarios, like Type II Seesaw. We also note here that in absence of dominant $t\bar{t}$ and Drell-Yan 
background for hadronically quiet four lepton events, one can utilise an upper MET cut or a small MET window to probe the low $\Delta M$ regions of our vector like DM model at LHC, which is difficult for hadronically quiet dilepton channel. 
 
The model naturally possess another novel signature: displaced vertex of the charged vector like lepton. It is however easily understood that the displaced vertex signature of the NLSP not only requires 
small singlet-doublet mixing, but also requires very small mass splitting $\Delta M$ between NLSP and DM. This is a natural outcome of the singlet doublet DM in absence of scalar triplet in DM allowed 
parameter space to respect direct search bound. So, while adding a scalar triplet, we enhance the possibility of seeing a dilepton signal excess at LHC through 
enlarging the mass splitting $\Delta M$ at certain resonance regions and particularly when the DM mass is close but larger than scalar triplet mass, the displaced vertex signature gets washed off in all those regions. 
While LHC favors large mass splitting between NLSP and DM for the dilepton signal to be segregated from SM background due to indomitable $t\bar{t}$ background, the `absence' of such a channel at ILC will 
favour the cases of small mass splitting $\Delta M$ to yield a signal excess over background. Thus, the model has a nice complementarity in its variety of signatures that can be probed at upcoming experiments.

\section*{Acknowledgements}

BB and PG would like to acknowledge discussions with Nivedita Ghosh regarding the collider analysis; BB would also like to acknowledge Abdessalam Arhrib and Alexander Belyaev for helping out with the model implementation. SB would like to acknowledge DST-INSPIRE faculty grant IFA 13 PH-57 at IIT Guwahati.

\appendix
\section{Appendix}
\subsection{Invisible Higgs and Z-decay}
\label{sec:invdecay}

Here we have shown that the BPs chosen for LHC analysis (Tab.~\ref{tab:bp}) are allowed by experimental bounds on invisible Higgs and $Z$-decays. The SM Higgs can decay to $\psi_1$ pairs. Now,  the combination of SM channels yields an observed (expected) upper limit on the Higgs branching fraction of 0.24 at 95 \%\ CL~\cite{Khachatryan:2016whc} with a total decay width $\Gamma=4.07\times 10^{-3}~\rm GeV$. On the other hand, SM $Z$ boson can also decay to DM pairs and hence constrained from observation: $\Gamma_{inv}^{Z}=499\pm 1.5~\rm MeV$~\cite{PhysRevD.98.030001}. So, if $Z$ is allowed to decay into $\psi_1\psi_1$ pair, the decay width should not be more than 1.5 MeV. 

\begin{table}[htb!]
\begin{center}
\begin{tabular}{|c|c|c|c|c|c|c|c|c|c|}
\hline
Benchmark & $Br_{inv}^{higgs}$ &  $\Gamma_{inv}^{Z}$ (MeV)  \\ [0.5ex] 
Point     &                    &             (MeV)  \\ [0.5ex] 
\hline\hline 

BP1 & $1.116\times 10^{-3}$ & {\it NA} \\
\hline
BP2 & $654.86\times 10^{-6}$ & {\it NA} \\
\hline
BP3 & $5.738\times 10^{-3}$ & 1.201  \\
\hline
BP4 & $80.24\times 10^{-6}$ & {\it NA} \\
\hline
\end{tabular}
\end{center}
\caption {Invisible Higgs branching ratio and invisible $Z$ decay width for different benchmark points tabulated in Tab.~\ref{tab:bp}. {\it NA} stands for `Not Applicable` for cases where $M_{\psi_1}>m_Z/2$.} 
\label{tab:invdecay}
\end{table}

Since $\Delta M>100~\rm GeV$ for all the BPs, hence Higgs or $Z$ can not decay to $\psi_2$'s. Therefore, the expressions for $H_1\to\psi_1\psi_1$ and $Z\to\psi_1\psi_1$ decay widths are given by:

\bea
\Gamma_{inv}^{higgs}\left(H_1\to\psi_1\psi_1\right) = \left(\frac{y_N^2\sin^4\theta\cos^2\theta_{0}}{8\pi}\right)m_{H_1}\left(1-\frac{4 M_{\psi_1}^2}{m_{H_1}^2}\right)^{3/2}
\eea

\bea
\Gamma_{inv}^{Z}\left(Z\to\psi_1\psi_1\right) = \frac{m_Z}{48\pi} \frac{e^2\sin^4\theta}{\sin^2\theta_W\cos^2\theta_W}\left(1+\frac{M_{\psi_1}^2}{m_Z^2}\right)\sqrt{1-\frac{4 M_{\psi_1}^2}{m_Z^2}}.
\eea

In Tab.~\ref{tab:invdecay} we have tabulated the Higgs branching ratio and $Z$-decay width for all the chosen benchmark points.  Constraint from invisible $Z$-decay is only applicable for BP1 and BP5 which correspond to $M_{\psi_1}=41~\rm GeV$ and $M_{\psi_1}=45~\rm GeV$ respectively, while invisible Higgs decay costraint is applicable for all the benchmarks.

\subsection{Lagrangian parametrs}
\label{sec:cplings}

One can express all the couplings appearing in the scalar potential~\ref{eq:pot} in terms of the physical masses. Apart from the parameters $\mu_H$ and $\mu_{\Delta}$ obtained through electroweak symmetry breaking condition (See Eq.~\ref{eq:minim1}), one can also determine the following parameters:

\bea
\begin{split}
\lambda_1 &= -\frac{2 m_A^2}{v_d^2+4 v_t^2}+\frac{4 m_{H^{\pm}}^2}{v_d^2+2 v_t^2}+\frac{\sin2\theta_0\left(m_{H_1}^2-m_{H_2}^2\right)}{2 v_d v_t},\nonumber\\
\lambda_2 &= \frac{1}{v_t^2}\left[\frac{1}{2}\left(\sin^2\theta_0 m_{H_1}^2+\cos^2\theta_0 m_{H_2}^2\right)+\frac{1}{2}.\frac{v_d^2 m_A^2}{v_d^2+4 v_t^2}-\frac{2 v_d^2 m_{H^{\pm}}^2}{v_d^2+2 v_t^2}+m_{H^{\pm\pm}}^2\right],
\nonumber\\
\lambda_3 &= \frac{1}{v_t^2}\left[-\frac{v_d^2 m_A^2}{v_d^2+4 v_t^2}+\frac{2 v_d^2 m_{H^{\pm}}^2}{v_d^2+2 v_t^2}-m_{H^{\pm\pm}}^2\right],
\nonumber\\
\lambda_4 &= \frac{4 m_A^2}{v_d^2+4 v_t^2}-\frac{4 m_{H^{\pm}}^2}{v_d^2+2 v_t^2},
\nonumber\\
\lambda &= \frac{2}{v_d^2}\left(\cos^2\theta_0 m_{H_1}^2+\sin^2\theta_0 m_{H_2}^2\right),
\nonumber\\
\mu &= \frac{\sqrt{2} v_t m_A^2}{v_d^2+4 v_t^2}.
\end{split}
\eea

\clearpage
\subsection{Annihilation and co-annihilation in presence of Higgs triplet}
\label{sec:tripdiagram}

Here we have gathered all the annihilation and co-annihilation graphs in presence of the triplet scalar. 

\begin{figure}[htb!]
$$
\includegraphics[scale=0.55]{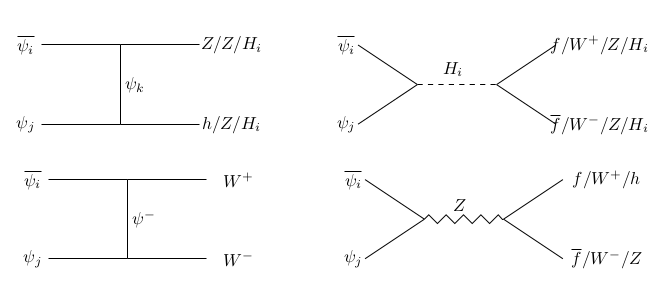}
$$
\caption{Annihilation ($i=j$) and co-annihilation ($i\neq j$) of vector-like fermion DM. Here $(i,j=1,2)$. }
\label{fd:an-coan}
 \end{figure}
\begin{figure}[htb!]
$$
\includegraphics[scale=0.55]{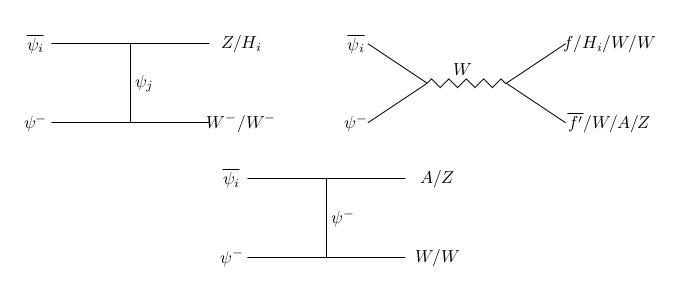}
$$
\caption{ Co-annihilation process of $\psi_i ~(i=1,2)$ with the charge component $\psi^-$ to SM particles. }
\label{co-ann-2}
 \end{figure}
\begin{figure}[htb!]
$$
\includegraphics[scale=0.55]{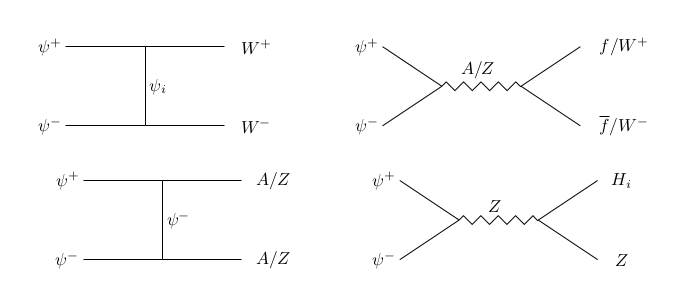}
$$
\caption{Co-annihilation process of charged fermions $\psi^\pm$ to SM particles in final states . }
\label{co-ann-3}
\end{figure}
 
 
\begin{figure}[htb!]
$$
\includegraphics[scale=0.55]{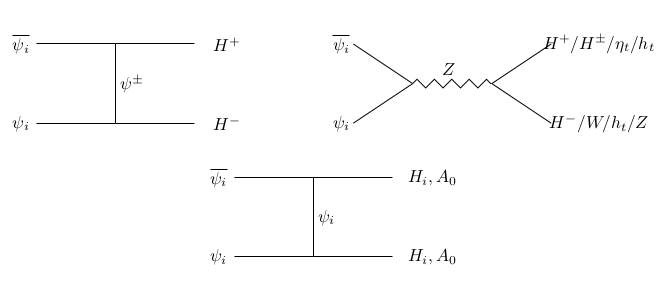}
$$
\caption{Additional annihilation $\psi_i \overline{\psi_i} $, in presence of scalar triplet. }
\label{fd:ann_ILD_triplet}
 \end{figure}

\begin{figure}[htb!]
$$
\includegraphics[scale=0.55]{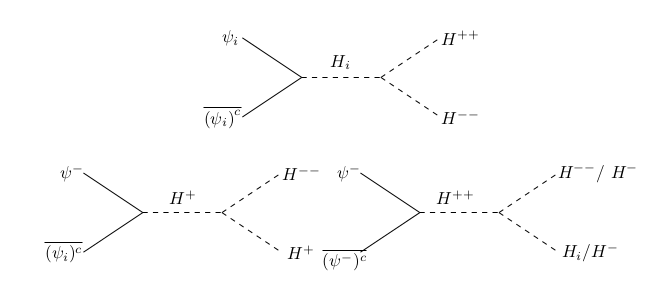}
$$
\caption{Dominant annihilation ($\psi_i \overline{(\psi_i)^c}$)and co-annihilation ($\psi^- \overline{(\psi_i)^c},~\psi^- \overline{(\psi^-)^c}$) processes of DM ($\psi_i$) to scalar triplet in final states.}
\label{fd:annco_ILD_triplet}
 \end{figure}
\begin{figure}[htb!]
$$
\includegraphics[scale=0.55]{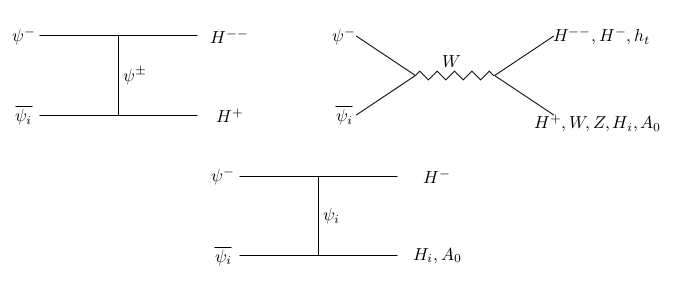}
$$
\caption{Co-annihilation channels of DM ($\psi_i$), with charged fermions $\psi^-$ in presence of scalar triplet. }
\label{fd:Chargedann_ILD_triplet}
 \end{figure}
\begin{figure}[htb!]
$$
\includegraphics[scale=0.55]{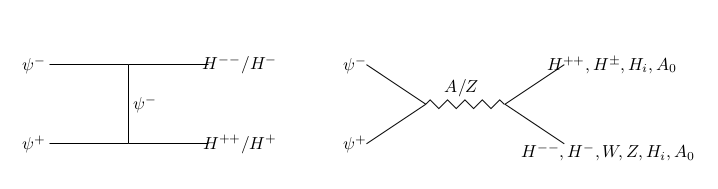}
$$
\caption{Co-annihilation processes involving only charged partner of DM, $\psi^\pm$ in presence of scalar triplet. }
\label{fd:coann_ILD_triplet}
 \end{figure}

\clearpage
\bibliography{ref}

\begin{thebibliography}{48}%
\makeatletter
\providecommand \@ifxundefined [1]{%
 \@ifx{#1\undefined}
}%
\providecommand \@ifnum [1]{%
 \ifnum #1\expandafter \@firstoftwo
 \else \expandafter \@secondoftwo
 \fi
}%
\providecommand \@ifx [1]{%
 \ifx #1\expandafter \@firstoftwo
 \else \expandafter \@secondoftwo
 \fi
}%
\providecommand \natexlab [1]{#1}%
\providecommand \enquote  [1]{``#1''}%
\providecommand \bibnamefont  [1]{#1}%
\providecommand \bibfnamefont [1]{#1}%
\providecommand \citenamefont [1]{#1}%
\providecommand \href@noop [0]{\@secondoftwo}%
\providecommand \href [0]{\begingroup \@sanitize@url \@href}%
\providecommand \@href[1]{\@@startlink{#1}\@@href}%
\providecommand \@@href[1]{\endgroup#1\@@endlink}%
\providecommand \@sanitize@url [0]{\catcode `\\12\catcode `\$12\catcode
  `\&12\catcode `\#12\catcode `\^12\catcode `\_12\catcode `\%12\relax}%
\providecommand \@@startlink[1]{}%
\providecommand \@@endlink[0]{}%
\providecommand \url  [0]{\begingroup\@sanitize@url \@url }%
\providecommand \@url [1]{\endgroup\@href {#1}{\urlprefix }}%
\providecommand \urlprefix  [0]{URL }%
\providecommand \Eprint [0]{\href }%
\providecommand \doibase [0]{http://dx.doi.org/}%
\providecommand \selectlanguage [0]{\@gobble}%
\providecommand \bibinfo  [0]{\@secondoftwo}%
\providecommand \bibfield  [0]{\@secondoftwo}%
\providecommand \translation [1]{[#1]}%
\providecommand \BibitemOpen [0]{}%
\providecommand \bibitemStop [0]{}%
\providecommand \bibitemNoStop [0]{.\EOS\space}%
\providecommand \EOS [0]{\spacefactor3000\relax}%
\providecommand \BibitemShut  [1]{\csname bibitem#1\endcsname}%
\let\auto@bib@innerbib\@empty
\bibitem [{\citenamefont {Jungman}\ \emph {et~al.}(1996)\citenamefont
  {Jungman}, \citenamefont {Kamionkowski},\ and\ \citenamefont
  {Griest}}]{Jungman:1995df}%
  \BibitemOpen
  \bibfield  {author} {\bibinfo {author} {\bibfnamefont {G.}~\bibnamefont
  {Jungman}}, \bibinfo {author} {\bibfnamefont {M.}~\bibnamefont
  {Kamionkowski}}, \ and\ \bibinfo {author} {\bibfnamefont {K.}~\bibnamefont
  {Griest}},\ }\href {\doibase 10.1016/0370-1573(95)00058-5} {\bibfield
  {journal} {\bibinfo  {journal} {Phys. Rept.}\ }\textbf {\bibinfo {volume}
  {267}},\ \bibinfo {pages} {195} (\bibinfo {year} {1996})},\ \Eprint
  {http://arxiv.org/abs/hep-ph/9506380} {arXiv:hep-ph/9506380 [hep-ph]}
  \BibitemShut {NoStop}%
\bibitem [{\citenamefont {Bertone}\ \emph {et~al.}(2005)\citenamefont
  {Bertone}, \citenamefont {Hooper},\ and\ \citenamefont
  {Silk}}]{Bertone:2004pz}%
  \BibitemOpen
  \bibfield  {author} {\bibinfo {author} {\bibfnamefont {G.}~\bibnamefont
  {Bertone}}, \bibinfo {author} {\bibfnamefont {D.}~\bibnamefont {Hooper}}, \
  and\ \bibinfo {author} {\bibfnamefont {J.}~\bibnamefont {Silk}},\ }\href
  {\doibase 10.1016/j.physrep.2004.08.031} {\bibfield  {journal} {\bibinfo
  {journal} {Phys. Rept.}\ }\textbf {\bibinfo {volume} {405}},\ \bibinfo
  {pages} {279} (\bibinfo {year} {2005})},\ \Eprint
  {http://arxiv.org/abs/hep-ph/0404175} {arXiv:hep-ph/0404175 [hep-ph]}
  \BibitemShut {NoStop}%
\bibitem [{\citenamefont {Conrad}(2014)}]{Conrad:2014tla}%
  \BibitemOpen
  \bibfield  {author} {\bibinfo {author} {\bibfnamefont {J.}~\bibnamefont
  {Conrad}},\ }in\ \href@noop {} {\emph {\bibinfo {booktitle} {{Interplay
  between Particle and Astroparticle physics (IPA2014) London, United Kingdom,
  August 18-22, 2014}}}}\ (\bibinfo {year} {2014})\ \Eprint
  {http://arxiv.org/abs/1411.1925} {arXiv:1411.1925 [hep-ph]} \BibitemShut
  {NoStop}%
\bibitem [{\citenamefont {Gaskins}(2016)}]{Gaskins:2016cha}%
  \BibitemOpen
  \bibfield  {author} {\bibinfo {author} {\bibfnamefont {J.~M.}\ \bibnamefont
  {Gaskins}},\ }\href {\doibase 10.1080/00107514.2016.1175160} {\bibfield
  {journal} {\bibinfo  {journal} {Contemp. Phys.}\ }\textbf {\bibinfo {volume}
  {57}},\ \bibinfo {pages} {496} (\bibinfo {year} {2016})},\ \Eprint
  {http://arxiv.org/abs/1604.00014} {arXiv:1604.00014 [astro-ph.HE]}
  \BibitemShut {NoStop}%
\bibitem [{\citenamefont {Hinshaw}\ \emph {et~al.}(2013)\citenamefont {Hinshaw}
  \emph {et~al.}}]{Hinshaw:2012aka}%
  \BibitemOpen
  \bibfield  {author} {\bibinfo {author} {\bibfnamefont {G.}~\bibnamefont
  {Hinshaw}} \emph {et~al.} (\bibinfo {collaboration} {WMAP}),\ }\href
  {\doibase 10.1088/0067-0049/208/2/19} {\bibfield  {journal} {\bibinfo
  {journal} {Astrophys. J. Suppl.}\ }\textbf {\bibinfo {volume} {208}},\
  \bibinfo {pages} {19} (\bibinfo {year} {2013})},\ \Eprint
  {http://arxiv.org/abs/1212.5226} {arXiv:1212.5226 [astro-ph.CO]} \BibitemShut
  {NoStop}%
\bibitem [{\citenamefont {Akrami}\ \emph {et~al.}(2018)\citenamefont {Akrami}
  \emph {et~al.}}]{Akrami:2018odb}%
  \BibitemOpen
  \bibfield  {author} {\bibinfo {author} {\bibfnamefont {Y.}~\bibnamefont
  {Akrami}} \emph {et~al.} (\bibinfo {collaboration} {Planck}),\ }\href@noop {}
  {\  (\bibinfo {year} {2018})},\ \Eprint {http://arxiv.org/abs/1807.06211}
  {arXiv:1807.06211 [astro-ph.CO]} \BibitemShut {NoStop}%
\bibitem [{\citenamefont {Kolb}\ and\ \citenamefont
  {Turner}(1990)}]{Kolb:1990vq}%
  \BibitemOpen
  \bibfield  {author} {\bibinfo {author} {\bibfnamefont {E.~W.}\ \bibnamefont
  {Kolb}}\ and\ \bibinfo {author} {\bibfnamefont {M.~S.}\ \bibnamefont
  {Turner}},\ }\href@noop {} {\bibfield  {journal} {\bibinfo  {journal} {Front.
  Phys.}\ }\textbf {\bibinfo {volume} {69}},\ \bibinfo {pages} {1} (\bibinfo
  {year} {1990})}\BibitemShut {NoStop}%
\bibitem [{\citenamefont {Hall}\ \emph {et~al.}(2010)\citenamefont {Hall},
  \citenamefont {Jedamzik}, \citenamefont {March-Russell},\ and\ \citenamefont
  {West}}]{Hall:2009bx}%
  \BibitemOpen
  \bibfield  {author} {\bibinfo {author} {\bibfnamefont {L.~J.}\ \bibnamefont
  {Hall}}, \bibinfo {author} {\bibfnamefont {K.}~\bibnamefont {Jedamzik}},
  \bibinfo {author} {\bibfnamefont {J.}~\bibnamefont {March-Russell}}, \ and\
  \bibinfo {author} {\bibfnamefont {S.~M.}\ \bibnamefont {West}},\ }\href
  {\doibase 10.1007/JHEP03(2010)080} {\bibfield  {journal} {\bibinfo  {journal}
  {JHEP}\ }\textbf {\bibinfo {volume} {03}},\ \bibinfo {pages} {080} (\bibinfo
  {year} {2010})},\ \Eprint {http://arxiv.org/abs/0911.1120} {arXiv:0911.1120
  [hep-ph]} \BibitemShut {NoStop}%
\bibitem [{\citenamefont {Hochberg}\ \emph {et~al.}(2014)\citenamefont
  {Hochberg}, \citenamefont {Kuflik}, \citenamefont {Volansky},\ and\
  \citenamefont {Wacker}}]{Hochberg:2014dra}%
  \BibitemOpen
  \bibfield  {author} {\bibinfo {author} {\bibfnamefont {Y.}~\bibnamefont
  {Hochberg}}, \bibinfo {author} {\bibfnamefont {E.}~\bibnamefont {Kuflik}},
  \bibinfo {author} {\bibfnamefont {T.}~\bibnamefont {Volansky}}, \ and\
  \bibinfo {author} {\bibfnamefont {J.~G.}\ \bibnamefont {Wacker}},\ }\href
  {\doibase 10.1103/PhysRevLett.113.171301} {\bibfield  {journal} {\bibinfo
  {journal} {Phys. Rev. Lett.}\ }\textbf {\bibinfo {volume} {113}},\ \bibinfo
  {pages} {171301} (\bibinfo {year} {2014})},\ \Eprint
  {http://arxiv.org/abs/1402.5143} {arXiv:1402.5143 [hep-ph]} \BibitemShut
  {NoStop}%
\bibitem [{\citenamefont {Akerib}\ \emph {et~al.}(2017)\citenamefont {Akerib}
  \emph {et~al.}}]{Akerib:2017kat}%
  \BibitemOpen
  \bibfield  {author} {\bibinfo {author} {\bibfnamefont {D.~S.}\ \bibnamefont
  {Akerib}} \emph {et~al.} (\bibinfo {collaboration} {LUX}),\ }\href {\doibase
  10.1103/PhysRevLett.118.251302} {\bibfield  {journal} {\bibinfo  {journal}
  {Phys. Rev. Lett.}\ }\textbf {\bibinfo {volume} {118}},\ \bibinfo {pages}
  {251302} (\bibinfo {year} {2017})},\ \Eprint
  {http://arxiv.org/abs/1705.03380} {arXiv:1705.03380 [astro-ph.CO]}
  \BibitemShut {NoStop}%
\bibitem [{\citenamefont {Zhang}\ \emph {et~al.}(2019)\citenamefont {Zhang}
  \emph {et~al.}}]{Zhang:2018xdp}%
  \BibitemOpen
  \bibfield  {author} {\bibinfo {author} {\bibfnamefont {H.}~\bibnamefont
  {Zhang}} \emph {et~al.} (\bibinfo {collaboration} {PandaX}),\ }\href
  {\doibase 10.1007/s11433-018-9259-0} {\bibfield  {journal} {\bibinfo
  {journal} {Sci. China Phys. Mech. Astron.}\ }\textbf {\bibinfo {volume}
  {62}},\ \bibinfo {pages} {31011} (\bibinfo {year} {2019})},\ \Eprint
  {http://arxiv.org/abs/1806.02229} {arXiv:1806.02229 [physics.ins-det]}
  \BibitemShut {NoStop}%
\bibitem [{\citenamefont {Aprile}\ \emph {et~al.}(2018)\citenamefont {Aprile}
  \emph {et~al.}}]{Aprile:2018dbl}%
  \BibitemOpen
  \bibfield  {author} {\bibinfo {author} {\bibfnamefont {E.}~\bibnamefont
  {Aprile}} \emph {et~al.} (\bibinfo {collaboration} {XENON}),\ }\href@noop {}
  {\  (\bibinfo {year} {2018})},\ \Eprint {http://arxiv.org/abs/1805.12562}
  {arXiv:1805.12562 [astro-ph.CO]} \BibitemShut {NoStop}%
\bibitem [{\citenamefont {Lowette}(2016)}]{Lowette:2014yta}%
  \BibitemOpen
  \bibfield  {author} {\bibinfo {author} {\bibfnamefont {S.}~\bibnamefont
  {Lowette}} (\bibinfo {collaboration} {CMS}),\ }\bibfield  {booktitle} {\emph
  {\bibinfo {booktitle} {{Proceedings, 37th International Conference on High
  Energy Physics (ICHEP 2014): Valencia, Spain, July 2-9, 2014}}},\ }\href
  {\doibase 10.1016/j.nuclphysbps.2015.09.074} {\bibfield  {journal} {\bibinfo
  {journal} {Nucl. Part. Phys. Proc.}\ }\textbf {\bibinfo {volume} {273-275}},\
  \bibinfo {pages} {503} (\bibinfo {year} {2016})},\ \Eprint
  {http://arxiv.org/abs/1410.3762} {arXiv:1410.3762 [hep-ex]} \BibitemShut
  {NoStop}%
\bibitem [{\citenamefont {Ahuja}(2018)}]{Ahuja:2018bbj}%
  \BibitemOpen
  \bibfield  {author} {\bibinfo {author} {\bibfnamefont {S.}~\bibnamefont
  {Ahuja}} (\bibinfo {collaboration} {CMS}),\ }\bibfield  {booktitle} {\emph
  {\bibinfo {booktitle} {{Proceedings, 6th Large Hadron Collider Physics
  Conference (LHCP 2018): Bologna, Italy, June 4-9, 2018}}},\ }\href {\doibase
  10.22323/1.321.0284} {\bibfield  {journal} {\bibinfo  {journal} {PoS}\
  }\textbf {\bibinfo {volume} {LHCP2018}},\ \bibinfo {pages} {284} (\bibinfo
  {year} {2018})}\BibitemShut {NoStop}%
\bibitem [{\citenamefont {Tanabashi}\ and\ \citenamefont
  {Hagiwara}(2018)}]{PhysRevD.98.030001}%
  \BibitemOpen
  \bibfield  {author} {\bibinfo {author} {\bibfnamefont {M.}~\bibnamefont
  {Tanabashi}}\ and\ \bibinfo {author} {\bibfnamefont {e.~a.}\ \bibnamefont
  {Hagiwara}} (\bibinfo {collaboration} {Particle Data Group}),\ }\href
  {\doibase 10.1103/PhysRevD.98.030001} {\bibfield  {journal} {\bibinfo
  {journal} {Phys. Rev. D}\ }\textbf {\bibinfo {volume} {98}},\ \bibinfo
  {pages} {030001} (\bibinfo {year} {2018})}\BibitemShut {NoStop}%
\bibitem [{\citenamefont {Ma}(2006)}]{Ma:2006km}%
  \BibitemOpen
  \bibfield  {author} {\bibinfo {author} {\bibfnamefont {E.}~\bibnamefont
  {Ma}},\ }\href {\doibase 10.1103/PhysRevD.73.077301} {\bibfield  {journal}
  {\bibinfo  {journal} {Phys. Rev.}\ }\textbf {\bibinfo {volume} {D73}},\
  \bibinfo {pages} {077301} (\bibinfo {year} {2006})},\ \Eprint
  {http://arxiv.org/abs/hep-ph/0601225} {arXiv:hep-ph/0601225 [hep-ph]}
  \BibitemShut {NoStop}%
\bibitem [{\citenamefont {Ma}(2018)}]{Ma:2018zuj}%
  \BibitemOpen
  \bibfield  {author} {\bibinfo {author} {\bibfnamefont {E.}~\bibnamefont
  {Ma}},\ }\href@noop {} {\  (\bibinfo {year} {2018})},\ \Eprint
  {http://arxiv.org/abs/1810.06506} {arXiv:1810.06506 [hep-ph]} \BibitemShut
  {NoStop}%
\bibitem [{\citenamefont {Bhattacharya}\ \emph {et~al.}(2016)\citenamefont
  {Bhattacharya}, \citenamefont {Sahoo},\ and\ \citenamefont
  {Sahu}}]{Bhattacharya:2015qpa}%
  \BibitemOpen
  \bibfield  {author} {\bibinfo {author} {\bibfnamefont {S.}~\bibnamefont
  {Bhattacharya}}, \bibinfo {author} {\bibfnamefont {N.}~\bibnamefont {Sahoo}},
  \ and\ \bibinfo {author} {\bibfnamefont {N.}~\bibnamefont {Sahu}},\ }\href
  {\doibase 10.1103/PhysRevD.93.115040} {\bibfield  {journal} {\bibinfo
  {journal} {Phys. Rev.}\ }\textbf {\bibinfo {volume} {D93}},\ \bibinfo {pages}
  {115040} (\bibinfo {year} {2016})},\ \Eprint
  {http://arxiv.org/abs/1510.02760} {arXiv:1510.02760 [hep-ph]} \BibitemShut
  {NoStop}%
\bibitem [{\citenamefont {Bhattacharya}\ \emph {et~al.}(2017)\citenamefont
  {Bhattacharya}, \citenamefont {Sahoo},\ and\ \citenamefont
  {Sahu}}]{Bhattacharya:2017sml}%
  \BibitemOpen
  \bibfield  {author} {\bibinfo {author} {\bibfnamefont {S.}~\bibnamefont
  {Bhattacharya}}, \bibinfo {author} {\bibfnamefont {N.}~\bibnamefont {Sahoo}},
  \ and\ \bibinfo {author} {\bibfnamefont {N.}~\bibnamefont {Sahu}},\ }\href
  {\doibase 10.1103/PhysRevD.96.035010} {\bibfield  {journal} {\bibinfo
  {journal} {Phys. Rev.}\ }\textbf {\bibinfo {volume} {D96}},\ \bibinfo {pages}
  {035010} (\bibinfo {year} {2017})},\ \Eprint
  {http://arxiv.org/abs/1704.03417} {arXiv:1704.03417 [hep-ph]} \BibitemShut
  {NoStop}%
\bibitem [{\citenamefont {Bhattacharya}\ \emph
  {et~al.}(2018{\natexlab{a}})\citenamefont {Bhattacharya}, \citenamefont
  {Ghosh},\ and\ \citenamefont {Sahu}}]{Bhattacharya:2018cgx}%
  \BibitemOpen
  \bibfield  {author} {\bibinfo {author} {\bibfnamefont {S.}~\bibnamefont
  {Bhattacharya}}, \bibinfo {author} {\bibfnamefont {P.}~\bibnamefont {Ghosh}},
  \ and\ \bibinfo {author} {\bibfnamefont {N.}~\bibnamefont {Sahu}},\
  }\href@noop {} {\  (\bibinfo {year} {2018}{\natexlab{a}})},\ \Eprint
  {http://arxiv.org/abs/1809.07474} {arXiv:1809.07474 [hep-ph]} \BibitemShut
  {NoStop}%
\bibitem [{\citenamefont {Bhattacharya}\ \emph
  {et~al.}(2018{\natexlab{b}})\citenamefont {Bhattacharya}, \citenamefont
  {Ghosh}, \citenamefont {Sahoo},\ and\ \citenamefont
  {Sahu}}]{Bhattacharya:2018fus}%
  \BibitemOpen
  \bibfield  {author} {\bibinfo {author} {\bibfnamefont {S.}~\bibnamefont
  {Bhattacharya}}, \bibinfo {author} {\bibfnamefont {P.}~\bibnamefont {Ghosh}},
  \bibinfo {author} {\bibfnamefont {N.}~\bibnamefont {Sahoo}}, \ and\ \bibinfo
  {author} {\bibfnamefont {N.}~\bibnamefont {Sahu}},\ }\href@noop {} {\
  (\bibinfo {year} {2018}{\natexlab{b}})},\ \Eprint
  {http://arxiv.org/abs/1812.06505} {arXiv:1812.06505 [hep-ph]} \BibitemShut
  {NoStop}%
\bibitem [{\citenamefont {Goh}\ \emph {et~al.}(2004)\citenamefont {Goh},
  \citenamefont {Mohapatra},\ and\ \citenamefont {Nasri}}]{Goh:2004fy}%
  \BibitemOpen
  \bibfield  {author} {\bibinfo {author} {\bibfnamefont {H.~S.}\ \bibnamefont
  {Goh}}, \bibinfo {author} {\bibfnamefont {R.~N.}\ \bibnamefont {Mohapatra}},
  \ and\ \bibinfo {author} {\bibfnamefont {S.}~\bibnamefont {Nasri}},\ }\href
  {\doibase 10.1103/PhysRevD.70.075022} {\bibfield  {journal} {\bibinfo
  {journal} {Phys. Rev.}\ }\textbf {\bibinfo {volume} {D70}},\ \bibinfo {pages}
  {075022} (\bibinfo {year} {2004})},\ \Eprint
  {http://arxiv.org/abs/hep-ph/0408139} {arXiv:hep-ph/0408139 [hep-ph]}
  \BibitemShut {NoStop}%
\bibitem [{\citenamefont {Caetano}\ \emph {et~al.}(2012)\citenamefont
  {Caetano}, \citenamefont {Cogollo}, \citenamefont {de~S.~Pires},\ and\
  \citenamefont {Rodrigues~da Silva}}]{Caetano:2012qc}%
  \BibitemOpen
  \bibfield  {author} {\bibinfo {author} {\bibfnamefont {W.}~\bibnamefont
  {Caetano}}, \bibinfo {author} {\bibfnamefont {D.}~\bibnamefont {Cogollo}},
  \bibinfo {author} {\bibfnamefont {C.~A.}\ \bibnamefont {de~S.~Pires}}, \ and\
  \bibinfo {author} {\bibfnamefont {P.~S.}\ \bibnamefont {Rodrigues~da
  Silva}},\ }\href {\doibase 10.1103/PhysRevD.86.055021} {\bibfield  {journal}
  {\bibinfo  {journal} {Phys. Rev.}\ }\textbf {\bibinfo {volume} {D86}},\
  \bibinfo {pages} {055021} (\bibinfo {year} {2012})},\ \Eprint
  {http://arxiv.org/abs/1206.5741} {arXiv:1206.5741 [hep-ph]} \BibitemShut
  {NoStop}%
\bibitem [{\citenamefont {Ghosh}\ \emph {et~al.}(2018)\citenamefont {Ghosh},
  \citenamefont {Ghosh}, \citenamefont {Saha},\ and\ \citenamefont
  {Shaw}}]{Ghosh:2017pxl}%
  \BibitemOpen
  \bibfield  {author} {\bibinfo {author} {\bibfnamefont {D.~K.}\ \bibnamefont
  {Ghosh}}, \bibinfo {author} {\bibfnamefont {N.}~\bibnamefont {Ghosh}},
  \bibinfo {author} {\bibfnamefont {I.}~\bibnamefont {Saha}}, \ and\ \bibinfo
  {author} {\bibfnamefont {A.}~\bibnamefont {Shaw}},\ }\href {\doibase
  10.1103/PhysRevD.97.115022} {\bibfield  {journal} {\bibinfo  {journal} {Phys.
  Rev.}\ }\textbf {\bibinfo {volume} {D97}},\ \bibinfo {pages} {115022}
  (\bibinfo {year} {2018})},\ \Eprint {http://arxiv.org/abs/1711.06062}
  {arXiv:1711.06062 [hep-ph]} \BibitemShut {NoStop}%
\bibitem [{\citenamefont {Agrawal}\ \emph {et~al.}(2018)\citenamefont
  {Agrawal}, \citenamefont {Mitra}, \citenamefont {Niyogi}, \citenamefont
  {Shil},\ and\ \citenamefont {Spannowsky}}]{Agrawal:2018pci}%
  \BibitemOpen
  \bibfield  {author} {\bibinfo {author} {\bibfnamefont {P.}~\bibnamefont
  {Agrawal}}, \bibinfo {author} {\bibfnamefont {M.}~\bibnamefont {Mitra}},
  \bibinfo {author} {\bibfnamefont {S.}~\bibnamefont {Niyogi}}, \bibinfo
  {author} {\bibfnamefont {S.}~\bibnamefont {Shil}}, \ and\ \bibinfo {author}
  {\bibfnamefont {M.}~\bibnamefont {Spannowsky}},\ }\href {\doibase
  10.1103/PhysRevD.98.015024} {\bibfield  {journal} {\bibinfo  {journal} {Phys.
  Rev.}\ }\textbf {\bibinfo {volume} {D98}},\ \bibinfo {pages} {015024}
  (\bibinfo {year} {2018})},\ \Eprint {http://arxiv.org/abs/1803.00677}
  {arXiv:1803.00677 [hep-ph]} \BibitemShut {NoStop}%
\bibitem [{\citenamefont {Choubey}\ \emph {et~al.}(2018)\citenamefont
  {Choubey}, \citenamefont {Khan}, \citenamefont {Mitra},\ and\ \citenamefont
  {Mondal}}]{Choubey:2017yyn}%
  \BibitemOpen
  \bibfield  {author} {\bibinfo {author} {\bibfnamefont {S.}~\bibnamefont
  {Choubey}}, \bibinfo {author} {\bibfnamefont {S.}~\bibnamefont {Khan}},
  \bibinfo {author} {\bibfnamefont {M.}~\bibnamefont {Mitra}}, \ and\ \bibinfo
  {author} {\bibfnamefont {S.}~\bibnamefont {Mondal}},\ }\href {\doibase
  10.1140/epjc/s10052-018-5785-1} {\bibfield  {journal} {\bibinfo  {journal}
  {Eur. Phys. J.}\ }\textbf {\bibinfo {volume} {C78}},\ \bibinfo {pages} {302}
  (\bibinfo {year} {2018})},\ \Eprint {http://arxiv.org/abs/1711.08888}
  {arXiv:1711.08888 [hep-ph]} \BibitemShut {NoStop}%
\bibitem [{\citenamefont {Dedes}\ and\ \citenamefont
  {Karamitros}(2014)}]{Dedes:2014hga}%
  \BibitemOpen
  \bibfield  {author} {\bibinfo {author} {\bibfnamefont {A.}~\bibnamefont
  {Dedes}}\ and\ \bibinfo {author} {\bibfnamefont {D.}~\bibnamefont
  {Karamitros}},\ }\href {\doibase 10.1103/PhysRevD.89.115002} {\bibfield
  {journal} {\bibinfo  {journal} {Phys. Rev.}\ }\textbf {\bibinfo {volume}
  {D89}},\ \bibinfo {pages} {115002} (\bibinfo {year} {2014})},\ \Eprint
  {http://arxiv.org/abs/1403.7744} {arXiv:1403.7744 [hep-ph]} \BibitemShut
  {NoStop}%
\bibitem [{\citenamefont {Arhrib}\ \emph {et~al.}(2011)\citenamefont {Arhrib},
  \citenamefont {Benbrik}, \citenamefont {Chabab}, \citenamefont {Moultaka},
  \citenamefont {Peyranere}, \citenamefont {Rahili},\ and\ \citenamefont
  {Ramadan}}]{Arhrib:2011uy}%
  \BibitemOpen
  \bibfield  {author} {\bibinfo {author} {\bibfnamefont {A.}~\bibnamefont
  {Arhrib}}, \bibinfo {author} {\bibfnamefont {R.}~\bibnamefont {Benbrik}},
  \bibinfo {author} {\bibfnamefont {M.}~\bibnamefont {Chabab}}, \bibinfo
  {author} {\bibfnamefont {G.}~\bibnamefont {Moultaka}}, \bibinfo {author}
  {\bibfnamefont {M.~C.}\ \bibnamefont {Peyranere}}, \bibinfo {author}
  {\bibfnamefont {L.}~\bibnamefont {Rahili}}, \ and\ \bibinfo {author}
  {\bibfnamefont {J.}~\bibnamefont {Ramadan}},\ }\href {\doibase
  10.1103/PhysRevD.84.095005} {\bibfield  {journal} {\bibinfo  {journal} {Phys.
  Rev.}\ }\textbf {\bibinfo {volume} {D84}},\ \bibinfo {pages} {095005}
  (\bibinfo {year} {2011})},\ \Eprint {http://arxiv.org/abs/1105.1925}
  {arXiv:1105.1925 [hep-ph]} \BibitemShut {NoStop}%
\bibitem [{\citenamefont {Kannike}(2012)}]{Kannike:2012pe}%
  \BibitemOpen
  \bibfield  {author} {\bibinfo {author} {\bibfnamefont {K.}~\bibnamefont
  {Kannike}},\ }\href {\doibase 10.1140/epjc/s10052-012-2093-z} {\bibfield
  {journal} {\bibinfo  {journal} {Eur. Phys. J.}\ }\textbf {\bibinfo {volume}
  {C72}},\ \bibinfo {pages} {2093} (\bibinfo {year} {2012})},\ \Eprint
  {http://arxiv.org/abs/1205.3781} {arXiv:1205.3781 [hep-ph]} \BibitemShut
  {NoStop}%
\bibitem [{\citenamefont {Aaboud}\ \emph {et~al.}(2018)\citenamefont {Aaboud}
  \emph {et~al.}}]{Aaboud:2017qph}%
  \BibitemOpen
  \bibfield  {author} {\bibinfo {author} {\bibfnamefont {M.}~\bibnamefont
  {Aaboud}} \emph {et~al.} (\bibinfo {collaboration} {ATLAS}),\ }\href
  {\doibase 10.1140/EPJC/S10052-018-5661-Z, 10.1140/epjc/s10052-018-5661-z}
  {\bibfield  {journal} {\bibinfo  {journal} {Eur. Phys. J.}\ }\textbf
  {\bibinfo {volume} {C78}},\ \bibinfo {pages} {199} (\bibinfo {year}
  {2018})},\ \Eprint {http://arxiv.org/abs/1710.09748} {arXiv:1710.09748
  [hep-ex]} \BibitemShut {NoStop}%
\bibitem [{\citenamefont {Das}\ and\ \citenamefont
  {Santamaria}(2016)}]{Das:2016bir}%
  \BibitemOpen
  \bibfield  {author} {\bibinfo {author} {\bibfnamefont {D.}~\bibnamefont
  {Das}}\ and\ \bibinfo {author} {\bibfnamefont {A.}~\bibnamefont
  {Santamaria}},\ }\href {\doibase 10.1103/PhysRevD.94.015015} {\bibfield
  {journal} {\bibinfo  {journal} {Phys. Rev.}\ }\textbf {\bibinfo {volume}
  {D94}},\ \bibinfo {pages} {015015} (\bibinfo {year} {2016})},\ \Eprint
  {http://arxiv.org/abs/1604.08099} {arXiv:1604.08099 [hep-ph]} \BibitemShut
  {NoStop}%
\bibitem [{\citenamefont {Melfo}\ \emph {et~al.}(2012)\citenamefont {Melfo},
  \citenamefont {Nemevsek}, \citenamefont {Nesti}, \citenamefont {Senjanovic},\
  and\ \citenamefont {Zhang}}]{Melfo:2011nx}%
  \BibitemOpen
  \bibfield  {author} {\bibinfo {author} {\bibfnamefont {A.}~\bibnamefont
  {Melfo}}, \bibinfo {author} {\bibfnamefont {M.}~\bibnamefont {Nemevsek}},
  \bibinfo {author} {\bibfnamefont {F.}~\bibnamefont {Nesti}}, \bibinfo
  {author} {\bibfnamefont {G.}~\bibnamefont {Senjanovic}}, \ and\ \bibinfo
  {author} {\bibfnamefont {Y.}~\bibnamefont {Zhang}},\ }\href {\doibase
  10.1103/PhysRevD.85.055018} {\bibfield  {journal} {\bibinfo  {journal} {Phys.
  Rev.}\ }\textbf {\bibinfo {volume} {D85}},\ \bibinfo {pages} {055018}
  (\bibinfo {year} {2012})},\ \Eprint {http://arxiv.org/abs/1108.4416}
  {arXiv:1108.4416 [hep-ph]} \BibitemShut {NoStop}%
\bibitem [{\citenamefont {Bhupal~Dev}\ and\ \citenamefont
  {Zhang}(2018)}]{Dev:2018kpa}%
  \BibitemOpen
  \bibfield  {author} {\bibinfo {author} {\bibfnamefont {P.~S.}\ \bibnamefont
  {Bhupal~Dev}}\ and\ \bibinfo {author} {\bibfnamefont {Y.}~\bibnamefont
  {Zhang}},\ }\href {\doibase 10.1007/JHEP10(2018)199} {\bibfield  {journal}
  {\bibinfo  {journal} {JHEP}\ }\textbf {\bibinfo {volume} {10}},\ \bibinfo
  {pages} {199} (\bibinfo {year} {2018})},\ \Eprint
  {http://arxiv.org/abs/1808.00943} {arXiv:1808.00943 [hep-ph]} \BibitemShut
  {NoStop}%
\bibitem [{\citenamefont {Ma}\ and\ \citenamefont {Sarkar}(1998)}]{Ma:1998dx}%
  \BibitemOpen
  \bibfield  {author} {\bibinfo {author} {\bibfnamefont {E.}~\bibnamefont
  {Ma}}\ and\ \bibinfo {author} {\bibfnamefont {U.}~\bibnamefont {Sarkar}},\
  }\href {\doibase 10.1103/PhysRevLett.80.5716} {\bibfield  {journal} {\bibinfo
   {journal} {Phys. Rev. Lett.}\ }\textbf {\bibinfo {volume} {80}},\ \bibinfo
  {pages} {5716} (\bibinfo {year} {1998})},\ \Eprint
  {http://arxiv.org/abs/hep-ph/9802445} {arXiv:hep-ph/9802445 [hep-ph]}
  \BibitemShut {NoStop}%
\bibitem [{\citenamefont {Ade}\ \emph {et~al.}(2016)\citenamefont {Ade} \emph
  {et~al.}}]{Ade:2015xua}%
  \BibitemOpen
  \bibfield  {author} {\bibinfo {author} {\bibfnamefont {P.~A.~R.}\
  \bibnamefont {Ade}} \emph {et~al.} (\bibinfo {collaboration} {Planck}),\
  }\href {\doibase 10.1051/0004-6361/201525830} {\bibfield  {journal} {\bibinfo
   {journal} {Astron. Astrophys.}\ }\textbf {\bibinfo {volume} {594}},\
  \bibinfo {pages} {A13} (\bibinfo {year} {2016})},\ \Eprint
  {http://arxiv.org/abs/1502.01589} {arXiv:1502.01589 [astro-ph.CO]}
  \BibitemShut {NoStop}%
\bibitem [{\citenamefont {Khachatryan}\ \emph {et~al.}(2017)\citenamefont
  {Khachatryan} \emph {et~al.}}]{Khachatryan:2016whc}%
  \BibitemOpen
  \bibfield  {author} {\bibinfo {author} {\bibfnamefont {V.}~\bibnamefont
  {Khachatryan}} \emph {et~al.} (\bibinfo {collaboration} {CMS}),\ }\href
  {\doibase 10.1007/JHEP02(2017)135} {\bibfield  {journal} {\bibinfo  {journal}
  {JHEP}\ }\textbf {\bibinfo {volume} {02}},\ \bibinfo {pages} {135} (\bibinfo
  {year} {2017})},\ \Eprint {http://arxiv.org/abs/1610.09218} {arXiv:1610.09218
  [hep-ex]} \BibitemShut {NoStop}%
\bibitem [{\citenamefont {Semenov}(2009)}]{Semenov:2008jy}%
  \BibitemOpen
  \bibfield  {author} {\bibinfo {author} {\bibfnamefont {A.}~\bibnamefont
  {Semenov}},\ }\href {\doibase 10.1016/j.cpc.2008.10.012} {\bibfield
  {journal} {\bibinfo  {journal} {Comput. Phys. Commun.}\ }\textbf {\bibinfo
  {volume} {180}},\ \bibinfo {pages} {431} (\bibinfo {year} {2009})},\ \Eprint
  {http://arxiv.org/abs/0805.0555} {arXiv:0805.0555 [hep-ph]} \BibitemShut
  {NoStop}%
\bibitem [{\citenamefont {Belanger}\ \emph {et~al.}(2002)\citenamefont
  {Belanger}, \citenamefont {Boudjema}, \citenamefont {Pukhov},\ and\
  \citenamefont {Semenov}}]{Belanger:2001fz}%
  \BibitemOpen
  \bibfield  {author} {\bibinfo {author} {\bibfnamefont {G.}~\bibnamefont
  {Belanger}}, \bibinfo {author} {\bibfnamefont {F.}~\bibnamefont {Boudjema}},
  \bibinfo {author} {\bibfnamefont {A.}~\bibnamefont {Pukhov}}, \ and\ \bibinfo
  {author} {\bibfnamefont {A.}~\bibnamefont {Semenov}},\ }\href {\doibase
  10.1016/S0010-4655(02)00596-9} {\bibfield  {journal} {\bibinfo  {journal}
  {Comput. Phys. Commun.}\ }\textbf {\bibinfo {volume} {149}},\ \bibinfo
  {pages} {103} (\bibinfo {year} {2002})},\ \Eprint
  {http://arxiv.org/abs/hep-ph/0112278} {arXiv:hep-ph/0112278 [hep-ph]}
  \BibitemShut {NoStop}%
\bibitem [{\citenamefont {Tucker-Smith}\ and\ \citenamefont
  {Weiner}(2001)}]{TuckerSmith:2001hy}%
  \BibitemOpen
  \bibfield  {author} {\bibinfo {author} {\bibfnamefont {D.}~\bibnamefont
  {Tucker-Smith}}\ and\ \bibinfo {author} {\bibfnamefont {N.}~\bibnamefont
  {Weiner}},\ }\href {\doibase 10.1103/PhysRevD.64.043502} {\bibfield
  {journal} {\bibinfo  {journal} {Phys. Rev.}\ }\textbf {\bibinfo {volume}
  {D64}},\ \bibinfo {pages} {043502} (\bibinfo {year} {2001})},\ \Eprint
  {http://arxiv.org/abs/hep-ph/0101138} {arXiv:hep-ph/0101138 [hep-ph]}
  \BibitemShut {NoStop}%
\bibitem [{\citenamefont {Duerr}\ \emph {et~al.}(2016)\citenamefont {Duerr},
  \citenamefont {Fileviez~Pérez},\ and\ \citenamefont
  {Smirnov}}]{Duerr:2015aka}%
  \BibitemOpen
  \bibfield  {author} {\bibinfo {author} {\bibfnamefont {M.}~\bibnamefont
  {Duerr}}, \bibinfo {author} {\bibfnamefont {P.}~\bibnamefont
  {Fileviez~Pérez}}, \ and\ \bibinfo {author} {\bibfnamefont {J.}~\bibnamefont
  {Smirnov}},\ }\href {\doibase 10.1007/JHEP06(2016)152} {\bibfield  {journal}
  {\bibinfo  {journal} {JHEP}\ }\textbf {\bibinfo {volume} {06}},\ \bibinfo
  {pages} {152} (\bibinfo {year} {2016})},\ \Eprint
  {http://arxiv.org/abs/1509.04282} {arXiv:1509.04282 [hep-ph]} \BibitemShut
  {NoStop}%
\bibitem [{\citenamefont {Durr}\ \emph {et~al.}(2016)\citenamefont {Durr} \emph
  {et~al.}}]{Durr:2015dna}%
  \BibitemOpen
  \bibfield  {author} {\bibinfo {author} {\bibfnamefont {S.}~\bibnamefont
  {Durr}} \emph {et~al.},\ }\href {\doibase 10.1103/PhysRevLett.116.172001}
  {\bibfield  {journal} {\bibinfo  {journal} {Phys. Rev. Lett.}\ }\textbf
  {\bibinfo {volume} {116}},\ \bibinfo {pages} {172001} (\bibinfo {year}
  {2016})},\ \Eprint {http://arxiv.org/abs/1510.08013} {arXiv:1510.08013
  [hep-lat]} \BibitemShut {NoStop}%
\bibitem [{\citenamefont {Achard}\ \emph {et~al.}(2001)\citenamefont {Achard}
  \emph {et~al.}}]{Achard:2001qw}%
  \BibitemOpen
  \bibfield  {author} {\bibinfo {author} {\bibfnamefont {P.}~\bibnamefont
  {Achard}} \emph {et~al.} (\bibinfo {collaboration} {L3}),\ }\href {\doibase
  10.1016/S0370-2693(01)01005-X} {\bibfield  {journal} {\bibinfo  {journal}
  {Phys. Lett.}\ }\textbf {\bibinfo {volume} {B517}},\ \bibinfo {pages} {75}
  (\bibinfo {year} {2001})},\ \Eprint {http://arxiv.org/abs/hep-ex/0107015}
  {arXiv:hep-ex/0107015 [hep-ex]} \BibitemShut {NoStop}%
\bibitem [{\citenamefont {Belyaev}\ \emph {et~al.}(2013)\citenamefont
  {Belyaev}, \citenamefont {Christensen},\ and\ \citenamefont
  {Pukhov}}]{Belyaev:2012qa}%
  \BibitemOpen
  \bibfield  {author} {\bibinfo {author} {\bibfnamefont {A.}~\bibnamefont
  {Belyaev}}, \bibinfo {author} {\bibfnamefont {N.~D.}\ \bibnamefont
  {Christensen}}, \ and\ \bibinfo {author} {\bibfnamefont {A.}~\bibnamefont
  {Pukhov}},\ }\href {\doibase 10.1016/j.cpc.2013.01.014} {\bibfield  {journal}
  {\bibinfo  {journal} {Comput. Phys. Commun.}\ }\textbf {\bibinfo {volume}
  {184}},\ \bibinfo {pages} {1729} (\bibinfo {year} {2013})},\ \Eprint
  {http://arxiv.org/abs/1207.6082} {arXiv:1207.6082 [hep-ph]} \BibitemShut
  {NoStop}%
\bibitem [{\citenamefont {Sjostrand}\ \emph {et~al.}(2006)\citenamefont
  {Sjostrand}, \citenamefont {Mrenna},\ and\ \citenamefont
  {Skands}}]{Sjostrand:2006za}%
  \BibitemOpen
  \bibfield  {author} {\bibinfo {author} {\bibfnamefont {T.}~\bibnamefont
  {Sjostrand}}, \bibinfo {author} {\bibfnamefont {S.}~\bibnamefont {Mrenna}}, \
  and\ \bibinfo {author} {\bibfnamefont {P.~Z.}\ \bibnamefont {Skands}},\
  }\href {\doibase 10.1088/1126-6708/2006/05/026} {\bibfield  {journal}
  {\bibinfo  {journal} {JHEP}\ }\textbf {\bibinfo {volume} {05}},\ \bibinfo
  {pages} {026} (\bibinfo {year} {2006})},\ \Eprint
  {http://arxiv.org/abs/hep-ph/0603175} {arXiv:hep-ph/0603175 [hep-ph]}
  \BibitemShut {NoStop}%
\bibitem [{\citenamefont {Placakyte}(2011)}]{Placakyte:2011az}%
  \BibitemOpen
  \bibfield  {author} {\bibinfo {author} {\bibfnamefont {R.}~\bibnamefont
  {Placakyte}},\ }in\ \href@noop {} {\emph {\bibinfo {booktitle} {{Proceedings,
  31st International Conference on Physics in collisions (PIC 2011): Vancouver,
  Canada, August 28-September 1, 2011}}}}\ (\bibinfo {year} {2011})\ \Eprint
  {http://arxiv.org/abs/1111.5452} {arXiv:1111.5452 [hep-ph]} \BibitemShut
  {NoStop}%
\bibitem [{\citenamefont {Alwall}\ \emph {et~al.}(2014)\citenamefont {Alwall},
  \citenamefont {Frederix}, \citenamefont {Frixione}, \citenamefont {Hirschi},
  \citenamefont {Maltoni}, \citenamefont {Mattelaer}, \citenamefont {Shao},
  \citenamefont {Stelzer}, \citenamefont {Torrielli},\ and\ \citenamefont
  {Zaro}}]{Alwall:2014hca}%
  \BibitemOpen
  \bibfield  {author} {\bibinfo {author} {\bibfnamefont {J.}~\bibnamefont
  {Alwall}}, \bibinfo {author} {\bibfnamefont {R.}~\bibnamefont {Frederix}},
  \bibinfo {author} {\bibfnamefont {S.}~\bibnamefont {Frixione}}, \bibinfo
  {author} {\bibfnamefont {V.}~\bibnamefont {Hirschi}}, \bibinfo {author}
  {\bibfnamefont {F.}~\bibnamefont {Maltoni}}, \bibinfo {author} {\bibfnamefont
  {O.}~\bibnamefont {Mattelaer}}, \bibinfo {author} {\bibfnamefont {H.~S.}\
  \bibnamefont {Shao}}, \bibinfo {author} {\bibfnamefont {T.}~\bibnamefont
  {Stelzer}}, \bibinfo {author} {\bibfnamefont {P.}~\bibnamefont {Torrielli}},
  \ and\ \bibinfo {author} {\bibfnamefont {M.}~\bibnamefont {Zaro}},\ }\href
  {\doibase 10.1007/JHEP07(2014)079} {\bibfield  {journal} {\bibinfo  {journal}
  {JHEP}\ }\textbf {\bibinfo {volume} {07}},\ \bibinfo {pages} {079} (\bibinfo
  {year} {2014})},\ \Eprint {http://arxiv.org/abs/1405.0301} {arXiv:1405.0301
  [hep-ph]} \BibitemShut {NoStop}%
\bibitem [{\citenamefont {Kamon}\ \emph {et~al.}(2017)\citenamefont {Kamon},
  \citenamefont {Ko},\ and\ \citenamefont {Li}}]{Kamon:2017yfx}%
  \BibitemOpen
  \bibfield  {author} {\bibinfo {author} {\bibfnamefont {T.}~\bibnamefont
  {Kamon}}, \bibinfo {author} {\bibfnamefont {P.}~\bibnamefont {Ko}}, \ and\
  \bibinfo {author} {\bibfnamefont {J.}~\bibnamefont {Li}},\ }\href {\doibase
  10.1140/epjc/s10052-017-5240-8} {\bibfield  {journal} {\bibinfo  {journal}
  {Eur. Phys. J.}\ }\textbf {\bibinfo {volume} {C77}},\ \bibinfo {pages} {652}
  (\bibinfo {year} {2017})},\ \Eprint {http://arxiv.org/abs/1705.02149}
  {arXiv:1705.02149 [hep-ph]} \BibitemShut {NoStop}%
\bibitem [{\citenamefont {Behnke}\ \emph {et~al.}(2013)\citenamefont {Behnke},
  \citenamefont {Brau}, \citenamefont {Foster}, \citenamefont {Fuster},
  \citenamefont {Harrison}, \citenamefont {Paterson}, \citenamefont {Peskin},
  \citenamefont {Stanitzki}, \citenamefont {Walker},\ and\ \citenamefont
  {Yamamoto}}]{Behnke:2013xla}%
  \BibitemOpen
  \bibfield  {author} {\bibinfo {author} {\bibfnamefont {T.}~\bibnamefont
  {Behnke}}, \bibinfo {author} {\bibfnamefont {J.~E.}\ \bibnamefont {Brau}},
  \bibinfo {author} {\bibfnamefont {B.}~\bibnamefont {Foster}}, \bibinfo
  {author} {\bibfnamefont {J.}~\bibnamefont {Fuster}}, \bibinfo {author}
  {\bibfnamefont {M.}~\bibnamefont {Harrison}}, \bibinfo {author}
  {\bibfnamefont {J.~M.}\ \bibnamefont {Paterson}}, \bibinfo {author}
  {\bibfnamefont {M.}~\bibnamefont {Peskin}}, \bibinfo {author} {\bibfnamefont
  {M.}~\bibnamefont {Stanitzki}}, \bibinfo {author} {\bibfnamefont
  {N.}~\bibnamefont {Walker}}, \ and\ \bibinfo {author} {\bibfnamefont
  {H.}~\bibnamefont {Yamamoto}},\ }\href@noop {} {\  (\bibinfo {year}
  {2013})},\ \Eprint {http://arxiv.org/abs/1306.6327} {arXiv:1306.6327
  [physics.acc-ph]} \BibitemShut {NoStop}%
\end{thebibliography}%


\end{document}